\documentclass[journal=jpcbfk,manuscript=article]{achemso}

\usepackage{amsmath}
\usepackage{soul}
\usepackage[colorlinks,allcolors=blue]{hyperref} 
\usepackage{braket}
\usepackage{numprint}

\newcommand{\im}{\textrm{i}}

\SectionNumbersOn

\author{Alessia Valzelli}
\affiliation[]{Dipartimento di Ingegneria dell’Informazione, Università degli Studi di Firenze, 50139 Firenze, Italy}
\email{alessia.valzelli@unifi.it}
\alsoaffiliation[]{Dipartimento di Fisica e Astronomia, Università degli Studi di Firenze e CSDC, 50019 Sesto Fiorentino, Italy}
\alsoaffiliation[]{Istituto Nazionale di Fisica Nucleare, Sezione di Firenze, 50019 Sesto Fiorentino, Italy}

\author{Francesco Mattiotti}
\affiliation[]{Theoretische Physik, Universit\"at des Saarlandes, D-66123 Saarbr\"ucken, Germany}

\author{Jianshu Cao}
\affiliation[]{Department of Chemistry, Massachusetts Institute of Technology, 77 Massachusetts Avenue, Cambridge, Massachusetts 02139, USA}

\author{Giuseppe Luca Celardo}
\affiliation[]{Dipartimento di Fisica e Astronomia, Università degli Studi di Firenze e CSDC, 50019 Sesto Fiorentino, Italy}
\alsoaffiliation[]{Istituto Nazionale di Fisica Nucleare, Sezione di Firenze, 50019 Sesto Fiorentino, Italy}
\alsoaffiliation[]{European Laboratory for Non-Linear Spectroscopy (LENS), Università degli Studi di Firenze, 50019 Sesto Fiorentino, Italy}

\title[Structure-functionality]{Relation between structure and functionality in photosynthetic antenna complex of green sulfur bacteria: efficiency under natural sunlight pumping}

\abbreviations{GSB,PB,BChl,2LS,RC,TDM,FMO,LHI,LHII,NHH,HH,DH}

\mciteErrorOnUnknownfalse
\begin{document}

\begin{tocentry}

    \centering
    \includegraphics{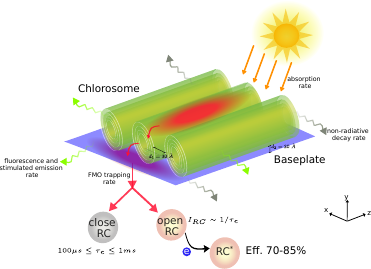}

    \label{For Table of Contents Only}

\end{tocentry}

\begin{abstract}
Large-scale simulations of  light-matter interaction in natural photosynthetic antenna complexes of the \textit{Chlorobium Tepidum} green sulfur bacteria (GSB) containing more than one hundred thousand chlorophyll molecules, comparable with natural size, have been performed. Here we have modeled the entire process of the exciton energy transfer, from sunlight absorption to exciton trapping in the reaction centers (RCs) in presence of a thermal bath. The energy transfer has been analyzed using  the radiative non-Hermitian Hamiltonian and solving the rate equations for the populations. Sunlight pumping has been modeled  as black-body radiation at $T=5800$~K, with an attenuation factor that takes the Sun-Earth distance into account. Cylindrical structures typical of GSB antenna complexes, and the dimeric baseplate have been considered. The maximal antenna size, comparable to natural size, includes three adjacent, 4-walls concentric cylinders $1485.7\ \mbox{\AA}$ long  arranged over a dimeric baseplate of dimensions $3075 \ \mbox{\AA}$  $\times$ $1148 \ \mbox{\AA}$, for  an overall number of molecules greater than $10^5$. Our analysis shows that under natural sunlight, in photosynthetic  antennae of GSB the number of excitations reaching the RC per unit time matches the RC closure rate and the internal efficiency shows values close to $\sim 80\%$.  We also considered cylindrical structures where the orientation of the dipoles does not reflect the natural one. Specifically, we vary continuously the angle of the transition dipole with respect to the cylinder main axis, focusing on  the case where all dipoles are parallel to the cylinder axis. We also consider the important case where the dipoles are randomly oriented. In all cases the light-harvesting efficiency is lower than in the natural structure, showing the high sensitivity of light harvesting to the specific orientation of the dipole moments.  
Our results allow for a better understanding of the relationship between structure and functionality in natural photosynthetic antennae of Green sulfur bacteria and could drive the design of efficient light-harvesting devices.    
\end{abstract}

\section{Introduction}
\label{m:intro}
Photosynthesis is a fundamental process able to  capture  sunlight and convert it into biochemical energy used to drive cellular processes~\cite{blankenship2021molecular}. In this manuscript we model the process of sunlight absorption and energy transfer in the entire antenna complex of a species of photosynthetic anaerobic bacteria: the \textit{Chlorobium Tepidum} green sulfur bacteria (GSB).
GSB use sunlight as their main source of energy. They are among the most efficient systems able to harvest sunlight in deep murky waters, where incident light levels are reduced much beyond the already dilute level on land~\cite{strumpfer2012quantum,schulten}. They are even able 
to perform photosynthesis with geothermal radiation from deep-sea hydrothermal vents at about 400$^\circ$C~\cite{greenPNAS}.   

The GSB antenna complexes are composed of a network of  bacteriochlorophyll molecules (BChls \textit{a}, \textit{c}, \textit{d} or \textit{e})~\cite{chew2007}.
In GSB BChl molecules are typically modeled as two-level systems (2LS). To each 2LS a transition dipole moment (TDM) is associated, which determines its coupling with both the electromagnetic field and other chlorophyll molecules. 
Photosynthesis in GSB involves chlorophyll pigments tightly packed in light-harvesting systems with (mostly) cylindrical shapes, known as chlorosomes. 
The geometry adopted for the GSB light-harvesting complexes is well-established in the literature. Specifically, the pigment organization and orientations within the GSB chlorosome have been extensively studied using infrared and resonance Raman spectroscopy, solid-state NMR, and cryo-EM. These studies reveal that pigments assemble into rod-like cylindrical aggregates characterized by lateral lamellae~\cite{linnanto,gunther2016,linnanto2008investigation,oostergetel2007long,pvsenvcik2004lamellar,pvsenvcik2006internal,psencik2009structure}. 
In nature these structures typically range in size from $1000$ to $2000 \ \mbox{\AA}$  in length, with widths and depths varying between $600$ and $1000 \ \mbox{\AA}$, and they can contain between $50000$ and $250000$ BChl \textit{c}~\cite{cohen1964fine,orf2013chlorosome,staehelin1978visualization}. 

Sunlight absorbed by the chlorosome is funneled  to other complexes, such as the baseplate (BPL) and the Fenna-Matthews-Olson (FMO) trimer
complexes, and finally to the reaction centers (RCs)~\cite{strumpfer2012quantum, guzik}, where charge separation occurs, a process which precedes and drives all other photosynthetic steps~\cite{strumpfer2012quantum,schulten,photo,photoT,grad1988radiative,mukamel,srfmo,srrc}. Once the RC receives an excitation, it produces a charge-separated state and electron transport through the RC begins. A reaction center is said to be in an open state before the excitation reaches it. Once an excitation reaches it and charge separation occurs, the RC goes in a closed state. The time needed for the RC to be open again is called the closure time (from $100~\mu\mbox{s}$ to few milliseconds~\cite{hauska2001,van1998transient}). The closure time defines the frequency at which each RC is able to process an excitation. 

\begin{figure}[!ht]
    \centering
    \includegraphics[width=\columnwidth]{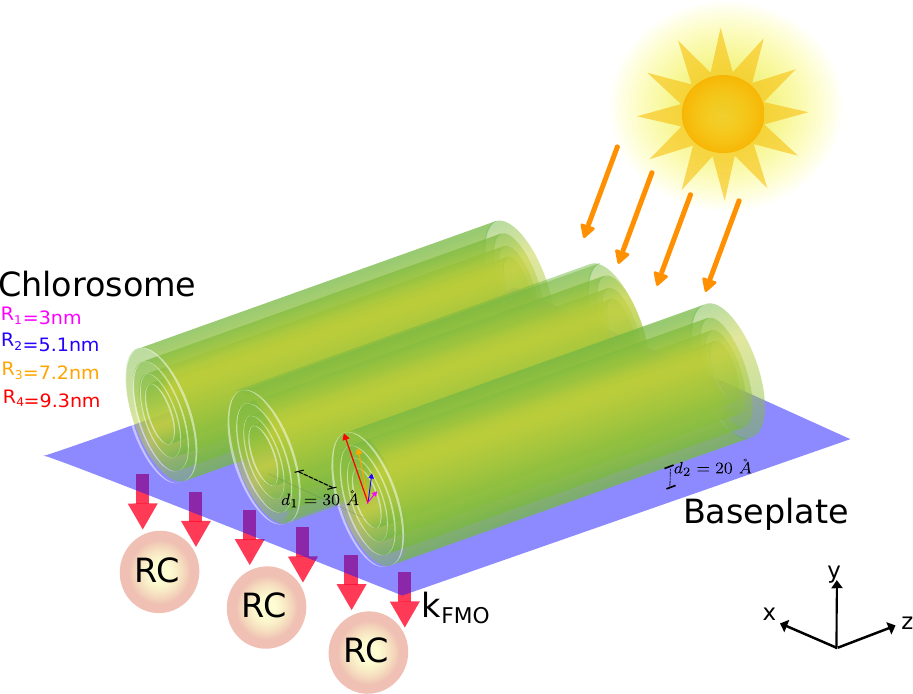}
    \caption{{\it Architecture of GSB light-harvesting unit under natural sunlight.}  The chlorosome of GSB comprising three adjacent concentric cylinders and the dimeric baseplate (BPL) have been represented. Each aggregate is made of four concentric MT model cylinders with radii of $30$, $51$, $72$ and $93 \ \mbox{\AA}$  and containing, respectively, $30$, $51$, $72$ and $93$ dipoles per ring.  The entire chlorosome contains $132840$ BChl \textit{c} with a length $L=1485.7 \ \mbox{\AA}$ and the distance between two adjacent concentric cylinders is $d_1 = 30 \ \mbox{\AA}$. Under the chlorosome a dimeric baseplate with dimesnsions $ 1147.5 \times 3075.3   \ \mbox{\AA}^2$ and comprising $3350$ BChl \textit{a} has been represented. The distance between the cylinder and the baseplate is set to $d_2=20  \ \mbox{\AA}$, according to Refs.~\citenum{guzik,linnanto}. Finally the energy transfer from the baseplate to the RCs is mediated by the $k_{FMO}$ rate, represented by red arrows connecting the baseplate to the RCs.}
    \label{chlorosome}
\end{figure}

Fig.~\ref{chlorosome} shows the model we used to analyze the entire light-harvesting unit in GSB. 
The energy transfer process from the light-harvesting
complex to the RC has a very high (near-$80\%$) internal efficiency in these complexes~\cite{chen2020architecture,guzik, dostal2016situ}.  A possible origin of this incredible ability to utilize weak sources of incoherent light and funnel the collected energy to specific molecular aggregates could be brought back to the high level of symmetry and hierarchical organization characterizing the antenna complexes of bacterial photosynthetic organisms~\cite{baghbanzadeh2016geometry,kassal2}.

The basic components of the photosynthetic antenna complexes of anaerobic bacteria have been widely studied
both theoretically and experimentally in Refs.~\citenum{saer2017light,eltsova2016effect,chew2007,linnanto,malina2021superradiance,molina2016superradiance,fujita2014theoretical}. 
Due to the symmetric arrangement of BChl molecules, these structures display bright (superradiant) and dark (subradiant) states in their single-excitation manifold~\cite{monshouwer1997superradiance,strumpfer2012quantum}. Bright states are characterized by a giant transition dipole moment (much larger than the single-molecule dipole moment), while dark states exhibit a significantly smaller transition dipole moment compared to that of a single molecule.

As already demonstrated in literature~\cite{strumpfer2012excited}, bright states robust to disorder arise due to the strong coupling between BChl molecules within the aggregates. Within each aggregate (chlorosome and baseplate), excitations are spread coherently due to the small distances between nearest-neighbor BChl molecules (a few $\mbox{\AA}$ngstroms), while between pigment groups at a distance of about a few nanometers, electronic excitation is shared incoherently through multi-chromophoric F\"orster resonant energy transfer (MC-FRET).

In this study, we thus conduct a large-scale analysis of sunlight absorption and exciton energy transfer in GSB photosynthetic complexes. Specifically, we examine a model of the GSB chlorosome comprising more than 10$^5$ chlorophyll molecules arranged on three adjacent concentric cylinders (the chlorosome)  above a two-dimensional dimeric baseplate. 
In order to understand the relation between shape and functionality, the orientations of BChl dipole moments in the cylindrical structures present in the chlorosome have been modified. The light-harvesting efficiency of the  natural geometry has been compared with the light-harvesting efficiency of these
mathematical models. Our findings shed new light on how the natural structure of the chlorosome is able to support an efficient energy transfer, even in presence of thermal noise and static disorder.

\section{Models: the geometry of the system}
\label{m:models}
\begin{figure}[!bht]
    \centering
    \includegraphics[width=\columnwidth]{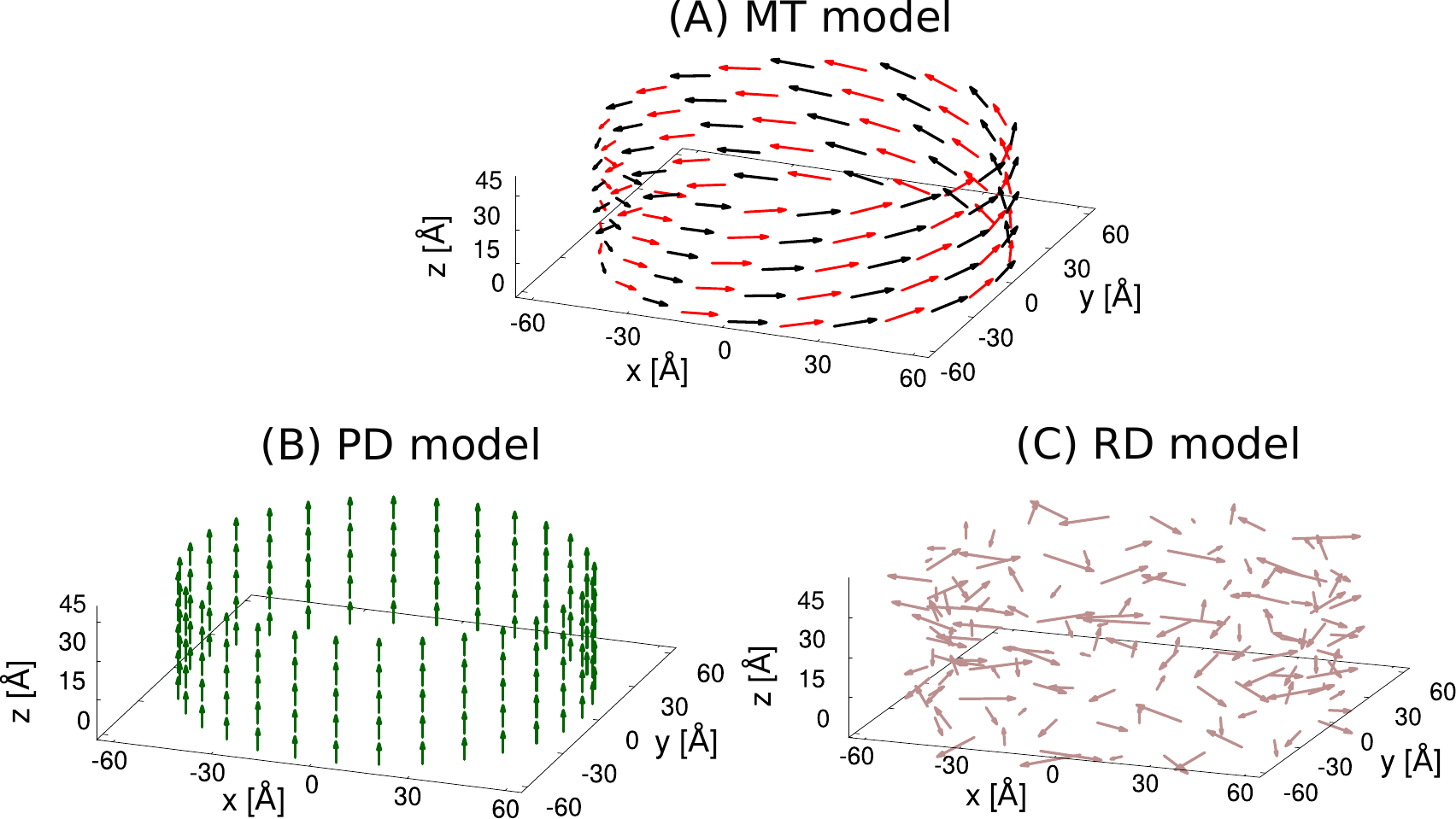}
    \caption{\emph{Section of the different cylindrical models.} Panel (A) shows the MT model, the structure derived by genetic modification of the natural wild-type model with a radius $R=60 \ \mbox{\AA}$. In panels  (B-C)  PD and RD cylinders made of a stack of rings with the same radius $R=60 \ \mbox{\AA}$ are presented.  For the sake of clarity we show only 30 dipoles per ring instead of 60 as we consider in this paper.  For more details about the geometry see Refs.~\citenum{macroscopic,valzelli2024large}.} 
    \label{3models}
\end{figure}

Here we analyze different models of the chlorosome. 
Together with the natural cylindrical structures present in the chlorosome, we also considered mathematical cylindrical structures  where both positions and orientations of the chlorophyll transition dipole moment have been changed with respect to the natural system.  Fig.~\ref{3models} shows the three cylindrical models we analyze.  The main difference between different models lies in the dipole orientation. 
\begin{itemize}
\item \textit{Chlorobium Tepidum} bchQRU triple mutant (MT).  The MT model  can be
     thought as organized into a stack of vertical rings~\cite{gunther2016,macroscopic}, each containing 60 BChl \textit{c} molecules represented by dipoles. In panel A of Fig.~\ref{3models}  the alternation between the colors of two consecutive
dipoles on the same ring (red and black) represents those dipoles pointing inward ($\alpha = + 4^\circ$) and outward
($\alpha = - 4^\circ$) with respect to the cylinder. See Ref.~\citenum{macroscopic} for more details. 
    \item Parallel dipoles cylinder (PD). The dipoles on the PD are arranged in circles, similarly to MT type, with their direction parallel to the cylinder main axis, see Fig.~\ref{3models} panel B.
    \item Random dipoles cylinder (RD). In the RD model, see Fig.~\ref{3models} panel C, the position of the dipoles are the same as in the PD model, while the dipole orientations are randomly chosen from the unit sphere. 
\end{itemize}

The maximum number of molecules we consider in a single cylinder (MT, PD and RD) is $N=6000$ BChl \textit{c} with a maximum length of $821.7  \ \mbox{\AA}$. More information on the geometry of the MT, PD and RD cylinders can be found in Refs.~\citenum{macroscopic,valzelli2024large}. 

The structure of natural GSB light-harvesting systems can vary depending on growing conditions. In general  the chlorosome is placed on the top of a two-dimensional dimeric baseplate. The chlorosome does not include only one cylindrical structure as antenna complex, it can contain several multi-wall cylindrical structures (usually four concentric cylinders) placed adjacently above the baseplate, see Fig.~\ref{chlorosome}.  Also BChl molecules aggregated in  lateral curved lamellae can be present, depending on the growing conditions~\cite{macroscopic,valzelli2024large,hohmann2005ultrastructure,oostergetel}.

In this manuscript, in order to model  the entire light-harvesting unit, we considered 
a  chlorosome  composed by three adjacent cylindrical aggregates with four concentric rolls each, containing  $132840$ BChl \textit{c} molecules and with a length of $L=1485.7 \ \mbox{\AA}$.    Each aggregate is made of four concentric MT model cylinders with radii of $30$, $51$, $72$ and $93 \ \mbox{\AA}$ and containing, respectively, $30$, $51$, $72$ and $93$ dipoles per ring.  The distance between two adjacent concentric cylinders is $d_1 = 30 \ \mbox{\AA}$, according to Ref.~\citenum{linnanto}.

\begin{figure}[!bht]
    \centering
    \includegraphics[width=\columnwidth]{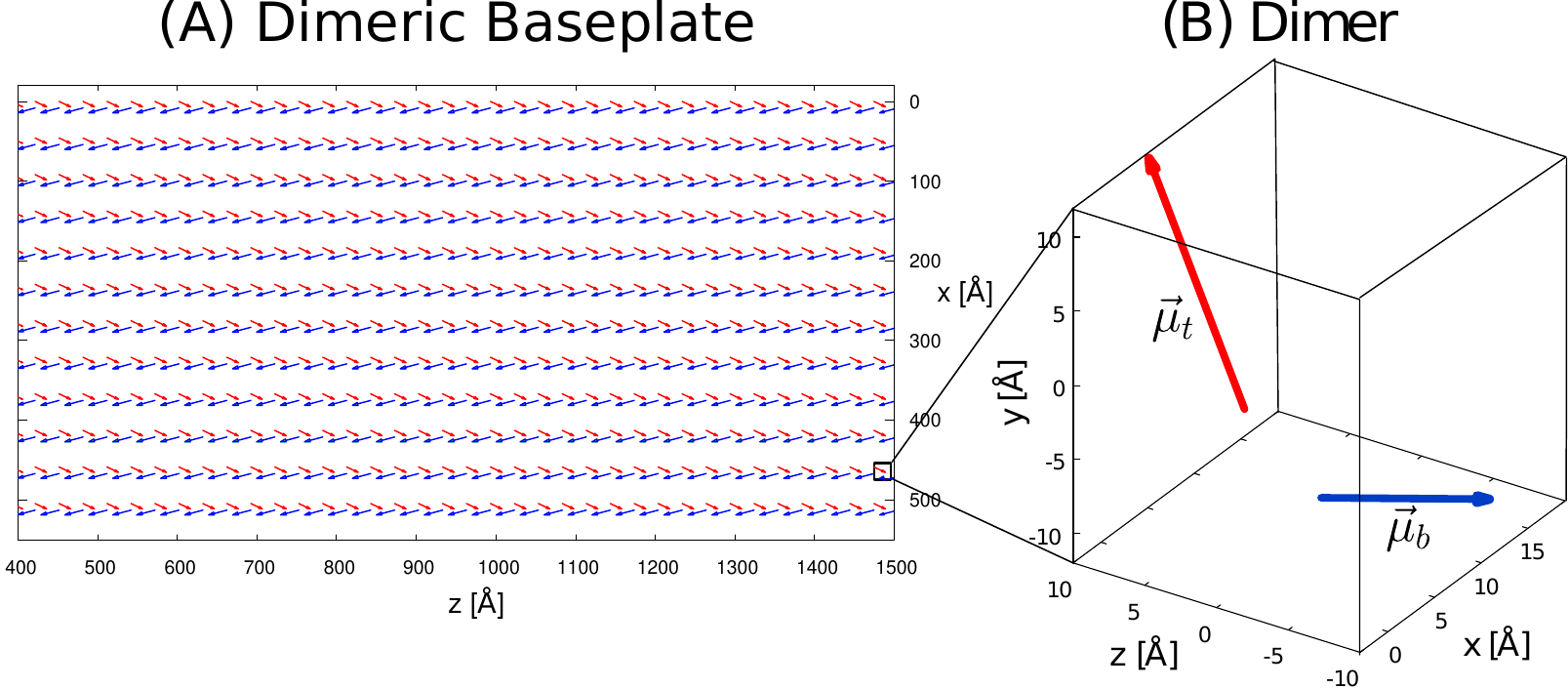}
    \caption{\textit{Representation of the dimeric baseplate (BPL) in GSB light-harvesting unit.} Panel (A): top view of a portion of the dimeric baseplate, a planar structure in the xz-plane containing  BChl \textit{a} molecules. The blue and red arrows represent the TDM associated to each BChl \textit{a} and belonging to the bottom and top layer respectively. Panel (B): zoomed-in view of the dimeric unit formed by a red arrow ($\mu_{t}$) and a blue one ($\mu_{b}$). Dipoles orientations have been found in Ref.~\citenum{guzik} and the corresponding unit vectors are given by $\hat{\mu}_{t}=(0.2795,0.7484,0.5982)$ and $\hat{\mu}_{b} =(0.2533,0.1607,-0.9533)$. The distance between two dipoles in the same dimeric unit cell is set to $12.8 \ \mbox{\AA}$ , while the distance along consecutive BChl \textit{a} on \textit{x} and \textit{z} axis are $45.9 \ \mbox{\AA}$  and $30.1 \ \mbox{\AA}$ respectively. For sake of clarity in panel B the dipole length is multiplied by a factor of $16$.  
    }
   \label{baseplate}
\end{figure}

Below the chlorosome, at a distance $d_2 = 20 \ \mbox{\AA}$, there is the dimeric baseplate.
The baseplate is a two-dimensional aggregate formed by BChl \textit{a} molecules that connects the chlorosome pigments to the RCs through the FMO proteins. It is located on the chlorosome envelope on the surface toward the cytoplasmic membranae~\cite{pedersen} and in nature its dimensions  were roughly estimated to be $500 \ \mbox{\AA}$  $\times$ $2000 \ \mbox{\AA}$~\cite{linnanto}.
Two kind of baseplates are found in literature: monomeric and dimeric~\cite{linnanto,pedersen}. The monomeric baseplate is typical of the FAP (filamentous anoxigenic phototrophs), represented by the \textit{Chloroflexux aurantiacus}, that do not show the FMO complexes~\cite{pedersen}, while the dimeric baseplate is tipical of GSB~\cite{linnanto,guzik}. Even if the microscopic structure of the baseplate has not yet been experimentally verified, a model for the dimeric baseplate lattice found in GSB has already been proposed in Refs.~\citenum{guzik,pedersen}. The lattice has two layers and two different transition dipole moments $\vec{\mu}_{t}$ and $\vec{\mu}_{b}$  are used for the top and bottom layers respectively.  
Panel A of Fig.~\ref{baseplate}  shows the arrangement of dimers in a portion of the baseplate, distinguishing between blue and red dipoles belonging respectively to the bottom and top layers, while panel B represents the dimeric unit cell comprising $\vec{\mu}_{t}$ and $\vec{\mu}_{b}$. 

After the excitation is absorbed by the BChl molecules in the cylindrical structures or in the baseplate, it  is driven to the RCs by the FMO complexes. In GSB light-harvesting systems we have 1 FMO trimer per 50~nm$^2$~\cite{guzik} and  1~RC per 100~nm$^2$ , since we have one RC every two FMO complexes.

In order to describe the light-harvesting process  in the entire photosynthetic complex, a system of incoherent rate equations has been derived, accounting for the interaction of the system with  natural sunlight, exciton energy transfer, the interaction with a thermal bath and the presence of dissipation through radiative and  non-radiative decay channels. 

To this purpose,  different levels of approximation have been considered. First, a Lindblad master equation approach has been developed for the populations of the eigenstates of the entire system (single cylinder + baseplate) assuming finite transfer time and finite thermalization time~\cite{baghbanzadeh2016geometry,mattiotti2021bio,mattiotti2022efficient} (see Sec.~\ref{model1}).  Then, two  approximations (see Sec.~\ref{model2} and Sec.~\ref{model3}), decreasing the computational costs, have been developed and compared to the Lindblad approach, finding good agreement. 

Before discussing these three approaches, a detailed description of the  Hamiltonian of the system is given. For the sake of clarity, in the entire manuscript, the indices $i,j$ are used to label the site basis (the BChl molecules), while the eigenstates are labeled by $n$ and $m$.  Finally, three figures of merit have been chosen in order to estimate the efficiency of the energy transfer in the systems here considered: (1) the trapped current to the RCs defined as the number of excitations reaching the RC per unit time under natural sunlight, (2) the internal and (3) external efficiencies, which are the number of excitations reaching the RC divided by the number of incoming photons (external efficiency) or by the number of absorbed photons (internal efficiency).

\section{The Hamiltonian of the system}
\label{m:ham}

Since photosynthetic antennae operate under natural sunlight, which is very dilute, the single-excitation approximation can be used, so that only states containing a single excitation have been considered. 
Choosing the basis states in  the single-excitation manifold, where $|i\rangle$ represents  a state in which the $i^{th}$ molecule is excited while all the others are in the ground state, the systems can be described 
 through a radiative non-Hermitian Hamiltonian (NHH) which accounts for the interaction between the molecules mediated by the electromagnetic field
(EMF)~\cite{mukamel,kaiser,grad1988radiative,gross1982superradiance}.  The radiative Hamiltonian reads:     
\begin{equation} 
  \hat{H}_{NHH}=\sum_{i=1}^N e_0|i\rangle \langle i|+\sum_{i\neq j}\Delta_{ij}|i\rangle \langle j|-\frac{\im}{2}\sum_{i,j=1}^{N}Q_{ij}|i\rangle \langle j| \, .
  \label{eq:ham}
\end{equation} 
where $e_0=\hbar \omega_0$ is the excitation energy of single emitter (BChl molecule in our case), with $\omega_0$ transition frequency. The terms $\Delta_{ij}$ and $Q_{ij}$ have a diagonal part  given  by: 
\begin{equation}
  \Delta_{jj} = 0 \, , \qquad
  Q_{jj} = \frac{4}{3} \mu^2 k_0^3 = \gamma \, , \label{eq:gamma}
\end{equation}
with $\mu=|\vec{\mu}|$ being the TDM and $k_0=\frac{2 \pi}{ \lambda_0}$, where $\lambda_0$ is the wavelength associated with the molecular transition. The off-diagonal part ($i \ne j$) is given  by
\begin{align}
  \Delta_{ij} = &\frac{3\gamma}{4} \left[ \left( -\frac{\cos (k_0 r_{ij})}{(k_0 r_{ij})} +
    \frac{\sin (k_0 r_{ij})}{(k_0 r_{ij})^2} + \frac{\cos (k_0 r_{ij})}{(k_0 r_{ij})^3} \right)
    \hat{\mu}_i \cdot \hat{\mu}_j\right. \nonumber \\
    &-\left. \left( -\frac{\cos (k_0 r_{ij})}{(k_0 r_{ij})} + 3\frac{\sin (k_0 r_{ij})}{(k_0 r_{ij})^2} +
    3\frac{\cos (k_0 r_{ij})}{(k_0 r_{ij})^3}\right) \left( \hat{\mu}_i \cdot \hat{r}_{ij}
    \right) \left( \hat{\mu}_j \cdot \hat{r}_{ij} \right) \right],
    \label{eq:d1}
\end{align}
\begin{align}    
  Q_{ij} = &\frac{3\gamma}{2} \left[ \left( \frac{\sin (k_0 r_{ij})}{(k_0 r_{ij})} +
    \frac{\cos (k_0 r_{ij})}{(k_0 r_{ij})^2} - \frac{\sin (k_0 r_{ij})}{(k_0 r_{ij})^3} \right)
    \hat{\mu}_i \cdot \hat{\mu}_j\right. \nonumber \\
  &-\left. \left( \frac{\sin (k_0 r_{ij})}{(k_0 r_{ij})} + 3\frac{\cos (k_0 r_{ij})}{(k_0 r_{ij})^2} -
    3\frac{\sin (k_0 r_{ij})}{(k_0 r_{ij})^3}\right) \left( \hat{\mu}_i \cdot \hat{r}_{ij}
    \right) \left( \hat{\mu}_j \cdot \hat{r}_{ij} \right) \right], 
    \label{eq:g1}
\end{align}
where $\hat{\mu}_i :=  \vec{\mu}_i  /
\mu$ is the unit dipole moment of the $i^{th}$ site and $\hat{r}_{ij} := \vec{r}_{ij}
/ r_{ij}$ is the unit vector joining the $i^{th}$ and the $j^{th}$ sites. See Tab.~\ref{tab1}  in the \textit{Supp.Info.} for the parameters we used for BChl \textit{a} and \textit{c}.

Diagonalizing  the Hamiltonian~(\ref{eq:ham}) we obtain the  complex eigenvalues $
\varepsilon_{n}=\mbox{E}_n-\im\frac{\Gamma_{n}}{2}$
where $\Gamma_{n}/\hbar$ is the radiative decay rate of the $n^{th}$ eigenstate in $\mbox{s}^{-1}$. In general $\Gamma_{n}/\hbar$ differs from the radiative decay rate of the single molecule $\gamma/\hbar$.   In particular, when the ratio $\Gamma_{n}/\gamma \gg 1$   we will talk about a ``superradiant state'' (SRS) or bright state, otherwise when $\Gamma_n/\gamma \ll 1$ the  state is called ``subradiant'' or dark. In other words, a SRS  can radiate much faster than a single molecule, while a subradiant  one radiates at a rate much slower than the single-molecule radiative decay.

The non-Hermitian part of the radiative Hamiltonian~(\ref{eq:ham}) can be treated as a perturbation whenever the decay widths are much smaller than the mean level spacing computed using the real part of the complex eigenvalues, see discussion in Sec.~\ref{SI-perturbation} in the {\it Supp.Info.}. When this criterion, known as \textit{resonance overlap criterion}~\cite{zelevinsky}, is valid, one can exclusively utilize the Hermitian part of the Hamiltonian. This reduction in complexity accelerates calculations. The Hermitian part of the Hamiltonian~(\ref{eq:ham}) is defined as follows:
\begin{equation} \label{eq:hreal}
\hat{H}_{HH}=\sum_{i=1}^N e_0|i\rangle \langle i|+\sum_{i\neq j}\Delta_{ij}|i\rangle \langle j|,
\end{equation}
where $\Delta_{i,j}$  is given in~(\ref{eq:d1}).
In Sec.~\ref{SI-perturbation} in the {\it Supp.Info.} a comparison between the radiative decay widths computed with the NHH model for a single cylinder (MT model) and the perturbation theory has been provided. Assuming that we are in the perturbative regime, where  the non-Hermitian term $Q_{ij}$ is much smaller than $\Delta_{ij}$, the radiative decay widths have been estimated as the expectation value of $Q_{ij}$ with the eigenstates of the Hermitian Hamiltonian. Our results have demonstrated that for a single cylinder $Q_{ij}$ can be considered a small perturbation and  the same values for the radiative decay widths have been obtained using both the radiative non-Hemrmitian Hamiltonian and the perturbation theory. These foundings have already been proved in Ref.~\citenum{valzelli2024large} by some of the authors of this manuscript, where the ratio between  all the radiative decay widths $\Gamma_{n}$ and the mean level spacing $\delta$ between the energies of the system has been computed for a single cylinder (MT model) and for the entire chlorosome. The results revealed that for the single cylinder  the ratio is always smaller than $1$, so the perturbative approach is still valid. For the entire chlorosome the maximum value of the ratio, corresponding to the most SRS, is almost $10^2$ and also other eigenstates show a ratio larger than $1$, proving that for larger systems the perturbative regime and the HH model fail, while the NHH model is the only way to describe the radiative response of the system.

If the non-Hermitian term $Q_{ij}$ can be considered a small perturbation and we are in the small-volume limit, when the system size is small compared to the
wavelength associated with the optical transition of the molecules ($L\ll \lambda_0$),
the radiative decay rate of an eigenstate can be estimated in terms of its dipole strength, computed using only the Hermitian part of the Hamiltonian (HH), see discussion at the end of Sec.~\ref{SI-perturbation} in the {\it Supp.Info.}.
Denoting the $n^{th}$ eigenstate of the Hermitian part of the Hamiltonian~(\ref{eq:ham}) with $|E_n\rangle$, we can expand it on the site basis, so that 
\begin{equation} \label{eq:expan}
|E_{n}\rangle=\sum_{i=1}^{N} C_{n}(i) \, |i\rangle.
\end{equation}
To each basis state $|i\rangle$, a dipole moment $\vec{\mu}_i$ is associated, corresponding to the TDM of the $i^{th}$ molecule. 
If $N$ is the total number of molecules, then we will express the TDM $\vec{D}_n$ associated with the $n^{th}$ eigenstate as follows: 
\begin{equation} \label{eq:dipst} 
\vec{D}_n=\sum_{i=1}^{N} C_{n}(i) \, \hat{\mu}_i. 
\end{equation} 
The dipole strength of the $n^{th}$ eigenstate is defined  by $|\vec{D}_n|^{2}$ (note that $\sum_{n=1}^{N} |\vec{D}_n|^{2}=N$~\cite{Note4}).  
Note that when  the system size is much smaller than $\lambda_0$ we have   $|\vec{D}_n|^2 \approx \Gamma_n/\gamma$.

Finally, we note that when resonances do not overlap and  the system size is much smaller than $\lambda_0$  (i.e. when $k_{0}r_{ij}\ll 1$), the Hermitian part of the radiative Hamiltonian reduces to  the standard dipole-dipole Frenkel Hamiltonian (DH): 

\begin{equation}
    \label{dip}
\hat{H}_{DH}=\sum_{i=1}^N e_0|i\rangle \langle i|+\sum_{i\neq j}\displaystyle \frac{\vec{\mu}_{i} \cdot \vec{\mu}_{j}-3(\vec{\mu}_{i} \cdot \hat{r}_{ij})(\vec{\mu}_{j} \cdot \hat{r}_{ij})}{r_{ij}^{3}}|i\rangle \langle j|.
\end{equation}

\begin{figure}[!ht]
    \centering
    \includegraphics[width=1\linewidth]{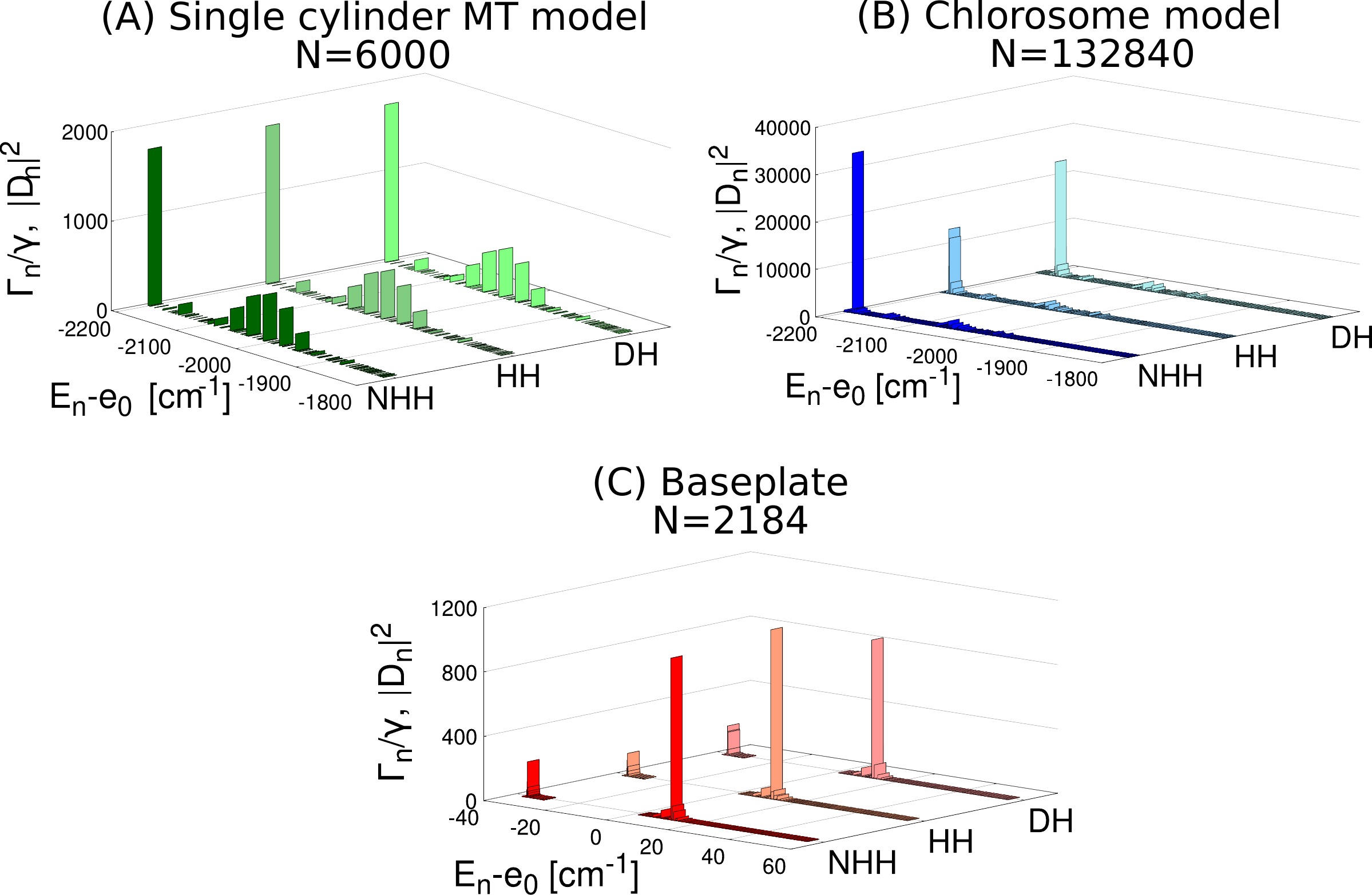}
    \caption{{\it Dipole strength $|D_n|^2$ (DH and HH models) and radiative decay rate $\Gamma_n/\gamma$ (NHH model) in cylindrical aggregates and baseplate.} Comparison between the dipole strength (DH and HH models) and radiative decay rate (NHH model) for (A) a single  MT cylinder, (B) the chlorosome and (C) the dimeric baseplate. Note that in Ref.~\citenum{valzelli2024large} for a single  MT cylinder (A)   the ratio $\Gamma_{n}/\delta$ is always less than $1$, while for the chlorosome (B) the maximum value of $(\Gamma_{n}/\delta)_{\text{max}}$  is almost $10^2$, proving that the perturbative regime for this aggregate fails and only the NHH model can be used to describe superradiance. The mean level spacing $\delta$ has been computed as the ratio between the energy spectral width and the total number of
eigenmodes for each complex. For the geometry of the system, refer to Tab.~\ref{table_size}. Panels (A-B) show only the lowest part of the energy spectrum, while panel C represents the entire energy spectrum. }
    \label{spectra}
\end{figure}

Here we compare the radiative decay widths of the eigenstates of different molecular aggregates computed with the three different Hamiltonian models introduced above:
\begin{enumerate}
    \item NHH: non-Hermitian radiative Hamiltonian~\eqref{eq:ham}.
    \item HH: Hermitian Hamiltonian~\eqref{eq:hreal} valid under the non-overlapping resonance criterion.
    \item DH: Dipole Hamiltonian~\eqref{dip} valid under the non-overlapping resonance criterion and when the system size is small compared to the wavelength associated with the optical transition of the molecules.
\end{enumerate}
In Fig.~\ref{spectra} the radiative decay widths of the eigenstates obtained by diagonalizing the NHH are compared with the dipole strength computed with the HH and DH models. Note that, both in the main text and in the \textit{Supp.Info.}, all energy values are given in cm$^{-1}$ units, \textit{i.e.} they are divided by $hc$. Panel A of Fig.~\ref{spectra}  shows that for the single cylinder, which is about $821.7\ \mbox{\AA}$ long, the three models give comparable results, because we are in the small-volume limit, defined as $L\ll \lambda_0$. For the whole chlorosome, see panel B of Fig.~\ref{spectra}, made of three adjacent concentric cylinders covering an area of $3075.3\times 1147.5$~\mbox{\AA}$^2$ we are neither in the small-volume limit nor in the perturbative one, see discussion in Ref.~\citenum{valzelli2024large} and in Sec.~\ref{SI-perturbation} in the {\it Supp.Info.}. Thus, the three Hamiltonians give very different results.  For the baseplate, see panel C of Fig.~\ref{spectra}, of size $2739.1\times 2739.1~\mbox{\AA}^2$, the NHH and the HH give very similar results, showing that we are in the perturbative limit with respect to the non-Hermitian interaction. On the other hand, some differences between the DH and the NHH model can be observed. Nevertheless, the discrepancies in the largest dipole strength are just $\approx 14 \%$, so that we can use the DH as a first approximation.

In the following we will use the DH model for the single cylinder and baseplate, while for the whole chlorosome the full NHH Hamiltonian will be used.

\section{Master equation}
\label{model1}
The whole light-harvesting process can be described by the following master equation for the density matrix $\hat{\rho}_S$~\cite{breuer2002theory}:
\begin{equation}
    \frac{d\hat{\rho}_S}{dt}=-\frac{\im}{\hbar}[\hat{H}_S,\hat{\rho}_S]+ \mathcal{L}_{fl}[\hat{\rho}_S] + \mathcal{L}_{Sun}[\hat{\rho}_S]  + \mathcal{L}_T[\hat{\rho}_S]
    \label{eq:me}
\end{equation}
that contains the following terms: $\hat{H}_S$ is the Hermitian part of the Hamiltonian of the system (according to the aggregate considered we will use either $\hat{H}_{DH}$ or $\hat{H}_{NHH}$, see dicussion above); $\mathcal{L}_{fl}$, $\mathcal{L}_{Sun}$ and $\mathcal{L}_T$ are Lindblad dissipators derived under the Born-Markov and secular approximations~\cite{breuer2002theory}. They describe respectively spontaneous  emission ($\mathcal{L}_{fl}$) and absorption and stimulated emission induced by sunlight ($\mathcal{L}_{Sun}$), while $\mathcal{L}_T$ is the dissipator modeling thermal relaxation and decoherence in presence of a thermal bath~\cite{mattiotti2021bio,mattiotti2022efficient}, see discussion in Sec.~\ref{SI:thermal} in the \textit{Supp.Info.}.

The Lindblad dissipators read explicitly: 
\begin{subequations}
\label{diss}
    \begin{align}
         \mathcal{L}_{fl}[\hat{\rho}_S]= & \sum_{ij}\frac{Q_{ij}}{\hbar}\left[\hat{a}_j \hat{\rho}_S \hat{a}_i^{\dagger} - \frac{1}{2}\{ \hat{a}_i^\dagger \hat{a}_j,\hat{\rho}_S \}  \right]\\
        \mathcal{L}_{Sun}[\hat{\rho}_S] = & \sum_{ij} f_s n_s\frac{Q_{ij}}{\hbar}\left[\hat{a}_j^\dagger \hat{\rho}_S \hat{a}_i-\frac{1}{2} \{ \hat{a}_i \hat{a}_j^{\dagger},\hat{\rho}_S \}\right]+\sum_{ij} f_s n_s\frac{Q_{ij}}{\hbar}\left[\hat{a}_j \hat{\rho}_S \hat{a}_i^{\dagger} -\frac{1}{2} \{ \hat{a}_i^{\dagger} \hat{a}_j,\hat{\rho}_S \}\right],   
    \end{align}
\end{subequations}
where the sums over $i,j$ run over all the system sites, $\hat{a}_i = \ket{0}\bra{i}$ ($\ket{0}$ is the ground state with no excitation, while $\bra{i}$ can be a cylinder or baseplate site), $Q_{ij}$ is given in~(\ref{eq:g1}) and in the small-volume limit $Q_{ij}\approx \gamma \hat{\mu_i} \cdot \hat{\mu_j}$.

Note that the absorption of sunlight photons is non-Markovian and leads to coherent oscillations, {\it i.e.} Fano coherence, even if the excitation is incoherent. The sunlight-induced coherence is particularly relevant as the dipole coupling in Eq.\eqref{eq:g1} is collective.  As demonstrated in an early calculation, the sunlight-induced coherence may not affect light-harvesting efficiency, as dephasing is typically faster than energy transfer.  Then, the Markov approximation assumed in the quantum master equation can be justified~\cite{olvsina2014can}.

Finally  $\mathcal{L}_T$ is
\begin{equation}
{\cal L}_T[\hat{\rho}_S] = \sum_{\omega} \gamma^{(p)}(\omega) \sum_{i} \left[ \hat{A}_i(\omega) \hat{\rho}_S \hat{A}_i^\dag(\omega) -\frac{1}{2} \left\{ \hat{A}_i^\dag(\omega) \hat{A}_i(\omega), \hat{\rho}_S \right\} \right] \, .
    \label{redfield}
\end{equation}
where 
\begin{equation}
\label{eq:gammap}
\gamma^{(p)}(\omega) = 2\pi\left[ J(\omega) (1+n_{BE}(\omega)) + J(-\omega)n_{BE}(-\omega)  \right]
\end{equation}
are the thermal rates, depending of the spectral density $J(\omega)$ and on the Bose distribution $n_{BE}(\omega)=(e^{\hbar\omega/k_BT} - 1)^{-1}$ of the phonons at room temperature ($T=300$~K) and 
\begin{equation}
    \hat{A}_i(\omega)= \sum_{E_m-E_n=\hbar \omega} C^*_m(i)C_n(i) \ket{E_n}\bra{E_m}.      
\end{equation}
More details about the Lindblad dissipators  can be found in Refs.~\citenum{mattiotti2021bio, mattiotti2022efficient}.

Eq.~(\ref{eq:me}), in its Lindblad form, is derived under the secular approximation. Moreover, it can be largely simplified under a well-motivated assumption: the excitons are transferred incoherently between the cylinder eigenstates and baseplate eigenstates; this is consistent with the approach of some recent works~\cite{mohseni2008environment,mattiotti2022efficient,mattiotti2021bio,bienaime2013cooperativity}, where a similar model to describe exciton dynamics in GSB antenna complexes has been already employed.  
Under these assumptions, the coupling with the thermal bath and natural sunlight are
well approximated by rate equations for the populations of the cylinder eigenstates ($P_{n\in C}$) and the populations of the baseplate eigenstates ($P_{n \in B}$), as we show in the following.

\subsection{Rate equations for the light-harvesting process: finite transfer time and finite thermalization time}
\label{m:model1}

The following set of full rate equations for the whole light-harvesting process have been derived starting from the master equation Eq.~\eqref{eq:me} (see also Sec.~\ref{SI:thermal} of the \emph{Supp.Info.}). They read: 
\begin{subequations}
\label{ILE}
    \begin{align}
        \label{ILE0}
        \frac{dP_0(t)}{dt}=&-\sum_n R_n P_0(t) + \sum_n (R_n +\Gamma_n/\hbar+\kappa_{NR})P_n(t) \nonumber \\
        &+\sum_{n\in B} \kappa P_n(t)~, \\
        \frac{dP_{n\in C}(t)}{dt}= &R_n P_0(t) - (R_n +\Gamma_n/\hbar+\kappa_{NR})P_n(t) \nonumber \\
        &+\sum_{m\in C} \left(T_{n,m}P_m(t) -T_{m,n} P_n(t) \right) \nonumber \\
        &+\sum_{m\in B} \left(K_{n,m}P_m(t) -K_{m,n} P_n(t) \right)~, \\
        \frac{dP_{n\in B}(t)}{dt}= &R_n P_0(t) - (R_n +\Gamma_n/\hbar+\kappa_{NR}+\kappa)P_n(t) \nonumber \\
        &+\sum_{m\in B} \left(T_{n,m}P_m(t) -T_{m,n} P_n(t) \right) \nonumber \\
        &+\sum_{m\in C}
        \left(K_{n,m}P_m(t) -K_{m,n} P_n(t) \right)~,
    \end{align}
\end{subequations}
where the summations $\sum_{n \in C}$ include all the cylinder  eigenstates, while $\sum_{n \in B}$ include all the baseplate  eigenstates and $\sum_n$ include all the eigenstates (cylinder and baseplate). 

The terms in Eq.~\eqref{ILE} are explained in detail in the following.

\paragraph{Radiative decay}
\label{m:rad}
The radiative decay $\Gamma_n/\hbar$ is given by the imaginary part of the NHH. When the DH is valid, it can be computed from the dipole strengths of the eigenstates as follows: 

\begin{equation} 
\label{eq:rad}
\frac{\Gamma_n}{\hbar}=\frac{4}{3}\frac{\mu^2 D_n^2 \omega_n^3}{ \hbar c^3},  
\end{equation}
 where $D_n$ is defined in~\eqref{eq:dipst} and $\omega_n = E_n /\hbar$ is the eigenstate transition frequency.

\paragraph{Absorption and stimulated emission induced by sunlight radiation}
\label{m:abs}
Sunlight induces absorption and stimulated emission to each eigenstate $\ket{E_n}$ with rates given by
\begin{equation}
\label{eq:R}
R_n=f_Sn_{S}(\omega_n)\frac{\Gamma_n}{\hbar}, 
\end{equation}
where
\begin{equation}
\label{m:bose}
    n_S(\omega_n)=\frac{1}{e^{ \hbar \omega_n/(k_BT_{S})}-1}
\end{equation}
is the Bose occupation of photons at the black-body temperature of the Sun, $T_S=5800$~K, while the factor
\begin{equation}
    \label{fac1Sun}
    f_S = \frac{\pi r_S^2}{4\pi R_{ES}^2} = 5.4\times 10^{-6} \, ,
\end{equation}
models how the absorption (and stimulated emission) rate is reduced by the Sun-Earth distance~\cite{mattiotti2021bio}. Specifically, $f_S$ is the fraction of the solid angle of the Sun as seen from the Earth, with $r_S$ being the radius of the Sun and $R_{ES}$ the Sun-Earth distance. See Sec.~\ref{SI:sun} in the \textit{Supp.Info.} for a more detailed discussion about the validity of the approximation of the solar spectrum with the black-body radiation theory and Sec.~\ref{SI:abs} in the \textit{Supp.Info.} for a comparison between different approaches used to compute the absorption rates of BChl molecules. Note that the solar spectrum  is not exactly described by black-body radiation, as the photon experiences multiple scatterings upon arriving on the Earth. However, the sunlight spectrum is sufficiently broad compared to the absorption of light-harvesting complexes and our approximation holds~\cite{olvsina2014can}.

\paragraph{Non-radiative decay}
\label{m:nr}
In addition to the radiative decay, we include non-radiative recombination processes on each $n$ eigenstate by adding a non-radiative rate $\kappa_{NR}=1 \ \mbox{ns}^{-1}$ for each eigenstate~\cite{baghbanzadeh2016geometry}. 

\paragraph{Trapping to RC}
\label{m:trapping}
We also consider an additional decay channel due to excitation transfer from the baseplate to the reaction centers (RCs) through the FMO trimers. We model this by adding a decay rate $\kappa$ on all the baseplate eigenstates. Specifically, the excitation is lost through the FMOs to the RCs with a rate $k_{FMO}\sim0.023-0.044$~ps$^{-1}$~\cite{dostal2016situ}.  $k_{FMO}$ is the reciprocal of the time required for an excitation to be lost in the RC through the FMO, and such time is the sum of four contributions,
\begin{equation}
    k_{FMO}^{-1} = \tau_{b} + \tau_{t} + \tau_{e}  + \tau_{cs}~,
\end{equation}
that represent explicitly:
\begin{itemize}
    \item the transfer time $\tau_{b}$ \emph{from the baseplate to the FMO}, which we estimate as the inverse of the FRET rate
    \begin{equation}
        \tau_{b} = \frac{\hbar^2\Gamma_\phi}{2J^2}
    \end{equation}
    between the two molecules closest to each other, one in the baseplate and one in the FMO complex; here, $\Gamma_\phi=1000$~cm$^{-1}$, where $\Gamma_\phi/\hbar$ is the dephasing rate, estimated from the cylinder-baseplate transition rate, as explained in Sec.~\ref{model2}; $J\approx \mu^2 / r^3$ is the dipole coupling between two BChl molecules (one belonging to the baseplate and the other one to the FMO), where $\mu\approx 6$~D is the transition dipole of a single BChl molecule and $r=20 \ \mbox{\AA}$ is the distance; therefore we have $J\approx 23$~cm$^{-1}$, from which $\tau_{b}\approx 5$~ps;
    \item the time for an excitation to go \emph{through the FMO} from edge to edge has been estimated in Ref.~\citenum{jianshufmo}, and it is $\tau_{t}\approx 3$~ps, in good agreement with the experimental data found in Ref.~\citenum{dostal2016situ} that report an energy relaxation within the FMO complex, which lies between 0.1 and 20 ps;
    \item the exit time $\tau_{e}$ \emph{from FMO to RC},  which is $\sim 17$ ps~\cite{dostal2016situ};  
    \item the \emph{charge-separation} time $\tau_{cs}$ at which the excitation is irreversibly lost in the RC is $\tau_{cs}\approx 1$~ps~\cite{strumpfer2012quantum,dostal2016situ}.
\end{itemize}
The sum of these three contributions gives  the typical range 22.5~ps~$<k_{FMO}^{-1}<$~43~ps, in agreement with the estimation given above.

Finally the trapping rate to the RC $\kappa$ can be estimated as follows. We have a molecular density of 1 molecule per $6.9$~nm$^2$ in the baseplate and a density of 1 FMO trimer per 50~nm$^2$.~\cite{guzik}. Since we have one RC every two FMO trimers~\cite{oostergetel,bina2016,hauska2001,wen2009}, we have a density of 1~RC per 100~nm$^2$. Therefore, we have a ratio $N_{FMO}/N_{BPL}=0.138$ between the number of FMO trimers and the number of baseplate molecules. So, we have modeled the FMO-RC trapping with a constant decay rate $\kappa=0.138~k_{FMO}$ from all the baseplate molecules (see Eqs.~\eqref{ILE}).

\paragraph{Thermal relaxation and energy transfer}

Finally, we include thermal relaxation rates $T_{n,m}$ between each pair of eigenstates within the cylinder and the baseplate, and F\"orster energy transfer rates $K_{n,m}$ between the cylinder and the baseplate. More details on this are given below.

The transfer rates $T_{m,n}$  model thermal relaxation inside the aggregate, see Sec.~\ref{SI:thermal} in \emph{Supp.Info.} for the derivation. They are detailed-balanced, namely
\begin{equation}
\label{th-rates}
     T_{m,n} = \frac{2\pi \Lambda_{mn} J[(E_m-E_n)/\hbar]}{1-e^{-(E_m-E_n)/(k_BT)}}
\end{equation}
where $\Lambda_{mn}=\sum_i |C_m(i)|^2 |C_n(i)|^2$, the phonon temperature is $T=300$~K and the spectral density is $J(\omega)=\kappa_{vib}\omega$. $\kappa_{vib}=0.3$ ensures thermal relaxation in a few-picoseconds timescale~\cite{mattiotti2021bio}.  
On the other hand, the transfer rates between eigenstates of different aggregates are the incoherent F\"orster rates
\begin{equation}
\label{forster}
      K_{m,n} =
      \begin{cases}
        \frac{2\Omega_{m,n}^2\Gamma_\phi}{\hbar\left[\Gamma_\phi^2+(E_m-E_n)^2\right]} & E_n \ge E_m \\
        \frac{2\Omega_{m,n}^2\Gamma_\phi}{\hbar\left[\Gamma_\phi^2+(E_m-E_n)^2\right]} e^{-(E_m-E_n)/(k_BT)} & E_n < E_m \\
      \end{cases}
\end{equation}
where $\Omega_{m,n}$ is the dipole-dipole coupling (matrix element) between the eigenstates of different aggregates computed by using~(\ref{dip}), while  $\Gamma_\phi=1000$~cm$^{-1}$, where $\Gamma_\phi/\hbar$ is the dephasing rate  for every $m-n$ transition and $E_n$ is the transition energy of the $n$th eigenstate. Note that the ``energy-upwards'' rates in Eq.~\eqref{forster} are multiplied by an exponential factor that ensures the detailed balance~\cite{kassal2}. The parameter $\Gamma_\phi$ has been tuned to match the transfer rates between the MT cylinder and the baseplate (tens of picoseconds~\cite{martiskainen2009excitation,guzik,martiskainen2012excitation}) and it corresponds to the sum of the HWHM of the emission spectrum of the donor and the absorption spectrum of the acceptor.  See Sec.~\ref{SI:fret}  for a detailed discussion about FRET and Sec.~\ref{SI:parameters}  in the \textit{Supp.Info.} for a clear validation of the parameters employed in this work, specifically regarding the choice of $\Gamma_\phi$, the intra-molecular coupling strength and the intrinsic radiative decay rate of the Bchl molecules.

\begin{figure}[!ht]
    \centering
    \includegraphics[width=0.9\columnwidth]{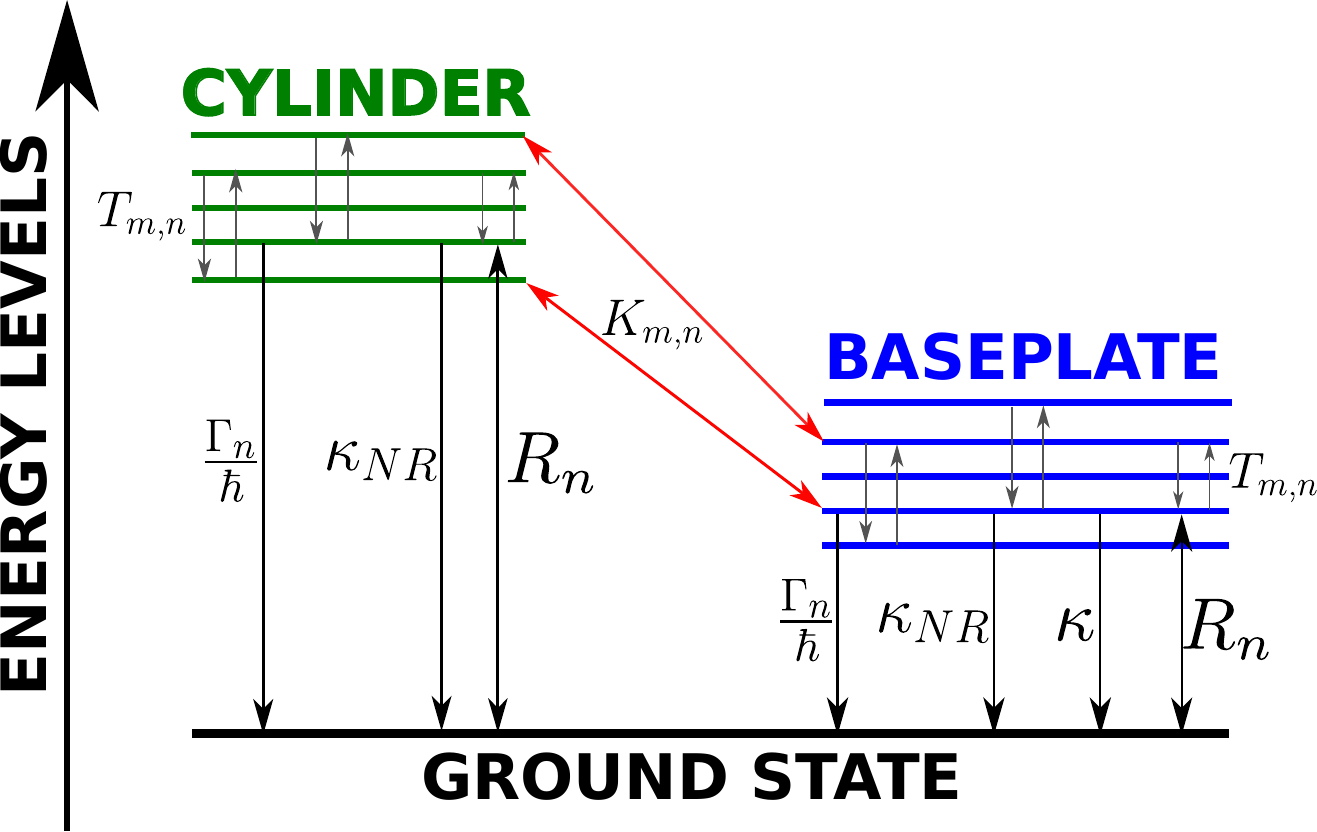}
    \caption{\emph{Rate equation scheme.} Scheme of the rate equations, Eqs.~\eqref{ILE}. Here, $\Gamma_n/\hbar$ is the radiative recombination rate, $\kappa_{NR}$ is the non-radiative recombination rate (equal for all levels), $R_n=f_Sn_S(\omega_n)\frac{\Gamma_n}{\hbar}$ are the sunlight absorption and stimulated emission rates, $T_{m,n}$ are the thermalization rates within each aggregate, see Eq.~\eqref{th-rates}, $K_{m,n}$ are the transfer rates between eigenstates of each aggregates, see Eq.~\eqref{forster}, and $\kappa$ is the trapping rate from the baseplate, through the FMOs, to the RC.}
    \label{fig-rate}
\end{figure}

The rate equations~\eqref{ILE} are linear and they can be solved at the steady state or as a function of time by numerical diagonalization, with a computational cost that increases as $N^3$, where $N$ is the total number of molecules in the cylinder+baseplate. A scheme of Eqs.~\eqref{ILE} is shown in Fig.~\ref{fig-rate}. 

From the steady-state solution of Eqs.~\eqref{ILE} we obtain the current trapped in the RCs,
\begin{equation}
\label{curr}
    I_{RCs} = \kappa \sum_{n \in B} P_n^{SS}~,
\end{equation}
where $P_n^{SS}$ are steady-state populations of the baseplate eigenstates, and we also compute the internal efficiency~\cite{kassal2},
\begin{equation}
\label{eff}
    \eta_{in} = \frac{\kappa \sum_{n \in B} P_n^{SS}}{\sum_{n} R_n P_0^{SS}}~,
\end{equation}
where $P_0^{SS}$ is the steady-state population of the ground state and the sum in the denominator runs over all the cylinder and baseplate states.

Here we introduce another figure of merit, the external efficiency:
\begin{equation}
    \label{ext-eff}
    \eta_{ext} = \frac{I_{RCs}}{I_{Sun}\times A_{BPL}}=\eta_{abs}\times\eta_{in},
\end{equation}
where $I_{Sun}$ is the photon flux coming from sunlight impinging the baseplate surface, $I_{RCs}$ is the total current trapped in the RCs, $\eta_{abs}$ and $\eta_{in}$ are the absorption and internal efficiencies, respectively.  An analytical expression for $\eta_{in}$ has been given in Eq.~\eqref{eff}, while the absorption efficiency $\eta_{abs}$, given by Eq.~\eqref{eff1} in the \textit{Supp.Info.}, is the ratio between the power absorbed by the system and the solar irradiance. Sec.~\ref{SI:ext} in the \textit{Supp.Info.} offers a deeper study  and detailed calculations of the absorption in the entire light-harvesting apparatus of GSB comprising the chlorosome and the baseplate.

\subsection{Thermal equilibrium within each aggregate}
\label{model2}

To simplify the rate equations~\eqref{ILE}, here we follow a widely used approach, presented {\it e.g.} in Refs.~\citenum{jianshumcfretnjp,jianshumcfretjcp}  and in Ref.~\citenum{strumpfer2012quantum}. In this approach, thermal equilibrium is assumed inside each aggregate (a cylinder, or the baseplate). 
The full rate equations shown in Eq.~\eqref{ILE} is the most accurate model, however, due to the large size of our systems, a less demanding approach, the partially thermalized model, is derived.  This approach is valid when thermal relaxation within each aggregate is the fastest time scales. Under this approximation the energy transfer process between aggregates is well captured by the MC-FRET theory, which is valid when the transfer rate between aggregates is much smaller than the thermalization rate within each aggregate. 
Note that light-harvesting systems are usually disordered and strongly coupled to protein environments, so the thermalization process may not lead to the Boltzmann distribution of the system Hamiltonian~\cite{moix2012equilibrium}. Instead, the light-harvesting system relaxes to non-canonical distribution due to the system-bath correlation and exhibit bath-induced coherence, which can affect fluorescence emission and F\"orster energy transfer.  The Master Equation in Eq.~\eqref{eq:me} assumes the factorization of the system-bath density matrix and thus implies the system Boltzmann distribution.  Possible corrections due to the system-bath correlation can be considered in the future.

Under the assumption of thermalization in each aggregate, the transfer rate from a donor aggregate ``D'' to an acceptor aggregate ``A'' is given by the generalized multi-chromophoric F\"orster resonance energy transfer rate (MC-FRET)
\begin{equation}
\label{mcfret}
    K_{D,A} = \sum_{m \in D} \sum_{n \in A} \frac{e^{-E_m/(k_BT)}}{\sum_{l\in D}e^{-E_l/(k_BT)}} K_{n,m}~,
\end{equation}
where $K_{n,m}$ is the transfer rate from the donor eigenstate $m$ to the acceptor eigenstate $n$, see Eq.~\eqref{forster}.  Sec.~\ref{SI:fret} in the \textit{Supp.Info.} gives a more detailed explanation of the MC-FRET.  Similarly, we assume thermal equilibrium within each aggregate to compute the decay rates, as explained in the following. We obtain the following 3-level rate equations,
\begin{subequations}
\label{3lev-rate}
\begin{align}
    \frac{dP_0(t)}{dt} &= -(R_C + R_B) P_0(t) + (\langle R \rangle_C + \langle \gamma \rangle_C)P_C(t) + (\langle R \rangle_B + \langle \gamma \rangle_B)P_B(t)+\kappa P_B(t) \notag \\  
    & \quad + \kappa_{NR} [P_C(t)+P_B(t)] \\
    \frac{dP_C(t)}{dt} &= R_C P_0(t) - (\langle R \rangle_C + \langle \gamma \rangle_C)P_C(t) + K_{B,C} P_B(t) - K_{C,B} P_C(t) - \kappa_{NR} P_C(t) \\
    \frac{dP_B(t)}{dt} &= R_B P_0(t) - (\langle R \rangle_B + \langle \gamma \rangle_B)P_B(t) + K_{C,B} P_C(t) - K_{B,C} P_B(t) -\kappa P_B(t) \notag \\
    & \quad - \kappa_{NR} P_B(t)
\end{align}
\end{subequations}
where $P_0(t)$, $P_C(t)$ and $P_B(t)$ are the populations of the ground state, the cylinder and the baseplate, respectively. $R_C=\sum_{n \in C}R_n$ and $R_B=\sum_{n \in B}R_n$ are the total absorption rates of the cylinder and the baseplate, respectively.
$(\langle R \rangle_C + \langle \gamma \rangle_C)$ and $(\langle R \rangle_B + \langle \gamma \rangle_B)$ are the thermal-averaged emission rates, given by
\begin{equation}
\langle R \rangle_{C,B} + \langle \gamma \rangle_{C,B}=\sum_{n\in C,B} (R_n+\Gamma_n/\hbar)p_n ,
\label{eq:emission}
\end{equation}
of the cylinder ($C$) and the baseplate ($B$) with the Boltzmann weights within each aggregate 
\begin{equation}
\label{eq:weights}
    p_n = \frac{e^{-E_n/(k_BT)}}{\sum_{m\in C,B} e^{-E_m/(k_BT)}}~,
\end{equation}
$K_{C,B}$, $K_{B,C}$ are donor-acceptor rates computed from Eq.~\eqref{mcfret} using the Lorentzian lineshapes as in Eq.~\eqref{forster}. 

The collective F\"orster energy transfer between two aggregates often exhibits simple scaling laws as a function of distance, site and orientation.  Using the generalized MC-FRET, we analyzed the scaling laws for interwire energy transfer~\cite{chuang2014scaling}. This approach can be applied to other geometries and was recently adopted for the chlorosome lamellae~\cite{zhong2023efficient,bondarenko2020comparison}. The reported results on energy transfer can be understood in this framework.

We numerically solve Eqs.~\eqref{3lev-rate} at the steady-state to obtain the number of excitations trapped in the RCs per unit time,
\begin{equation}
\label{curr3}
    I_{RCs} = \kappa P_B^{SS}~.
\end{equation}
We also compute the internal efficiency as the ratio between the excitations trapped in the RC per unit time and the excitations absorbed per unit time~\cite{kassal2},
\begin{equation}
\label{eff3}
    \eta_{in} = \frac{\kappa P_B^{SS}}{(R_C+R_B) P_0^{SS}}~.
\end{equation}

\subsection{Thermal equilibrium among all the aggregates}
\label{model3}
If the transfer rate between the aggregates is very fast, we can make a further approximation and assume that all the aggregates are at thermal equilibrium between one another. In such case, let us recall the rate equation for the population of the ground state, Eq.~\eqref{ILE0}
\begin{align}
  \label{rate}
  \frac{dP_0(t)}{dt} = &- \sum_n R_n P_0(t) + \sum_n R_n P_n(t) \nonumber \\
                 &+ \sum_n \frac{\Gamma_n}{\hbar} P_n(t) + \sum_n \kappa_{NR} P_n(t) + \sum_{n\in B} \kappa P_n(t) \, ,
\end{align}
where $P_0$ is the population of the ground state and $P_n$ is the population of the $n$-th excitonic eigenstate.

Then, let us assume that the whole single-excitation subspace is at thermal equilibrium with a temperature $T=300$~K. This approximation is reasonable in the case where the thermalization process among all the aggregates (between cylinders, or between cylinders and the baseplate) happens faster than the superradiant radiative decay of the cylinders. In this case, we define the population in the excited subspace of the whole system at time $t$ as
\begin{equation}
  P_e(t) = \sum_{n=1}^N P_n(t) \, ,
\end{equation}
so that the trace preservation condition can be written as
\begin{equation}
  \label{trace}
  P_0(t) + P_e(t) = 1 \, ,
\end{equation}
and, crucially, we assume that
\begin{equation}
  \label{therm}
  P_n(t) = P_e(t) p_n \quad \text{with} \quad p_n = \frac{e^{-E_n/(k_BT)}}{\sum_m e^{-E_m/(k_BT)}}~.
\end{equation}
By substituting \eqref{therm} into \eqref{rate} we have
\begin{align}
  \label{rate2}
  \frac{dP_0(t)}{dt} = &- \left(\sum_n R_n\right) P_0(t) + \left(\sum_n R_n p_n \right) P_e(t) \nonumber \\
                 &+ \left(\sum_n \frac{\Gamma_n}{\hbar} p_n \right) P_e(t) + \kappa_{NR} \left(\sum_n  p_n \right) P_e(t) \nonumber \\
                 &+ \kappa \left(\sum_{n\in B}  p_n \right) P_e(t) \, .
\end{align}
Eq.~\eqref{rate2} can be expressed in terms of the thermal averages of the parameters,
\begin{equation}
    \langle R \rangle = \sum_n R_n p_n~, \qquad \langle \gamma \rangle = \sum_n \frac{\Gamma_n}{\hbar} p_n~,
\end{equation}
noting that $\sum_n p_n=1$, defining the fraction of the excited population on the baseplate $p_B=\sum_{n\in B}p_n$,
and by defining the total absorption rate as $R_{TOT} = \sum_n R_n$ we get
\begin{align}
  \frac{dP_0(t)}{dt} = - R_{TOT} P_0(t) + \left(\left\langle R \right\rangle + \left\langle \gamma \right\rangle + \kappa_{NR} + \kappa p_B \right) P_e(t) \, .
\end{align}
The current trapped at the steady state is obtained by using Eq.~\eqref{trace}:
\begin{equation}
  \label{curr2}
  I_{RCs}=\kappa p_B P_e^{SS} = \frac{\kappa p_B R_{TOT}}{R_{TOT} + \left\langle R \right\rangle + \left\langle \gamma \right\rangle + \kappa_{NR} + \kappa p_B } \, ,
\end{equation}
while the internal efficiency is
\begin{equation}
\label{eff2}
    \eta_{in} = \frac{\kappa p_B P_e^{SS}}{R_{TOT} P_0^{SS}} = \frac{\kappa p_B}{\left\langle R \right\rangle + \left\langle \gamma \right\rangle + \kappa_{NR} + \kappa p_B }~.
\end{equation}
Here, $P_e^{SS}$ represents the total excitonic population in the system at the steady state, under natural sunlight illumination, assuming thermal equilibrium in the whole system and accounting for radiative and non-radiative recombination.
This method only requires the knowledge of the radiative and non-radiative decay rates of the eigenstates and of their energies.

Note that, since the baseplate is well gapped below the cylinder, we have $p_B\approx 1$. Moreover, for $\kappa_{NR}=1$~ns$^{-1}$, we also typically have $\braket{R}+\braket{\gamma} \ll \kappa_{NR}$. Therefore, the internal efficiency has the approximate expression: 
\begin{equation}
\label{eff2app}
    \eta_{in} \approx \frac{\kappa}{\kappa_{NR} + \kappa}~.
\end{equation}

\section{Results and discussions}
\label{m:results}

\subsection{Exciton energy transfer in GSB light-harvesting systems}
In this section the main results of this study are presented comparing the natural structure with the light-harvesting systems obtained by modifying the orientation of the  dipoles in the cylindrical aggregates. 

First we consider a single-wall nanotube coupled to a baseplate where the excitation can be trapped with a trapping rate $k_{FMO}$. For the single-wall nanotubes, we consider three different types of cylindrical aggregates (MT, PD and RD, see Sec.~\ref{m:models}). All single-wall cylinders are  composed of $6000$ BChl molecules and they are $821.7  \ \mbox{\AA}$ long. The baseplate coupled to the cylindrical structures is composed of $2184$ BChl molecules and has an area of $550.8 \times 2739.1 \ \mbox{\AA}^2$.  In addition to the single-wall nanotubes, we also consider the full model of the GSB photosynthetic antenna (chlorosome) composed of $132840$ BChl molecules distributed among three MT adjacent concentric cylinders (four wall each), see Sec.~\ref{m:models}.  Each concentric cylinder includes $44280$ BChl molecules, and the  baseplate for the whole chlorosome is composed of $3350$ molecules and with an area of $1145.7 \times 3075.3 \ \mbox{\AA}^2$, see Tab.~\ref{table_size} in Sec.~\ref{SI:table} of the {\it Supp.Info.} where  a more detailed explanation of the geometrical features of the cylindrical aggregates and baseplate is given. 

\begin{figure}[ht!]
    \centering
    \includegraphics[width=\columnwidth]{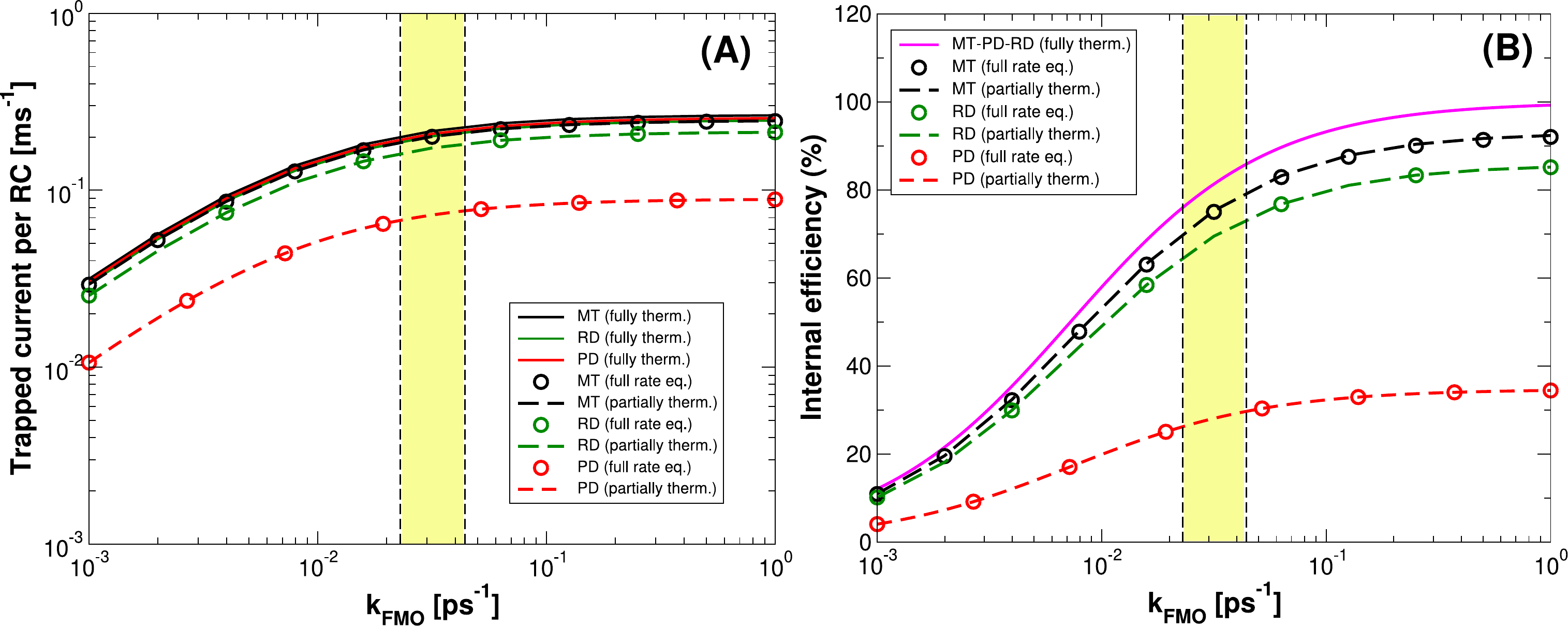}
    \caption{  \emph{ Trapped current and internal efficiency {\it vs.} transfer rate $k_{FMO}$ from baseplate to RCs for single-cylinder models.}  Trapped current per RC (panel A) and internal efficiency (panel B) for the single-cylinder models (MT, PD, RD coupled to the dimeric baseplate) comprising $6000$ BChl \textit{c} in the cylinder and $ 2184$ BChl \textit{a} in the baseplate. More details about the dimensions of the systems can be found in Tab.~\ref{table_size} of the \textit{Supp.Info.}. Black, green and red dashed lines have been obtained assuming thermalization in each aggregate and solving Eq.~\eqref{3lev-rate} in MT, RD and PD models, respectively. Open circles represents the results obtained by solving the full rate equations given in Eq.~\eqref{ILE}. Continuous lines represent the trapped current and internal efficiency assuming thermalization among all aggregates, obtained from Eq.~\eqref{rate}. According to Eq.~\eqref{eff2app} the internal efficiency has a universal expression for all the models. The yellow box between the two vertical dashed lines represents the regime in which FMO complexes typically work ($0.023 \ \mbox{ps}^{-1}<k_{FMO}<0.044\ \mbox{ps}^{-1}$). Trapped current and internal efficiency for the RD model have been computed averaging over $10$ realizations of random dipole orientations. } 
    \label{f:cyl}
\end{figure}

\begin{figure}[ht!]
    \centering
    \includegraphics[width=\columnwidth]{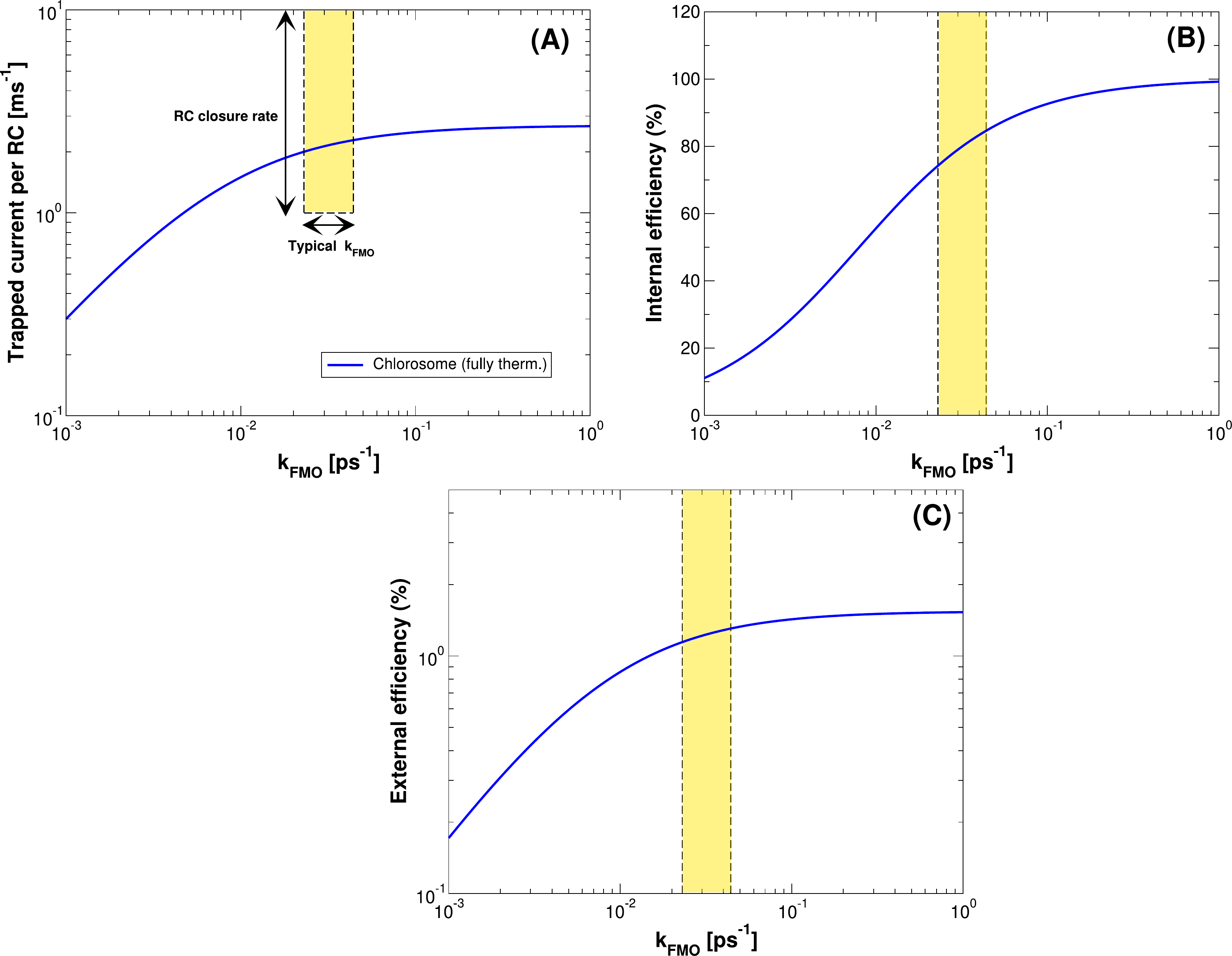}
    \caption{  \emph{ Trapped current, internal and external efficiency {\it vs.} transfer rate $k_{FMO}$ from baseplate to RCs for the chlorosome coupled to a dimeric baseplate.}  Trapped current per RC (panel A), internal efficiency (panel B) and external efficiency (panel C) for the entire chlorosome coupled to a dimeric baseplate computed assuming  thermalizazion among all the aggregates. Both the trapped current and the efficiency have been obtained by solving the rate equation system shown in Eq.~\eqref{rate} at the steady state. The system contains $132840$ BChl {\it c} molecules distributed among three MT adjacent concentric cylinders, each including $44280$ BChl molecules, on a baseplate composed of $3350$ BChl {\it a} molecules and with an area of $1145.7 \times 3075.3 \ \mbox{\AA}^2$. See Tab.~\ref{table_size} in Sec.~\ref{SI:table} of the {\it Supp.Info.} for more details about the geometry. The yellow box in panels (A-C) between the two vertical dashed lines represents the regime in which FMO complexes typically work ($0.023 \ \mbox{ps}^{-1}<k_{FMO}<0.044\ \mbox{ps}^{-1}$). Only for panel A the height of the yellow box represents the range where the RCs closure rate  is typically found, which is between $ 1 $ and $10\  \mbox{ms}^{-1}$ ~\cite{hauska2001,van1998transient}. }
    \label{f:chlorosome}
\end{figure}

For each light-harvesting complex we computed the trapped current (Fig.~\ref{f:cyl} and Fig.~\ref{f:chlorosome} panel A) and the internal efficiency (Fig.~\ref{f:cyl} and Fig.~\ref{f:chlorosome} panel B). The external efficiency has been computed  for the entire chlorosome coupled to a baseplate (Fig.~\ref{f:chlorosome} panel C), and  for all the other single-cylinder models, see  Sec.~\ref{SI:ext} of the {\it Supp.Info.}.

In Fig.~\ref{f:cyl}, the trapped current and the internal efficiency have been computed using the three different approaches described in Sec.~\ref{model1} (full rate equations, partially thermalized rate equations, and fully thermalized rate equations) as a function of the $k_{FMO}$ trapping rate for the single-wall cylindrical models (MT, PD and RD) coupled to a dimeric baseplate. 
Circular open symbols stand for the full rate equations model (see Eqs.~\eqref{curr},~\eqref{eff}  in Sec.~\ref{model1})
dashed line for the partially thermalized model (Eqs.~\eqref{curr3} and~\eqref{eff3} in Sec.~\ref{model2}) and the continuous line is given by Eqs.~\eqref{curr2} and~\eqref{eff2} in Sec.~\ref{model3}, referring to the fully thermalized model.

The first two approaches, full rate equations and partially thermalized rate equations, yield very similar results, see open circles and dashed lines in Fig.~\ref{f:cyl} panels (A-B), providing an accurate description of the excitation energy dynamics. The validity of the partially thermalized rate equations is further supported by the fact that thermalization within these aggregates occurs in few ps, while exciton transfer between aggregates is slower and occurs on a timescale two orders of magnitude longer, tens of picoseconds.
The fully thermalized model does not provide equivalent results, see the continuous lines in Fig.~\ref{f:cyl}. It overestimates the efficiencies and trapped current and it is independent of the specific kind of cylinder considered (MT, PD, RD), see Eq.~\eqref{eff2app}. However for the MT model, which exhibits the largest values of trapped current and internal efficiency, the results computed by using this approach are close to the ones obtained by using the full rate equations and the partially thermalized model. 
Therefore, due to the large size of the biggest aggregate considered here, the chlorosome coupled to a baseplate comprising more than $10^5$ Bchl molecules, only the fully thermalized model has been considered.

Fig.~\ref{f:chlorosome} shows the trapped current (panel A), the internal efficiency (panel B) and the external efficiency (panel C) as a function of the $k_{FMO}$ trapping rate for the entire chlorosome obtained by using the fully thermalized model. 
One of the most interesting results is that the trapped current per RC (see panel A), which is of the order of 
$\sim 1 \ \mbox{ms}^{-1}$  in the realistic  range of values for $k_{FMO}$, matches the closure rate range of the RCs, see yellow box in Fig.~\ref{f:chlorosome}. This suggests that natural complexes may tend to optimize the number of excitations which arrive on the RC in order to match its operational time. Indeed, a lower number of excitations per second arriving on the RC would leave the RC in the open state for longer times, thus decreasing the rate of charge separation, while a larger number of excitation per second would not be used, since the RC could be in the closed state. 
In this model the RC closure rate has not been treated  dynamically and saturation effects have not been included, since these natural systems operate under low light intensity. In natural biological systems, once charge separation occurs, the RC can no longer undergo excitation, leading to the dissipation of any additional excitation energy.  While a full description of the RC dynamics is beyond the scope of this manuscript, it would be certainly interesting in the future to further analyze this issue. 
Panel B shows the internal efficiency that can reach values between $70-85\%$ for the entire chlorosome  for $0.023\ \mbox{ps}^{-1}<k_{FMO}<0.044 \ \mbox{ps}^{-1}$, the typical range of FMO  (see the yellow window). Note that this value of internal efficiency is in good agreement with Refs.~\citenum{chen2020architecture,guzik, dostal2016situ}.
On the other hand, the external efficiency (see panel C) has a value close to $\sim 1\%$ in agreement with our theoretical prediction, see also Sec.~\ref{SI:ext} in the \textit{Supp.Info.}. 

From previous results shown in Figs.~\ref{f:cyl} and~\ref{f:chlorosome}, we can understand that accounting for the entire chlorosome is essential to accurately capture the optimal operating conditions of GSB light-harvesting complexes. Indeed, only the chlorosome architecture can process a trapped current per RC  that matches the RC closure rate, thereby optimizing  the energy transfer process, see panel (A) in Fig.~\ref{f:chlorosome}. Conversely, the single MT cylinder fails to reproduce this optimal  condition. Nevertheless,  both models, single cylinder and entire chlorosome, yield similar values for the internal efficiency, close to the ones found in literature.  A more detailed justification for the necessity of the  chlorosome model, along with a systematic comparison to the single cylindrical MT model, is provided in Sec.~\ref{SI:size-dependence} of the \textit{Supp.Info.}.

\subsection{Study of the relationship between geometry and efficient energy transfer}
\begin{figure}[ht!]
    \centering
    \includegraphics[scale=0.3]{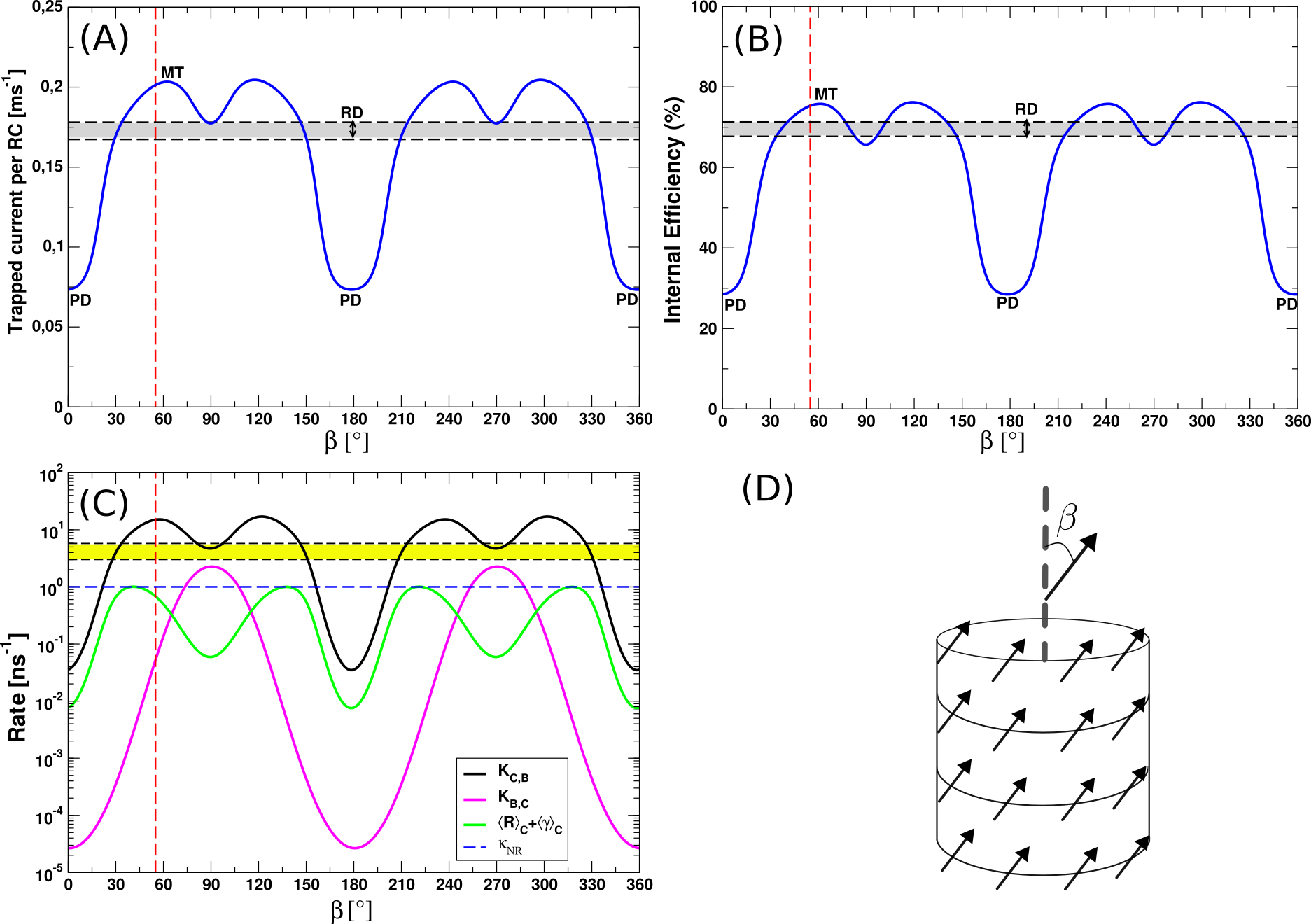}
    \caption{{\it Study of the relationship between trapped current and internal efficiency in single-wall aggregates and the orientation of TDMs.} Panels (A-B) show the trapped current and the internal efficiency as a function of the $\beta$ angle of each TDM and the main axis of the cylindrical aggregate. Each model contains $6000$ BChl {\it c} molecules in the cylinder and $2184$ BChl {\it a} molecules in the baseplate. The results have been obtained by assuming thermalization inside each aggregate, see Eqs.~\eqref{curr3} and~\eqref{eff3}.  The red dashed line represents $\beta=55^\circ$, the typical angle found for the MT model. The gray window between the two black dashed lines represents  the interval for the trapped current and efficiency computed for the RD model and averaging over $10$ realizations for random dipole orientations. The two black dashed lines in panels (A-B) represent the average value of trapped current and internal efficiency $\pm$ one standard deviation.    Panel (C): the rates used in Eq.~\eqref{3lev-rate} have been shown as a function of $\beta$ angle. The black and magenta curves are the MC-FRET rates from cylinder to baseplate and viceversa, respectively. The green curve is the  radiative decay rate from the cylinder, accounting for both fluorescence and  stimulated emission due to sunlight. The blue dashed line represents the non-radiative decay rate $\kappa_{NR}$. Finally the yellow window between the black dashed lines represents the average trapping rate from baseplate to RCs $\kappa=0.138 \cdot k_{FMO}$ that ranges from $\sim 3.17-6$~ns$^{-1}$. Panel (D): the orientation of the TDMs in a cylindrical structure is represented. $\beta$ is the angle between each TDM and the main axis of the cylinder. The positions of the TDM are the same used for the MT model, but we vary continuously the $\beta$ angle and we keep $\alpha$, the alternating angle of each dipole with respect to the tangent plane of the cylinder, equal to zero, see Ref.~\citenum{macroscopic} for more details about the geometry.  }
    \label{fig:beta}
\end{figure}

In order to explain the better efficiency of natural structures (MT model with a single-wall cylinder and the entire chlorosome) to transfer the excitation to the RCs with respect to the other single-wall cylinder models (RD and PD), we now analyze the dependence of the efficiency on the orientation of the TDMs in the cylinder and  the coupling strength between different cylinder models and the baseplate. 
In Fig.~\ref{fig:beta} (panels (A-B)) the trapped current and the internal efficiency  are computed for a single cylindrical aggregate containing $6000$ BChl {\it  c} coupled to a dimeric baseplate with $2184$ BChl {\it a} as a function of the TDMs orientation with respect to the cylinder main axis, given by the $\beta$ angle, see panel D of Fig.~\ref{fig:beta}. For $\beta=0$ we have the PD model, for $\beta=55^\circ$ we have the MT model~\cite{macroscopic,valzelli2024large,ganapathy2009,gunther2016}, for $\beta=\pi/2$ we have the TD model (tangent dipoles) discussed in Ref.~\citenum{macroscopic}.    Panels (A-B) show that among all the possible orientations of the TDMs, for $\beta=55^\circ$, which is the angle found for the MT model (see the red dashed line), the two figures of  merit reach the largest values. We also included the RD model (see the gray area between the horizontal black dashed lines), which has large values of both trapped current and internal efficiency, but always smaller than the MT model.

A possible explanation for the higher efficiencies of the natural model (MT) can be found analyzing the MC-FRET transfer rates, see Sec.~\ref{model2}: $K_{C,B}$ from cylinder to baseplate and $K_{B,C}$ from baseplate to cylinder.
Panel C in Fig.~\ref{fig:beta} shows the MC-FRET rates $K_{C,B}$ and $K_{B,C}$ (black and magenta curves respectively) given by Eq.~\eqref{mcfret}   and the thermal-averaged emission rate from the cylinder $\braket{R}_C+\braket{\gamma}_C$ (green curve), see Eq.~(\ref{eq:emission}) used in  Eq.~\eqref{3lev-rate} as a function of $\beta$ angle. From panel C of Fig.~\ref{fig:beta} we can see that the MC-FRET rate from cylinder to baseplate $K_{C,B}$ is  faster than the backward  rate and the thermal-averaged emission rate, ensuring most of the excitation in the cylinder is funneled to the baseplate. $K_{C,B}$  has the same behavior as the trapped current and the internal efficiency,  reaching the largest values for $\beta=55^\circ$, which is about $15 \ \mbox{ns}^{-1}$ in agreement with Ref.~\citenum{guzik}. Once the excitation reaches the baseplate, it is transferred to the RCs  through the FMO complexes by the average FMO trapping rate $\kappa$ (see the horizontal yellow window), that typically ranges from  $3.17$  to $6 \ \mbox{ns}^{-1}$   in GSB light-harvesting aggregates.  Other radiative and non-radiative processes are present in these systems (see the continuous green curve and the blue dashed line respectively), but for the MT model ($\beta=55^\circ$) these quantities are more than one order of magnitude less than the MC-FRET rate from cylinder to baseplate.   

These foundings explain that only specific geometries can exploit an efficient exciton energy transfer under natural sunlight even in presence of thermal dephasing comparable to room temperature energy and confirm that  natural models are the only ones capable to present a geometrical arrangement of TDMs such that both the trapped current and the internal efficiency are maximized with respect to the other mathematical models. Furthermore, the results shown Sec.~\ref{SI:cone} in the {\it Supp.Info.}, where the TDMs are randomly and uniformly distributed in a cone around their original direction, clearly show that the system can  maintain a degree of robustness against moderate angular fluctuations,  while it is optimized for the $\beta$ close to $55^\circ$.

See also Sec.~\ref{SI:comp} in the {\it Supp.Info.}, where the results for another natural model, the wild type (WT), has been provided and compared to the MT model. The values of the trapped current and internal efficiency found for the WT are close to the ones found for the MT model and higher than all the other mathematical models. 

\begin{figure}[ht!]
    \centering
    \includegraphics[width=\columnwidth]{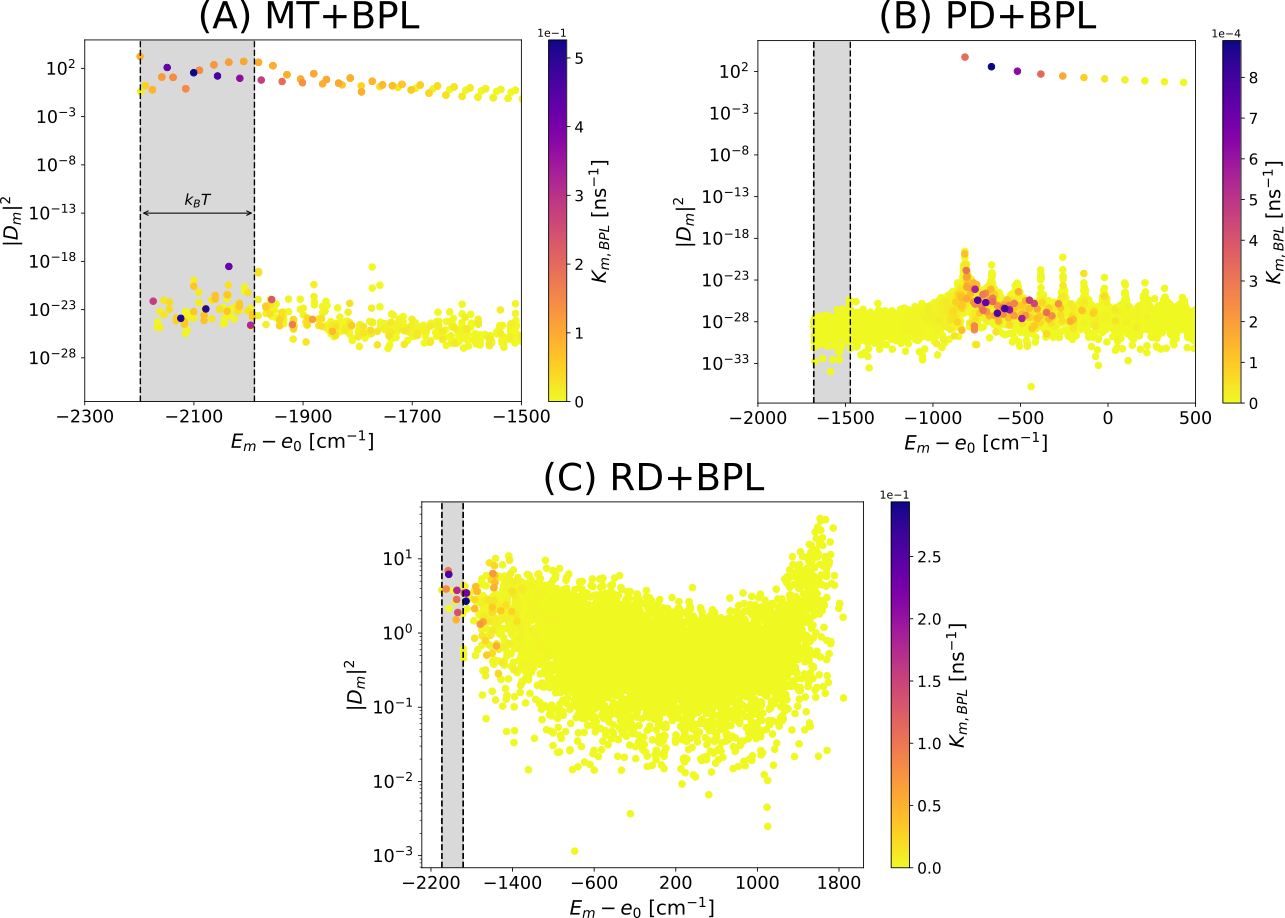}
    \caption{{\it Transfer rates $K_{m,BPL}$ between each eigenstate of the cylindrical  models and baseplate.} Panels (A-C): transfer rates computed  between each single eigenstate of the cylinder  and all the eigenstates of the baseplate for  MT and PD and RD models coupled to a dimeric baseplate. The rates $K_{m,BPL}=\sum_{n \in BPL} p_m K_{nm}$ are represented as a function of the dipole strength $|D_m|^2$ and the eigenvalues $E_m-e_0$ of the cylinder (see Eq.~\eqref{mc-fret} in Sec.~\ref{SI:fret} in the \textit{Supp.Info.}). Note that the dipole strength has been represented in logarithmic scale. The system contains $6000$ BChl {\it c} molecules in the cylinder and $2184$ BChl {\it a} molecules in the baseplate. The indices $m$ and $n$ refers to the eigenstates of the cylinder and baseplate respectively, while  $e_0$ is the excitation energy of BChl {\it c}. The gray area between the two  dashed lines represents the energy region between the lowest eigenvalue $E_1$ of the cylinder and $E_1+k_BT$ computed at room temperature ($T=300$ K). Panels (A-B): $K_{m,BPL}$ rates for MT and PD models, respectively. In these cases only the lowest portion of the spectrum, where the SRSs and the most relevant rates are present, is represented. Here the maximal dipole strength reaches $1760$ and $2638$ for MT and PD, respectively. Panel (C): $K_{m,BPL}$ rates for a single realization of random dipoles. In this case the whole spectrum has been represented. For RD model the maximal value of the dipole strength is about $\sim 50 $. }
    \label{fig:mc-fret}
\end{figure}

The origin of such a fast  MC-FRET rate $K_{C,B} $ in the MT model cylinder is investigated by comparing the spectral properties of the different models (MT, PD and RD) involved in the calculation of $K_{C,B} $. MC-FRET rates describe the incoherent excitation energy transfer from a donor to an acceptor unit. In our case the cylinder plays the role of the donor, while the baseplate is the acceptor unit. As already demonstrated in Sec.~\ref{SI:fret} in the \textit{Supp.Info.}, the MC-FRET   rate $K_{C,B} $ is strictly related to  the F\"orster rates computed between all the possible pairs of eigenstates of cylinder and baseplate, weighted on the Boltzmann factor for the donor unit (the cylinder). F\"orster rates given in Eq.~\eqref{forster} depends on the squared coupling strength between the TDMs associated to the eigenstates of the donor and acceptor units.  
In Fig.~\ref{fig:mc-fret}  the MC-FRET rates $K_{m,BPL}=\sum_{n\in BPL} p_m K_{nm}$ between each eigenstate of  the cylinder, indicated by the index $m$, and all the eigenstates of the baseplate is shown for MT, PD and RD models (panels (A-C)) as a function of the dipole strength and the eigenvalues of the cylinder (see Eq.~\eqref{mc-fret} in Sec.~\ref{SI:fret} in the \textit{Supp.Info.}). The most relevant terms that mainly contribute to the MC-FRET rate $K_{C,B}=\sum_{m \in C} K_{m,BPL}$ arise from eigenstates in the energy window between  the lowest eigenvalue of the cylinder $E_1$ and $E_1+k_BT$ ($T=300$ K), as dictated by the Boltzmann factor shown in Eq.~\eqref{mcfret}, see the gray window in panels (A-C) between the two dashed lines. Panel A shows that for the MT model the states with the largest MC-FRET rate lie in the lowest part of the spectrum within the gray area. These states dominate the total rate $K_{C,B}$, resulting in an overall  fast MC-FRET from the cylinder to the baseplate. Note that not only some superradiant states have a large MC-FRET but also some subradiant states. Indeed for such closeby aggregates it is not the dipole moment of the eigenstates which determined the efficiency of the energy transfer. 
On the other hand, for the PD model (see panel B) the eigenstates with the highest MC-FRET are far from the gray area. As a consequence, for the PD model MC-FRET is significantly lower than in the MT model. Note that for the PD model the $K_{m,BPL}$ rates are three orders of magnitude lower than the ones found for the MT model.
Finally for the RD model (panel C) a few eigenstates included in the gray area can still show  $K_{m,BPL}$ rate  comparable to those ones found for the MT model, however the trapped current and the efficiency are worse (see panels (A-B) in Fig.~\ref{fig:beta}).

\subsection{How static disorder affects exciton energy transfer}
\begin{figure}[ht!]
    \centering
    \includegraphics[width=\columnwidth]{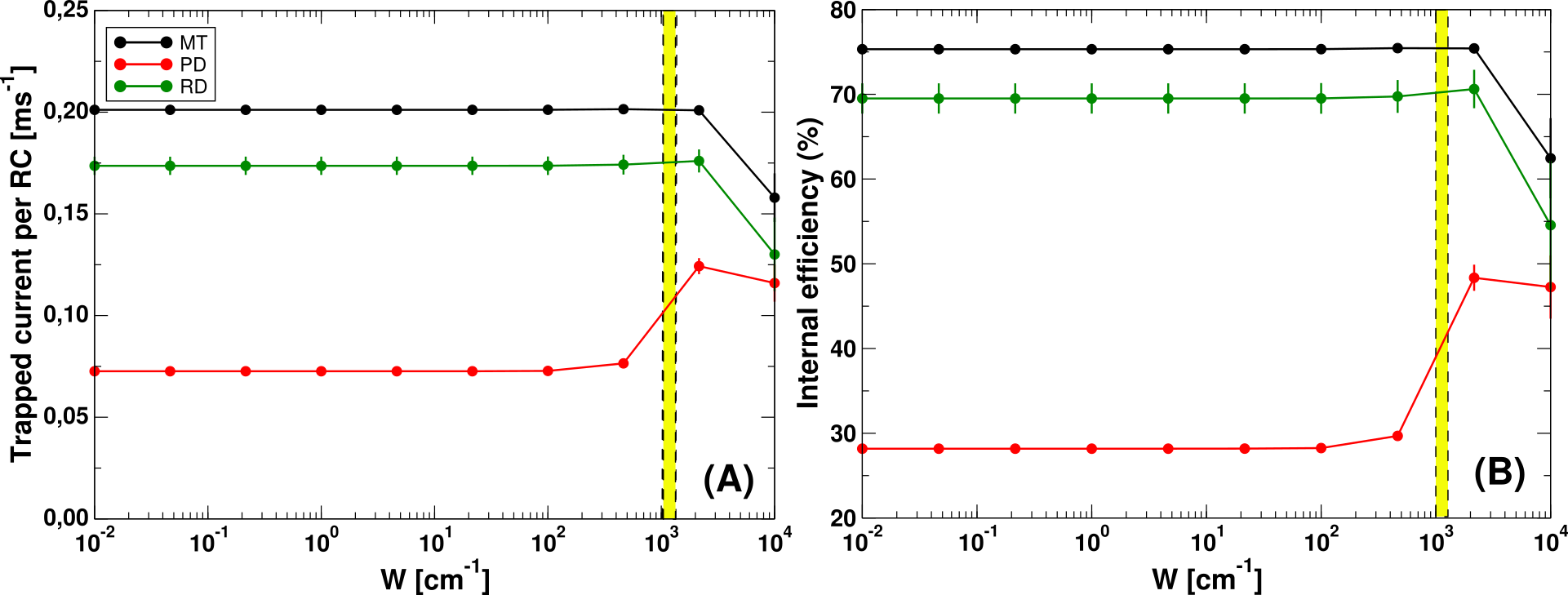}
    \caption{{\it Static disorder in single cylindrical complexes coupled to a dimeric baseplate: trapped current and internal efficiency.} Panels (A-B): trapped current per RC and internal efficiency as a function of static disorder strength $W$ for all the single cylinder models MT, PD and RD comprising $6000$ BChl \textit{c} coupled to a dimeric baseplate with $ 2184$ BChl \textit{a}. In all panels all the simulations have been run by assuming thermalization inside each aggregate. Eq.~\eqref{curr3} and~\eqref{eff3} in the main text have been used to compute respectively the trapped current per RC and the internal efficiency.  Numerical results  have been obtained by averaging over $100$ realizations for MT and PD models, while for the RD model $10 \times 10$ realizations have been done by changing both the randomness and the disorder strength. The yellow window between the two dashed vertical lines represent the typical disorder strength found in literature for GSB light-harvesting systems ($1014 \ \mbox{cm}^{-1} < W < 1368 \ \mbox{cm}^{-1}$)~\cite{guzik,molina2016superradiance,dostal2012two,pvsencik1998fast,dostal2016situ}.}
    
    \label{fig:dis}
\end{figure}

Considering the effect of disorder to model the environment is an important issue in light-harvesting complexes. Here we propose a more realistic study of the energy transfer in the single cylindrical aggregates (MT, PD and RD) by adding static  disorder. Static disorder is  modeled as space-dependent and time-independent fluctuations of the site energies, keeping the couplings between the molecules constant. This approach has been widely used in literature~\cite{fujita2014theoretical,celardo2014cooperative,klinger2023living,guzik,valzelli2024large,macroscopic}. The fluctuations which occur on a time scale much larger than the time scale of the dynamics are usually described as static disorder. 
Specifically, we consider energy fluctuations that are uniformly distributed around the excitation energy of the molecules $e_0$, between $e_0-W/2$ and $e_0+W/2$, where $W$ represents the disorder strength . In this study the trapped current and internal efficiency are computed as a function of static disorder for the three single cylindrical aggregates (MT-PD-RD models) by solving the rate equations given in Eq.~\eqref{3lev-rate} and assuming thermal equilibrium within each aggregate, see Fig.~\ref{fig:dis}. Static disorder affects both BChl \textit{a} in the baseplate and BChl \textit{c} in the  cylindrical aggregates and the values of  current and internal efficiency have been computed averaging over $100$ realizations of static disorder for MT and PD models and $10 \time 10$ realizations of disorder and random orientations of TDMs for the RD model respectively. Fig.~\ref{fig:dis} shows the trapped current (panel A) and internal efficiency (panel B) as a function of the disorder strength for the single cylinder models (MT, PD and RD) comprising $6000$ BChl \textit{c}  coupled to a dimeric baseplate with $ 2184$ BChl \textit{a}. The main results found in Fig.~\ref{fig:dis} demonstrate that all the models (MT,PD and RD) are robust to static disorder, showing constant values of both trapped current and internal efficiency up to a disorder strength $W$ close to $10^3 \ \mbox{cm}^{-1}$. For larger values of $W$ both figures of merit  in MT and  RD models start to drop, while for the PD model they  increase, reaching a common value for all the models. The yellow window between the two black dashed lines represents the realistic values of static disorder found in the literature~\cite{guzik,molina2016superradiance,dostal2012two,pvsencik1998fast,dostal2016situ} for GSB light-harvesting complexes ($1014 \ \mbox{cm}^{-1} < W < 1368 \ \mbox{cm}^{-1}$).

\clearpage
\section{Conclusions}
\label{m:conclusions}

In this manuscript we have studied the excitation energy transfer process in the entire light-harvesting apparatus of GSB, comprising more that $100000$ BChl molecules distributed in the chlorosome and in the baseplate, from  solar light absorption  to  the trapping of excitation in the RCs.  Sunlight is modeled through its  black body spectrum, taking into account the Earth-Sun distance. Also the coupling to room temperature thermal bath in presence of static disorder has been taken into account.  We  have developed three rate equations approaches in order to describe the process of the energy transfer: full rate equations model, with the highest computational costs, partially thermalized and fully thermalized models. The partially thermalized model  requires less numerical efforts and it gives results in agreement with the full rate equation approach.
All the approaches used to model excitation energy transfer (full rate equations, partially thermalized and fully thermalized rate equations) rely on the secular approximation and the assumption that the effects of multiple environments, the electromagnetic field and the phonon bath, can be treated independently in the Lindblad master equation. The problem of the interplay of multiple environments is very delicate,  Ref.~\citenum{giusteri2017interplay}: both the assumption that the environments can be treated independently and the secular approximation, might not be  justified in presence of resonances overlapping.  Indeed, the secular Redfield is valid if the energy level spacing is considerably larger than the Redfield rate. If this condition is not obeyed, one can use the full Redfield equation without the secular approximation for the nearly degenerate states, see Ref.~\citenum{cao1997phase}. Ideally, one could avoid secular approximation in the subspace of the closely spaced states and apply it outside the subspace. In perspective it would be very useful to analyze further how the interplay of different environments  affects energy transfer in antenna complexes.

The chlorosome has been modeled as three adjacent concentric cylindrical aggregates coupled to a dimeric baseplate. 
The trapped current per RC and the internal efficiency have been chosen as the two main figures of merit and we have demonstrated that for the chlorosome coupled to the dimeric baseplate the trapped current matches the RC closure rate, while the internal efficiency is about $\sim80\%$. In order to investigate the high efficiency of natural systems, we have considered  smaller systems composed of a  single wall cylindrical aggregate coupled to a baseplate:  the MT model, used to build up the entire chlorosome, and other mathematical models, obtained by changing the $\beta$ angle of each TDM with respect to the main axis of the cylinder. In particular we focused our study on two mathematical models: the RD model, where all the TDMs have random orientation in the space,  and the PD model, where all the TDMs are parallel to the cylinder axis ($\beta=0^\circ$). The main foundings presented in this paper  confirm that  natural models show the largest values of trapped current and internal efficiency and their behavior is strictly related to their geometry. In fact, natural models (single cylinder MT model and the entire chlorosome coupled to the dimeric baseplate) are characterized by the typical orientation of their TDMs with respect to the cylinder axis ($\beta=55^\circ$), that ensures a fast MC-FRET from the cylinder to the baseplate, faster than all the other radiative and non-radiative processes.

The emergence of fast energy transfer only in the natural models has been investigated by studying the MC-FRET from each single cylinder eigenstate to the baseplate $K_{m,BPL}$, where $m$ stands for an eigenstate of the cylinder. The main results  show that only the MT model can support rapid MC-FRET rates  within the $k_BT$ energy region at room temperature. These fast transfer rates  correspond   to both  superradiant states of the cylinder  and  subradiant ones, revealing that for such closely spaced aggregates, energy transfer efficiency  is not  determined by the net dipole moment of the eigenstates alone. To determine the general conditions which allow  efficient energy transfer in natural models remains a pivotal question in quantum biology and represents a compelling direction for future research.

Our results demonstrate the non trivial interplay of geometry and functionality in realistic light-harvesting systems, showing that the specific symmetry present in natural complexes is optimal for energy transfer. 
It should be noted, however, that natural architectures are subjected to complex environmental and conformational constraints. While we have addressed random dipoles and site energy disorder here, demonstrating the robustness of the natural model to the presence of static disorder comparable to room temperature energy and identifying a tolerance range for dipole orientations, exploring correlated positional and orientational disorder remains an interesting direction for future perspective.
Proving the sensibility of energy transfer on the specific dipole disposition and orientation, our analysis will inspire the design of artificial light-harvesting systems.

\begin{acknowledgement}
We acknowledge useful discussions with Massimo Trotta, Anna Painelli, Rafael A. Molina, Fausto Borgonovi and Julian Wiercinski.
We also acknowledge Alice Boschetti for her contribution in the realization of part of Fig.~\ref{chlorosome}.
This project has received funding from the European Union’s Horizon Europe Research and Innovation Programme under Grant Agreement No 101161312. 
\end{acknowledgement}
\clearpage

\bibliography{bibliography}

\clearpage
\newpage
\onecolumn

\begin{suppinfo}
\section*{Supplementary Information}
\renewcommand{\thesection}{S\arabic{section}}
\renewcommand{\thefigure}{S\arabic{figure}}
\renewcommand{\thetable}{S\arabic{table}}
\renewcommand{\theequation}{S\arabic{equation}}
\setcounter{section}{0}
\setcounter{figure}{0}
\setcounter{table}{0}
\setcounter{equation}{0}

Supporting Information (PDF): 
\begin{itemize}
    \item Solar irradiance
    \item Optical properties of BChl \textit{a} and \textit{c} 
    \item External efficiency of the GSB light-harvesting complex
    \item Table of the models
    \item Regime of validity of the Hamiltonian models
    \item General derivation of Lindblad Master Equations for independent phononic baths
    \item Multichromophoric transfer rates MC-FRET
    \item Parameters of the model
    \item Size-dependent energy transfer efficiency and trapped current in GSB light-harvesting architectures
    \item Analysis of TDMs distribution in a cone
    \item Comparison between MT and WT models
\end{itemize}

\section{Solar irradiance}
\label{SI:sun}
In this section a study of the solar irradiance has been given by using different approaches. 
The solar spectrum measured on the Earth is analyzed and compared with the Planck's law for a  black body at a finite temperature $T_S=5800$ K, typical of sunlight. In reality, the sunlight radiation is not exactly described by black-body radiation, as the photon experiences multiple scatterings upon arriving on the earth.  Yet, the measured spectrum of sunlight is sufficiently broad compared with the absorption spectrum of light-harvesting systems (see the absorption spectra of BChl molecules in Sec.~\ref{graph}), such that the predictions of our model calculation are not sensitive to the precise functional form of the sunlight spectrum~\cite{olvsina2014can}.

From the black body  theory, sunlight radiation is modeled as a photon bath at a given temperature $T_S$. Due to the fact that photons are boson particles, they can be studied with the Bose-Einstein statistics, that describes how a collection of non-interacting  identical particles may occupy a set of available discrete energy levels at thermodynamic equilibrium. Here this approach has been  used and  the occupation number of photons at a given frequency $\omega$ and temperature has been determined. In this case the Bose-Einstein distribution $n_S(\omega)$ reads:
\begin{equation}
n_S(\omega)= \dfrac{1}{e^{\hbar \omega/(k_BT_S)}-1}.
    \label{eq:bose}
\end{equation}
 
Fig.~\ref{Solar-irradiance} shows the comparison between experimental and theoretical models for solar irradiance as a function of the wavelength. The function here represented by the black line is the  well known Planck's law at finite temperature $T_S$  multiplied by a coefficient $\Omega_{S}$, see Eq.~\eqref{SI:planck}, that accounts for  the angle under which the Sun can be seen from Earth.  The Planck's law for a black body at a temperature $T_S$  at a given wavelength $\lambda$ reads:\\
\begin{equation}
\label{SI:planck}
g(\lambda,\Omega_{S})= \Omega_{S} \frac{2 h c^2}{\lambda^5} \frac{1}{\mbox{exp}\left( \frac{h c}{k_{B} T_S \lambda} \right)-1},
\end{equation}
where $h$ is the Planck's constant, $c$ the speed of light, $k_B$ the Boltzman's constant and $\Omega_{S}$ the solid angle under which the Sun is seen from the Earth.

\begin{figure}[!ht]
    \centering
    \includegraphics[width=\columnwidth]{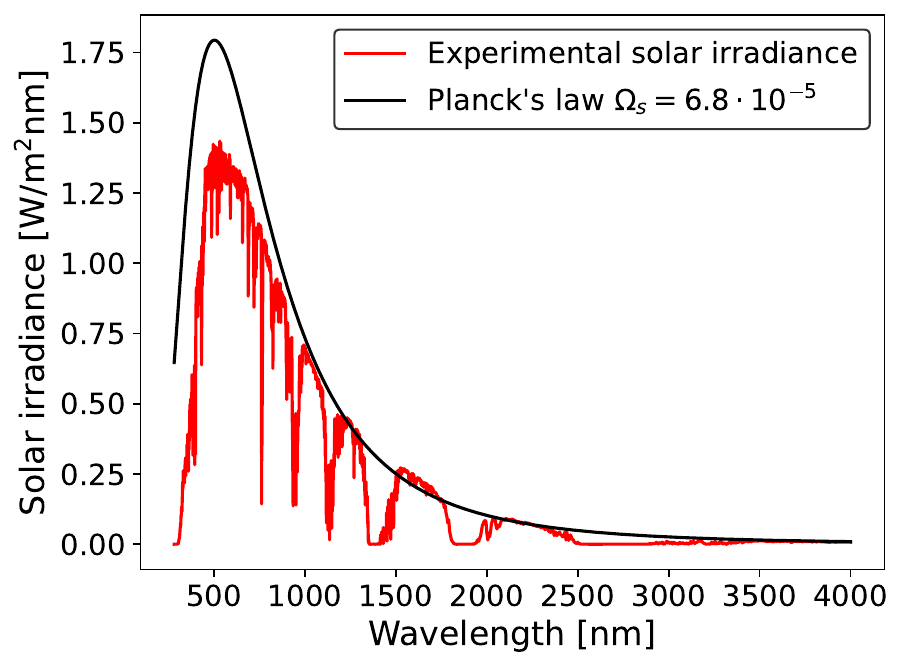}
  \caption{{\it Solar irradiance: comparison between experimental data measured on the Earth surface and black body theory.} Comparison between an experimental solar spectrum (red continuous line) and the Planck's law $g(\lambda,\Omega_{S})$ (black continuous line, see Eq.~\eqref{SI:planck}) for a black body at the Sun temperature $T_S=5800$~K multiplied by  $\Omega_{S}=6.8 \cdot 10^{-5}$,  the solid angle under which Sun is seen from the Earth.}
    \label{Solar-irradiance}
\end{figure}

In Ref.~\citenum{wurfel2016physics} the concentration of the solar radiation on the Earth has already been computed. Viewed from the Earth, the Sun has an angular diameter of $\alpha_S=0^\circ \  32'$, corresponding to a solid angle $\Omega_S$ given by
\begin{equation}
\label{SI:solid}
\Omega_S= 2 \pi \int_{0}^{\alpha_S/2} \mbox{sin}(\theta)  \,  \mbox{d}\theta = 6.8 \cdot 10^{-5}\ .
\end{equation}
Because of the small value of the solid angle $\Omega_S$, the sunlight energy current density measured on the Earth is reduced with respect to the total energy flux emitted by the Sun.
Integrating the experimental solar irradiance $\Phi(\lambda)$ (red curve in Fig.~(\ref{Solar-irradiance})), we obtain $P_{Sun}=\int_{\lambda_i}^{\lambda_f} \Phi(\lambda) d\lambda = 901.24 \ \mbox{W}/\mbox{m}^2$ (with cut-offs $\lambda_i = 280 \ \mbox{nm}$ and $\lambda_f = 4000 \ \mbox{nm}$), which is close to $1353\ \mbox{W}/\mbox{m}^2$ found in Ref.~\citenum{wurfel2016physics} referring to the solar irradiance measured outside the atmosphere.  
The small discrepancy between the two data is due to the Earth atmosphere and it strongly depends on the latitude, the weather conditions and the incident angle of sunlight with respect to the normal to the Earth surface. 

Our findings show that modeling sunlight through the black body radiation is a good approximation for sunlight. The spectrum computed using the black body theory can be compared to the experimental spectrum, minus a factor $\Omega_S$ that takes the Sun-Earth distance into account.  This approach has already been done in literature. In Ref.~\citenum{mattiotti2021bio} the Sun is seen as a black body at finite temperature $T_S$. The authors used the parameter  $f_S$, which represents the fraction of solid angle under which Sun is seen from Earth, to compute the occupation number of photons at a given wavelength and temperature.  $f_S$ found in Ref.~\citenum{mattiotti2021bio} is related to the solid angle $\Omega_S$ by the following relationship:
\begin{equation}
\label{SI:fs}
f_{S}=\Omega_S / 4\pi=\frac{\pi r_s^2}{4 \pi R_{ES}^2} = 5.4 \cdot 10^{-6},
\end{equation}
depending on the radius of the Sun $r_s$ and the Sun-Earth distance $R_{ES}$.

In reality, the sunlight radiation is not exactly described by black-body radiation, as the photon experiences multiple scatterings upon arriving on the earth.  Yet, the measured spectrum of sunlight is sufficiently broad compared with the absorption spectrum of light-harvesting systems (see the absorption spectra of BChl molecules in Sec.~\ref{graph}), such that the predictions of our model calculation are not sensitive to the precise functional form of sunlight spectrum~\cite{olvsina2014can}.

\clearpage

\section{Optical properties of BChl \textit{a} and \textit{c} }

In this section the optical properties (emission and absorption) for BChl \textit{a} and \textit{c} molecules typical of  GSB and PB light-harvesting complexes are computed by using different methods.

First a subsection for the emission rate is given. Then  the absorption rate is computed through the black body approach, where each BChl molecule is seen as a TLS and only the main absorption transition frequency and its transition dipole moment are taken into account. Then a more accurate estimation of the absorption is given by integration of the overlap between the absorption cross section and the solar irradiance for all the transition frequencies.

\subsection{Emission}
\label{SI:fluo}
The radiative decay rate of a molecule, modeled as a TLS, can be estimated as the inverse of the fluorescence lifetime $\tau_{fl}$ and reads:
\begin{equation}
\frac{\gamma}{\hbar}=\frac{1}{\tau_{fl}}=\frac{4\mu_0^2 \omega_0^3}{3 \hbar c^3},
    \label{gamma}
\end{equation}
where $\omega_0=\frac{2\pi c}{\lambda_0}$ is the transition frequency, while $\mu_0$ is its corresponding transition dipole moment.
Tab.~\ref{tab1} shows all the parameters for BChl \textit{a} and \textit{c}.

\begin{table}[!ht]
    \centering
    \begin{tabular}{|c|c|c|}
    \hline
       & BChl \textit{a} & BChl \textit{c} \\
       \hline
      TDM $\mu_0$ [Debye]~\cite{Note2} & 10 & 5.6 \\
      \hline
      Trans. wavelength $\lambda_0$ [nm] & 780 & 670 \\
      \hline
      $\gamma/\hbar \ [\mbox{s}^{-1}]$& $7\cdot 10^7$ & $3.2\cdot 10^7$ \\
      \hline
      $\tau_{fl}$ [ns] & 15  & 30 \\
      \hline
    \end{tabular}
    \caption{{\it Emission in BChl molecules.} The table shows the emission parameters for BChl \textit{a} and \textit{c} molecules found in literature~\cite{macroscopic,linnanto,connolly1982effects,blankenship2021molecular}.}
    \label{tab1}
\end{table}

\subsection{Absorption}
\label{SI:abs}
\paragraph{Black body theory} 
\label{bb}
Here we compute the absorption rate for BChl \textit{a} and \textit{c} through the black body theory and assuming that each BChl molecule can be treated as a TLS with excitation energy $e_0$ and TDM $\vec{\mu_0}$, given in Tab.~\ref{tab1}. We can express the absorption rate as follows,
\begin{equation}
P_{abs}=\frac{\gamma}{\hbar} n_S(\omega) f_S e_0~,
    \label{abs-rate}
\end{equation}
where $e_0=hc/\lambda_0$ is the excitation energy of single BChl molecules  (see Tab.~\ref{tab1} for the values of $\lambda_0$ for BChl \textit{a} and \textit{c}), $n_S(\omega)$ represents the  occupancy of photons at the Sun temperature $T_S=5800$~K for a given frequency $\omega$ (see Eq.~\eqref{eq:bose}), while $f_S$ is the fraction of solid angle from which sunlight is seen from the Earth.  

Using the parameters shown in Tab.~\ref{tab1}, the absorption rate for BChl \textit{a} and \textit{c} is computed, showing good agreement with the data found in literature~\cite{blankenship2021molecular}:
\begin{itemize}
    \item BChl \textit{a}: $P_{abs}=4\cdot 10^{-18}$ W, corresponding to $N_{abs}=15.8 \ \mbox{s}^{-1}$ photons per second;
    \item BChl \textit{c}: $P_{abs}=1.3\cdot 10^{-18}$ W, corresponding to $N_{abs}=4.6 \ \mbox{s}^{-1}$ photons per second.
\end{itemize}

\paragraph{Absorption cross section}
\label{graph}
Refs.~\citenum{borrego1999molar,connolly1982effects,malina2021superradiance,pvsenvcik2014chlorosomes,stanier1960chlorophylls} report the extinction coefficient $\epsilon$ for BChl molecules in units of $\mbox{M}^{-1} \mbox{cm}^{-1}$. Through an easy calculation already found in Ref.~\citenum{blankenship2021molecular}, the absorption cross section in $\mbox{cm}^2$ can be obtained as
\begin{equation}
\sigma = \frac{\ln(10)\epsilon c_m}{n} = \frac{\ln(10)10^3 \epsilon}{N_{av}},
    \label{eq:cross}
\end{equation}
where $c_m$ and \textit{n} are respectively the concentration of the solution in molarity [M] and in [$\mbox{cm}^{-3}$], $N_{av}$ is the Avogadro's number and $c_m = n \frac{10^3}{N_{av}}$ is the relationship between $c_m$ and \textit{n}.

Fig.~\ref{fig:abs-cross} shows the solar irradiance $\Phi(\lambda)$ (black line) and the absorption cross section for both BChl \textit{a} ($\sigma_a$ in red line) and \textit{c} ($\sigma_c$ in blue line). 

\begin{figure}
    \centering
    \includegraphics[width=\columnwidth]{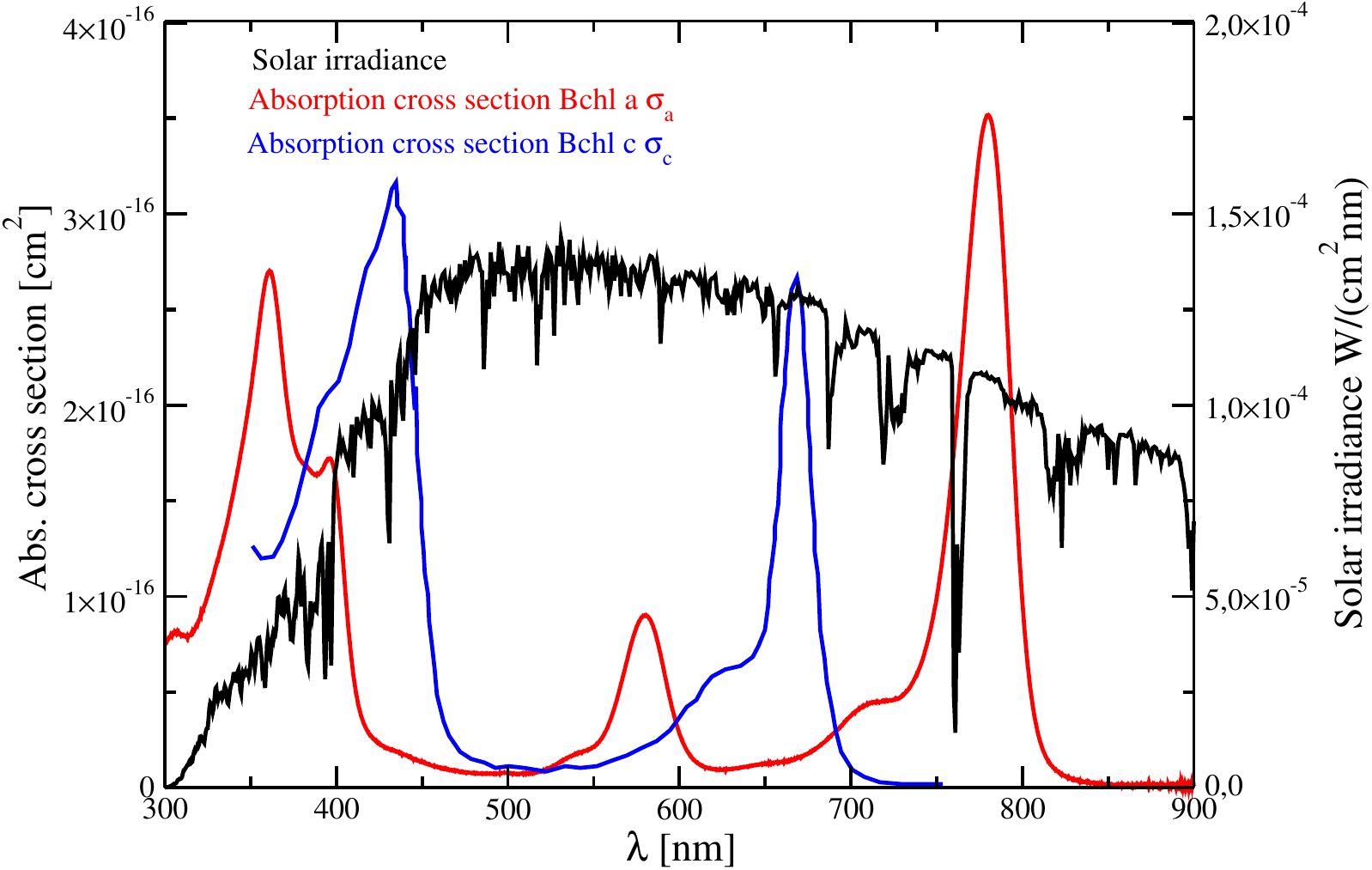}
    \caption{{\it Solar irradiance and absorption cross section for BChl a and c.} Solar irradiance (black line) and absorption cross section for BChl \textit{a} (red  line) and \textit{c} (blue line) are represented as a function of the transition wavelength. The absorption cross section has been computed by Eq.~\eqref{eq:cross} from experimental data of the extinction coefficient found in literature~\cite{borrego1999molar,connolly1982effects,malina2021superradiance,pvsenvcik2014chlorosomes,stanier1960chlorophylls}. }
    \label{fig:abs-cross}
\end{figure}

Finally the number of photons absorbed by each BChl has been computed by computing numerically the following integral:
\begin{equation}
N_{abs} = \int_{\lambda_i}^{\lambda_f} d\lambda\, \Phi(\lambda)\, \sigma_{a,c}(\lambda)\, \frac{\lambda}{hc}~.
    \label{int}
\end{equation}

The number of photons absorbed by a BChl is given by
\begin{itemize}
    \item $N_{abs}=12.5 \ \mbox{s}^{-1}$ for BChl \textit{a}
    \item $N_{abs}=8.6 \ \mbox{s}^{-1}$ for BChl \textit{c}
\end{itemize}
in good agreement with the previous calculations obtained by using the black body approach  and the experimental data reported in literature~\cite{blankenship2021molecular}.

\section{External efficiency of the GSB light-harvesting complex}
\label{SI:ext}
Here we compute the external efficiency for the entire light-harvesting complex of GSB, comprising the chlorosome and the baseplate. Chlorosomes contain mainly BChl \textit{c}, while the baseplate is formed by BChl \textit{a} molecules. We assume that the photons flux on Earth from sunlight is 
\begin{equation}
I_{Sun}= \int_{\lambda_i}^{\lambda_f} d\lambda\, \Phi(\lambda)\, \frac{\lambda}{hc}=1.75 \cdot 10^{21} \ \mbox{s}^{-1}\mbox{m}^{-2}~,  
\end{equation}
obtained by integration of the solar irradiance given in Fig.~\ref{fig:abs-cross} from $\lambda_i=280  \ \mbox{nm}$ to $\lambda_f = 4000\ \mbox{nm}$ and in agreement with the results provided in Ref.~\citenum{blankenship2021molecular}. 
Using the results found in the previous sections, the total absorption rate in the GSB light-harvesting system is computed.
The total absorption rate in the chlorosome and in the baseplate is given respectively by:
\begin{itemize}
    \item $N_{chl}=132840$ BChl \textit{c} and $N_{abs}^{c}=132840 \cdot 8.6 = 1184299 \ \mbox{s}^{-1}$,
    \item $N_{BPL}=3350$ BChl \textit{a} and $N_{abs}^{a}=3350 \cdot 12.5 = 41875 \ \mbox{s}^{-1}$,
\end{itemize}
where $3350$ and $132840$ are the total number of BChl molecules in the baseplate and in the chlorosome respectively.
If we assume the baseplate area $A_{BPL}=3.5 \cdot 10^{-14} \ \mbox{m}^2$, the absorption rate per unit area in the GSB reads: 
\begin{equation}
\label{eq:iabs}
I_{abs}=\frac{N_{abs}^c + N_{abs}^a}{A_{BPL}}=3.4 \cdot 10^{19} \ \mbox{s}^{-1} \mbox{m}^{-2}.
\end{equation}

Finally the absorption efficiency for the GSB can be estimated as:
\begin{equation}
\eta_{abs} = \frac{I_{abs}}{I_{Sun}}= 1.9\%.
    \label{eff1}
\end{equation}

For light-harvesting complexes three main quantities have been determined: the internal efficiency $\eta_{int}=\dfrac{I_{RCs}}{I_{abs}}$, the external efficiency $\eta_{ext}=\dfrac{I_{RCs}}{I_{Sun}}$ and the absorption efficiency $\eta_{abs}$ defined above in Eq.~\eqref{eff1}. In our theoretical model  we have found values of the internal efficiency between $70\%<\eta_{int}<85\%$ for the chlorosome, close to the values found in literature~\cite{kassal2}. Assuming the absorption efficiency is about $\sim 1.9\%$ for the chlorosome,
the external efficiency has been determined by\begin{equation}
\eta_{ext}=\eta_{abs} \times \eta_{int} 
    \label{eq:ext}
\end{equation}
and we found values between $1.3-1.6\%$.

These calculations represent an approximate estimate of the external efficiency in the GSB antenna complexes  in good agreement with the numerical results found by using the rate equations approach. Fig.~\ref{num-ext-eff} shows the external efficiency obtained by using Eq.~\eqref{ext-eff} in the main text as a function of the $k_{FMO}$ trapping rate  for the chlorosome (see the blue continuous line). The results show that the theoretical prediction given above is in agreement with numerical simulations, considering realistic values for the $k_{FMO}$ trapping rate  represented in the yellow window in Fig.~\ref{num-ext-eff}. 
Fig.~\ref{num-ext-eff} also shows  the external efficiency computed by numerical simulations for all the three single cylinder models (MT, PD and RD) coupled to the baseplate. Here systems formed by a single cylindrical aggregate  with $6000$ BChl \textit{c} and a length of $821.7 \ \mbox{\AA} $ and a dimeric baseplate with $2184$ BChl \textit{a}   and an area of $2739.1 \times 550.8 \ \mbox{\AA}^2$ have been considered.  For single cylinder aggregates lower values of the external efficiency have been found. Also these numerical results can be compared to the theoretical prediction presented in Tab.~\ref{tab:eff}, where   a calculation of the external efficiency in the systems formed by a single cylindrical aggregate  has been done following the same approach used for the chlorosome. 

The total absorption rate in the cylinders and in the baseplate are given respectively by:
\begin{itemize}
    \item $N_{abs}^c = 51600 \ \mbox{s}^{-1}$, 
    \item $N_{abs}^a = 27300 \ \mbox{s}^{-1}$.
\end{itemize}
The absorption rate per unit area in all the models (MT, PD and RD) is given by Eq.~\eqref{eq:iabs} and reads $I_{abs}= 5.26\cdot 10^{13} \ \mbox{s}^{-1} \mbox{m}^{-2}$, while the absorption efficiency is around $\eta_{abs}\approx0.3\%$ for MT, PD and RD models.  
The following table shows the external efficiency calculated  for the systems comprising a single cylinder  and a baseplate. 

\begin{table}[!ht]
    \centering
    \begin{tabular}{ccc}
    \hline
    \textbf{Model} & \textbf{Int. Efficiency}  & \textbf{Ext. Efficiency} \\
    \hline
      MT + BPL   & $70\%-80\%$ & $0.21\%-0.24\%$ \\
      PD + BPL   & $25\%-30\%$ &  $0.08\%-0.09\%$ \\
      RD + BPL   & $65\%-73\%$ &  $0.2\%$ \\
      \hline
    \end{tabular}
    \caption{Internal and external efficiencies computed for the single cylindrical models (MT, PD and RD) with $6000$ BChl \textit{c} and a length of $821.7 \ \mbox{\AA} $ placed above a dimeric baseplate with with $2184$ BChl \textit{a}   and an area of $2739.1 \times 550.8 \ \mbox{\AA}^2$. The internal efficiency has been taken from Fig.~\ref{f:cyl} in the main text assuming $0.023 \ \mbox{ps}^{-1}<k_{FMO}<0.044\ \mbox{ps}^{-1}$, while the external efficiency has been computed following Eq.~\eqref{eq:ext}.  }
    \label{tab:eff}
\end{table}

\begin{figure}[ht!]
    \centering
    \includegraphics[width=0.8\linewidth]{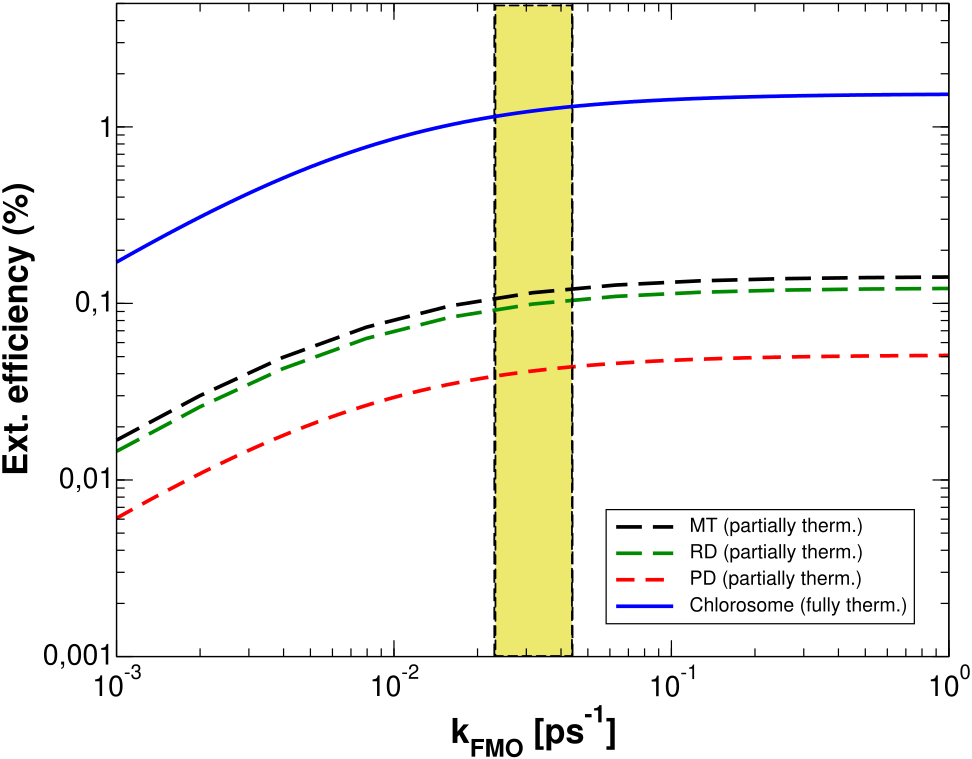}
    \caption{{\it External efficiency in the chlorosome and in the single cylinder models coupled to a dimeric baseplate.} The figure shows the external efficiency as a function of the $k_{FMO}$ trapping rate for different systems: the chlorosome  and single cylinder models (MT-PD-RD) coupled to a baseplate. The results for the entire chlorosome (blue continuous line) have been obtained by assuming thermalization among all the aggregates, while for the single cylinders model the partially thermalized rate equations approach has been used  (see the black, red and green dashed lines). For the RD model an average over 10 realizations of random TDMs orientations has been computed. The yellow window represents the typical range for $k_{FMO}$ trapping rate.  }
    \label{num-ext-eff}
\end{figure}
\clearpage

\section{Table of the models}
\label{SI:table}
Here a table comprising all the models taken into account in this manuscript is given. Tab.~\ref{table_size} contains the total number of molecules and the dimensions for both cylindrical aggregates and baseplate.
\begin{table}[!bht]
     \centering
     \includegraphics[scale=0.7]{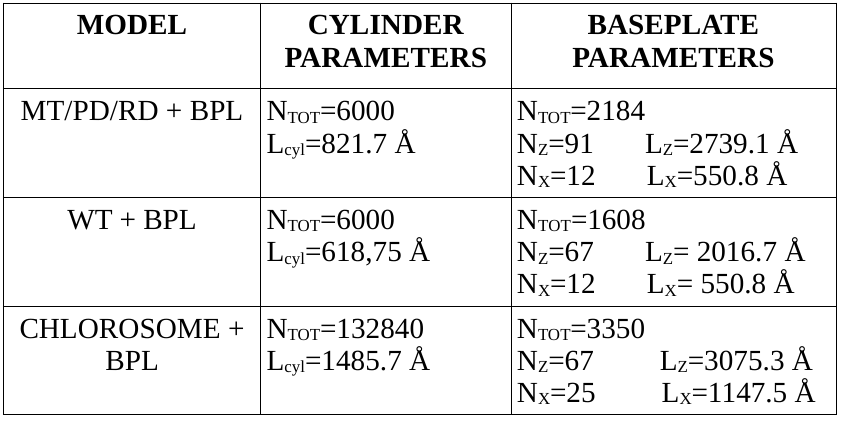}
     \caption{\emph{Cylinder and dimeric baseplate parameters.} The table shows the sizes and  the total number of BChl molecules in each aggregate (single cylinder models (MT, PD, RD and WT) and the entire chlorosome coupled to a baseplate). $N_{TOT}$ is the total number of molecules in the cylinder or baseplate. $L_{cyl}$ refers to the cylinders length, while $L_{z}$ and  $L_{x}$ are respectively the length and the width of the dimeric baseplate. $N_{z}$ and $N_{x}$ represent the number of dimers of the baseplate along the \textit{x} and \textit{z} directions. For the baseplate the total number of BChl molecules is given by $N_{TOT}=2\times N_{x}\times N_{z}$. Ref.~\citenum{guzik} has been taken into account to determine the relative dimensions between cylindrical aggregates and the dimeric baseplate.}
     \label{table_size}
\end{table}  

\clearpage

\section{Regime of validity of the Hamiltonian models}
\label{SI-perturbation}
In this section a deeper study of the three Hamiltonian models used in the main text has been addressed (see discussion in Sec.~\ref{m:ham} in the main text). In particular here we have studied the regime of validity of three Hamiltonian models (DH, HH and NHH) and we have compared our results of the dipole strength and radiative decay width obtained diagonalizing the three Hamiltonian models with the expectation value of the non-Hermitian part of the full Hamiltonian (NHH), according to the perturbation theory.  

As already described in the main text, the perturbed Hamiltonian is given by the NHH model and it reads:
\begin{equation}
\hat{H}_{NHH}=\sum_{i=1}^N e_0|i\rangle \langle i|+\sum_{i\neq j}\Delta_{ij}|i\rangle \langle j|-\frac{\mbox{i}}{2}\sum_{i,j=1}^{N}Q_{ij}|i\rangle \langle j| \ , 
  \label{si-eq-ham}
\end{equation} 
where $e_0=\hbar \omega_0$ is the excitation energy of the single emitter, while  $\Delta_{ij}$ and $Q_{ij}$ are the out-of diagonal terms with $Q_{ij}\ll \Delta_{ij}$.

The diagonal part of $\Delta_{ij}$ and $Q_{ij}$ are  given  by: 
\begin{equation}
  \Delta_{jj} = 0 \, , \qquad
  Q_{jj} = \frac{4}{3} \mu^2 k_0^3 = \gamma \, , \label{eq:gamma}
\end{equation}
where $\gamma$ is the radiative decay width associated to each emitter.

The non-Hermitian Hamiltonian gives complex eigenvalues $\varepsilon_{n}=\mbox{E}_n-\mbox{i}\frac{\Gamma_{n}}{2}$
where $\Gamma_{n}/\hbar$ is the radiative decay rate of the $n^{th}$ eigenstate in $\mbox{s}^{-1}$. 

In our systems the out-of diagonal term $Q_{ij}$ can be treated as a perturbation, since $Q_{ij}\ll \Delta_{ij}$. 
Let us define the unperturbed Hamiltonian, which coincides with the Hermitian part of the NHH shown in Eq.~\eqref{si-eq-ham} and reads:

\begin{equation} \label{si-eq-zero}
\hat{H}_{HH}=\sum_{i=1}^N e_0|i\rangle \langle i|+\sum_{i\neq j}\Delta_{ij}|i\rangle \langle j|,
\end{equation}
whose unperturbed eigenvalues and eigenstates are  $E^0_n$ and $\ket{E^0_n}$, respectively.~\footnote{Note that only in this section we introduce this notation to indicate  eigenvalues and eigenstates of the unperturbed Hamiltonian (HH), while in the main text we use  $E_n$ and $\ket{E_n}$. The reason of this choice is to distinguish clearly between the eigenstates and eigenvalues of the perturbed and unperturbed Hamiltonians.}

According to the non-degenerate perturbation theory,  the first-order energy shifts~\footnote{Note that here we use the non-degenerate perturbation theory to compute the energy-shift. This approach is valid only when the  eigenstates of the unperturbed system are non-degenerate. In this case the unperturbed system shows degeneracies that have been broken by adding static disorder $W=1 \ \mbox{cm}^{-1}$.  Such value of static disorder does not affect the spectrum and does not destroy superradiance, but it is enough to remove the degeneracy. }, which are the expectation values of the perturbation Hamiltonian $Q_{ij}$ while the system is in the unperturbed eigenstate $\ket{E^0_n}$, have been computed and compared to the radiative decay widths obtained by diagonalizing the full non-Hermitian Hamiltonian.

Fig.~\ref{fig:HH-DH-NHH} shows the maximal dipole strength obtained  diagonalizing the DH (black circles) and HH (red squares) models and the maximal decay width (blue stars) obtained with the NHH model as a function of the length of the MT cylinder rescaled over $\lambda_0$ (see Tab.~\ref{tab1} for the value of $\lambda_0$ for BChl {\it c}). Finally  the first-order energy shifts $\braket{E^0_n|Q|E^0_n}/\gamma$ (magenta diamonds) computed with the perturbation theory are compared with the previous results. In all models we added a source of static disorder ($W=1~\mbox{cm}^{-1}$, a single realization), such that degeneracies have been broken while superradiance is preserved. 
The geometrical model that we consider is the single-wall MT cylinder with $6$ TDMs on each ring with a length $L$ that goes from $ 74.7$ to $8291.7 \ \mbox{\AA}$. 
\begin{figure}[ht!]
    \centering
    \includegraphics[width=\columnwidth]{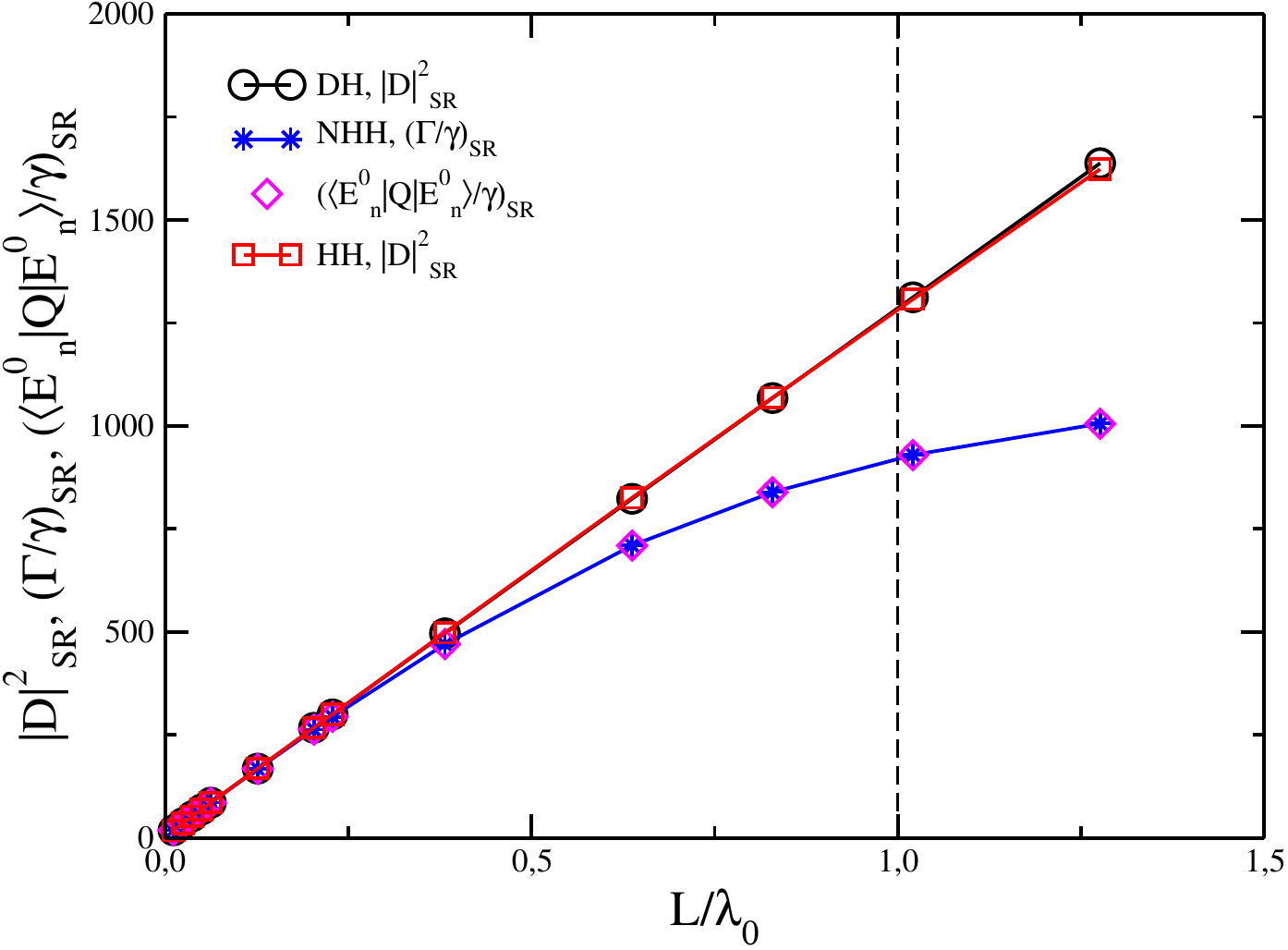}
    \caption{{\it Regime of validity of the three hamiltonian models and comparison with the perturbation theory approach in a single wall MT cylinder.}  Maximal dipole strength obtained by diagonalizing the DH (black cirlces) and HH (red squares) models and the maximal decay width (blue stars) obtained with the NHH model as a function of the length of the MT cylinder rescaled over $\lambda_0$. The first-order energy shifts $\braket{E^0_n|Q|E^0_n}/\gamma$ (magenta diamonds) computed with the perturbation theory is shown. The geometrical model considered is the single wall MT cylinder with $6$ TDMs on each ring with a length  $ 74.7 \le L\le 8291.7 \ \mbox{\AA}$. The total number of TDMs goes from $60$ to $60000$ for the longest cylinder that we consider here. The vertical dashed line ($L=\lambda_0$) stands for a cylinder length comparable to the excitation wavelength of the single emitter.  In all models a source of static disorder ($W=1 \mbox{cm}^{-1}$, a single realization) has been added in order to break the degeneracies in the unperturbed system, without changing the spectral properties of the system and the superradiance response. }
    \label{fig:HH-DH-NHH}
\end{figure}

In Fig.~\ref{fig:HH-DH-NHH} the maximal dipole strength computed with the DH and HH models increases linearly with the length of the system (see black circles and red squares in Fig.~\ref{fig:HH-DH-NHH}), while the radiative decay width (see blue stars) tends to saturate. Finally a comparison  with the perturbation theory is provided. Since the imaginary part $Q_{ij}$ is a small perturbation of the non-Hermitian Hamiltonian, the first-order energy shifts have been computed as the expectation value of $Q_{ij}$ with the unperturbed eigenstates $\ket{E^0_n}$ of the Hermitian Hamiltonian given by Eq.~\eqref{si-eq-zero}. 
The results show that the maximal value of  $\braket{E^0_n|Q|E^0_n}/\gamma$ (see magenta diamonds in Fig.~\ref{fig:HH-DH-NHH}) is in agreement with all the other three methods (DH, HH, NHH) when $L<\lambda_0$. When the system size increases the dipole strength description is no longer valid, while the perturbation approach is still in agreement with the NHH model, showing a tendency of the superradiant decay width to saturate with the length of the cylinder. 

From our analysis, we can conclude that all the  models (dipole strength computed with DH and HH, radiative decay width computed with NHH and perturbative approach) are in good agreement when the system size is small compared to the transition wavelength of the single emitter $\lambda_0$.  When the system size becomes comparable to $\lambda_0$ and the small volume approximation starts to fail, the dipole strengths computed with the DH and HH Hamiltonians are no longer good approximations to describe superradiance, but only the NHH model and the perturbative approach are still valid ($\Gamma_{SR}<\delta$, where $\delta$ is the mean energy level spacing). As already proved in Ref.~\citenum{valzelli2024large} by some of the authors of this manuscript, if we consider larger aggregates, such as the entire chlorosome, also the perturbation approach fails, because resonances start to overlap ($\Gamma_{SR}>\delta$) and the NHH becomes the only   model that can describe superradiance.  

It is known that disorder and thermal noise induce localization of the excitation. For one-dimensional systems the most relevant parameter is the localization length, instead of the system length $L$, and the small-volume limit refers to the regime in which the localization length is smaller than $\lambda_0$, regardless of the system size $L$. If we assume a disorder strength that induces a localization length smaller than $\lambda_0$, for one-dimensional systems the three Hamiltonian models  give always good agreement. In three-dimensional systems, however, due to the stronger influence of  long-range interactions, the behavior is not so trivial. Therefore, in this case the system size has been taken as the upper bound for the localization length.

\clearpage
\section{General derivation of  Lindblad Master Equations for independent phononic baths}
\label{SI:thermal}

\subsection{Hamiltonian and Master Equation for the whole system}

Let us consider a system made of $N$ coupled sites representing the BChl molecules in the cylinder and baseplate. All across this section  
for the sake of clarity we use indices $i$ and $j$ to indicate sites and $m$ and $n$ for the eigenstates, as already done in the main text. Each site is connected to an independent thermal bath. The $N$ baths have the same temperature and the same spectral density. The full Hamiltonian is
\begin{equation}
  \hat{H} = \hat{H}_S + \hat{H}_B + \hat{H}_I \, .
\end{equation}
Here  $\hat{H}_S$ is the system Hamiltonian (cylinder + baseplate) already described in Sec.~\ref{m:ham} in the main text, while the Hamiltonian term for the $N$ independent baths is
\begin{equation}
  \hat{H}_B = \sum_{k,i} \hbar\omega_k \hat{b}_{k,i}^\dag \hat{b}_{k,i}
\end{equation}
where the summation runs over the modes $k$ of the baths and the sites $i=1,\dots,N$
and the creation/annihilation operators follow the commutation rules $[\hat{b}_{k,j}, \hat{b}_{k',i}^\dag] = \delta_{k,k'} \delta_{j,i}$. Finally, the interaction is
\begin{equation}
  \hat{H}_I = \sum_{k,i} \hbar g_k \ket{i}\bra{i} \left( \hat{b}_{k,i}^\dag + \hat{b}_{k,i} \right)
\end{equation}
where $\hbar g_k$ is the coupling between a site and the position of the $k$-th mode of the bath and $\ket{i}$ is a state  in which the $i^{th}$ molecule is excited while all the others are in the ground state. Note that the derivation presented in this section does not rely on the form of $\hat{H}_S$, but only on the form of the interaction $\hat{H}_I$.

The dynamics of the full system is described by the Liouville Master Equation which, in the interaction picture, reads
\begin{equation}
  \label{fullme}
  \frac{d\hat{\rho}(t)}{dt} = -\frac{\mbox{i}}{\hbar} \left[ \hat{H}_I(t), \hat{\rho}(t) \right]
\end{equation}
where $\hat{\rho}(t)$ is the density matrix in the interaction picture and it is related to the density matrix in the Schrödinger picture $\hat{\rho}$ by
\begin{equation}
  \hat{\rho}(t) = e^{\mbox{i}(\hat{H}_S + \hat{H}_B)t/\hbar} \, \hat{\rho} \, e^{-\mbox{i}(\hat{H}_S + \hat{H}_B)t/\hbar} \, .
\end{equation}
Note that also the density matrix $\hat{\rho}$ is time-dependent both in the Schrödinger and in the interaction picture~\cite{breuer2002theory}.

The interaction Hamiltonian in the interaction picture can be factorized as
\begin{subequations}
  \label{hint}
  \begin{align}
    \hat{H}_I (t) &= \sum_{\omega} \sum_i e^{-\mbox{i}\omega t} \hat{A}_i(\omega) \otimes \hat{B}_i(t) \\
                  &= \sum_{\omega} \sum_i e^{\mbox{i}\omega t} \hat{A}_i^\dag(\omega) \otimes \hat{B}_i(t)
  \end{align}
\end{subequations}
where $\hat{A}_i(\omega)$ are operators acting on the system,
\begin{equation}
  \label{Asigma}
  \hat{A}_i(\omega) = \sum_{\substack{m,n\\(E_n-E_m)/\hbar=\omega}}  C^*_m(i) C_n(i) \ket{E_m}\bra{E_n} \, , \qquad  C_m(i)= \braket{i|E_m}
\end{equation}
with $\ket{E_m}$ and $\ket{E_n}$ being eigenstates of the system Hamiltonian ($\hat{H}_S\ket{E_m} = E_m\ket{E_m}$), and $\hat{B}_i(t)$ are hermitian operators acting only on the baths, given by
\begin{align}
  \label{Bj}
  \hat{B}_i(t) = \sum_k \hbar g_k \left( \hat{b}_{k,i}^\dag e^{\mbox{i}\omega_k t} + \hat{b}_{k,i} e^{-\mbox{i}\omega_k t} \right)
\end{align}

\subsection{Born and Markov approximations}

Eq.~\eqref{fullme} can be integrated from $0$ to $t$ to obtain
\begin{equation}
  \hat{\rho}(t) = \hat{\rho}(0) -\frac{\mbox{i}}{\hbar}\int_0^t dt' \left[ \hat{H}_I(t'), \hat{\rho}(t') \right]
\end{equation}
which, substituted back into \eqref{fullme}, gives the integro-differential equation
\begin{equation}
  \label{intdiff}
  \frac{d\hat{\rho}(t)}{dt} = -\frac{\mbox{i}}{\hbar} \left[ \hat{H}_I(t), \hat{\rho}(0) \right] -\frac{1}{\hbar^2}\int_0^t dt' \left[ \hat{H}_I(t), \left[ \hat{H}_I(t'), \hat{\rho}(t') \right] \right] .
\end{equation}

Now, let us perform the \emph{Born approximation}: the interaction between the system and the bath is assumed to be weak, so that it does not change relevantly the status of the bath. Formally, it translates in approximating the density matrix as
\begin{equation}
  \hat{\rho}(t) \approx \hat{\rho}_S(t) \otimes \hat{\rho}_B ,
\end{equation}
where the bath part $\hat{\rho}_B$ is also assumed to be a steady state of the bath, i.e. $[\hat{H}_B,\hat{\rho}_B]=0$.
Making the change of variable $\tau=t-t'$ into the integral, tracing over the degrees of freedom of the bath and assuming that the average value of the positions of the bath oscillators vanish
\begin{equation}
  \label{ave0}
  \text{tr}_B \left\{ \left[ \hat{H}_I(t), \hat{\rho}(0) \right] \right\} = 0 ,
\end{equation}
we have
\begin{equation}
  \frac{d\hat{\rho}_S(t)}{dt} = 
  -\frac{1}{\hbar^2}\int_0^t d\tau \, \text{tr}_B \left\{ \left[ \hat{H}_I(t), \left[ \hat{H}_I(t-\tau), \hat{\rho}_S(t-\tau)\otimes \hat{\rho}_B \right] \right] \right\} \, .
\end{equation}

Now let us perform the \emph{Markov approximation}: the memory effects between the system and the bath are neglected, i.e. we approximate the density matrix as
\begin{equation}
  \hat{\rho}_S(t-\tau) \approx \hat{\rho}_S(t)
\end{equation}
(first Markov approximation) and we extend the integration to $\infty$ (second Markov approximation). This gives the Redfield Master Equation
\begin{equation}
  \label{redfield}
  \frac{d\hat{\rho}_S(t)}{dt} = 
  -\frac{1}{\hbar^2}\int_0^\infty d\tau \, \text{tr}_B \left\{ \left[ \hat{H}_I(t), \left[ \hat{H}_I(t-\tau), \hat{\rho}_S(t)\otimes \hat{\rho}_B \right] \right] \right\} \, .
\end{equation}

\subsection{Secular approximation }

Let us now rewrite Eq.~\eqref{redfield} more explicitly. We use the notation
\begin{equation}
  \left\langle \hat{C} \right\rangle_B = \text{tr}_B \left\{ \hat{C} \hat{\rho}_B \right\}
\end{equation}
to indicate average value of some operator $\hat{C}$ on the bath and h.c. for the Hermitian conjugate. Thus, Eq.~\eqref{redfield} becomes
\begin{equation}
  \label{redhc}
  \frac{d\hat{\rho}_S(t)}{dt} = \frac{1}{\hbar^2} \int_0^\infty d\tau \, \left[ \left\langle \hat{H}_I(t-\tau) \hat{\rho}_S(t)\hat{H}_I(t) \right\rangle_B \right. 
  \left. - \left\langle \hat{H}_I(t) \hat{H}_I(t-\tau) \hat{\rho}_S(t) \right\rangle_B \right] + \text{h.c.}
\end{equation}
Now we use Eq.~\eqref{hint} to write the interaction Hamiltonian explicitly as
\begin{subequations}
  \begin{align}
    \hat{H}_I (t) &= \sum_{\omega'} \sum_j e^{\mbox{i}\omega' t} \hat{A}_j^\dag (\omega') \otimes \hat{B}_j(t) \\
    \hat{H}_I (t-\tau) &= \sum_{\omega} \sum_i e^{-\mbox{i}\omega (t-\tau)} \hat{A}_i(\omega) \otimes \hat{B}_i(t-\tau)
  \end{align}
\end{subequations}
and, substituting into \eqref{redhc}, we have
\begin{equation}
  \frac{d\hat{\rho}_S(t)}{dt} = \sum_{\omega,\omega'} e^{\mbox{i}(\omega'-\omega)t} \sum_{j,i} \Gamma_{ji}(\omega,t) \left[ \hat{A}_i(\omega) \hat{\rho}_S(t) \hat{A}_j^\dag(\omega') \right.
  \left. - \hat{A}_j^\dag(\omega') \hat{A}_i(\omega) \hat{\rho}_S(t) \right] + \text{h.c.} \label{redfield2}
\end{equation}
where we have defined the rates $\Gamma_{ji}(\omega,t)$ as the half-sided Fourier transformed correlators of the bath
\begin{equation}
  \Gamma_{ji}(\omega,t) =\frac{1}{\hbar^2} \int_0^\infty d\tau \, e^{\mbox{i}\omega\tau} \left\langle \hat{B}_j(t) \hat{B}_i(t-\tau) \right\rangle_B \, .
\end{equation}
Since we assumed $\hat{\rho}_B$ stationary, the rates $\Gamma_{ji}(\omega)$ are independent of time, namely
\begin{equation}
  \label{gij}
  \Gamma_{ji}(\omega) = \frac{1}{\hbar^2}\int_0^\infty d\tau \, e^{\mbox{i}\omega\tau} \left\langle \hat{B}_j(\tau) \hat{B}_i(0) \right\rangle_B \, .
\end{equation}

Now let us perform the \emph{secular approximation}: we neglect the oscillating terms, assuming that the energy spacings between the system states induces Rabi oscillations which are faster than the relaxation time so that, in a coarse-grained time scale, these oscillations average to 0. So, keeping just the non-oscillating terms ($\omega=\omega'$), we get a Master Equation in the Lindblad form:

\begin{equation}
  \frac{d\hat{\rho}_S(t)}{dt} = \sum_{\omega} \sum_{j,i} \Gamma_{ji}(\omega) \left[ \hat{A}_i(\omega) \hat{\rho}_S(t) \hat{A}_j^\dag(\omega) \right. 
  \left. - \hat{A}_j^\dag(\omega) \hat{A}_i(\omega) \hat{\rho}_S(t) \right] + \text{h.c.} \label{lindblad}
\end{equation}

It is important to stress that the positivity of populations is guaranteed only by the secular approximation. Redfield Eq.~\eqref{redfield} can have solutions with negative populations.

\subsection{Explicit calculation of the rates}

Here we compute the rates $\Gamma_{ji}(\omega)$ for the specific case of identical independent thermal baths under consideration. Note that these rates appear both in the Redfield Master Equation~\eqref{redfield2} and in the Lindblad Master Equation~\eqref{lindblad}. By substituting the expressions~\eqref{Bj} into \eqref{gij} we have

  \begin{align}
    \label{gije}
    \Gamma_{ji}(\omega) = \int_0^\infty d\tau \, e^{\mbox{i}\omega\tau}   \sum_{k,j,k',i} g_k g_{k'} &\left[ e^{-\mbox{i}\omega_k\tau} \left\langle \hat{b}_{k,j} \hat{b}_{k',i}\right\rangle_B + e^{-\mbox{i}\omega_k\tau} \left\langle \hat{b}_{k,j} \hat{b}_{k',i}^\dag\right\rangle_B \right. \nonumber \\
													 &\left. + e^{\mbox{i}\omega_k\tau} \left\langle \hat{b}_{k,j}^\dag \hat{b}_{k',i}\right\rangle_B + e^{\mbox{i}\omega_k\tau} \left\langle \hat{b}_{k,j}^\dag \hat{b}_{k',i}^\dag\right\rangle_B \right] .
  \end{align}

Now we assume that the thermal baths are at thermal equilibrium, i.e.
\begin{equation}
  \hat{\rho}_B = \frac{e^{-\beta\hat{H}_B}}{\text{tr}_B\left\{ e^{-\beta\hat{H}_B} \right\}}
\end{equation}
where $\beta=1/(k_BT)$ is the inverse temperature. In this case one can show that the correlators in~\eqref{gije} are
\begin{subequations}
  \begin{align}
    \left\langle \hat{b}_{k,j} \hat{b}_{k',i}\right\rangle_B &= 0 \\
    \left\langle \hat{b}_{k,j}^\dag \hat{b}_{k',i}^\dag\right\rangle_B &= 0 \\
    \left\langle \hat{b}_{k,j} \hat{b}_{k',i}^\dag\right\rangle_B &= \delta_{k,k'} \delta_{j,i} \left( 1 + N_{BE}(\omega_k)\right) \\
    \left\langle \hat{b}_{k,m}^\dag \hat{b}_{k',n}\right\rangle_B &= \delta_{k,k'} \delta_{j,i} N_{BE}(\omega_k)
  \end{align}
\end{subequations}
where we have defined the Bose-Einstein function
\begin{equation}
  N_{BE}(\omega_k) = \frac{1}{e^{\beta\hbar\omega_k}-1} \, .
\end{equation}
Thus we have
\begin{equation}
  \Gamma_{ji}(\omega) = \delta_{j,i} \int_0^\infty d\tau \, e^{\mbox{i}\omega\tau}   \sum_k g_k^2 
  \left[ e^{-\mbox{i}\omega_k\tau} \left( 1 + N_{BE}(\omega_k) \right) + e^{\mbox{i}\omega_k\tau} N_{BE}(\omega_k)\right].
\end{equation}
As regards the sum over $k$, we take the continuum limit
\begin{equation}
  \sum_{k} g_k^2 f(\omega_k) \rightarrow \int_0^\infty d\omega_k \, J(\omega_k) f(\omega_k) \, ,
\end{equation}
where $J(\omega_k)$ is the spectral density and $f(\omega_k)$ is the function in the squared brackets.

Now we perform the integral over $\tau$ using the relation
\begin{equation}
  \int_0^\infty d\tau \, e^{\mbox{i}\omega\tau} = \pi \delta(\omega) + \mbox{i} \text{P} \frac{1}{\omega}
\end{equation}
where P is the Cauchy principal value. So, we can split the rates into their real and an imaginary parts,
\begin{equation}
  \Gamma_{ji}(\omega) = \frac{1}{2}\gamma_{ji}(\omega) + \mbox{i}S_{ji}(\omega)
\end{equation}
which are, respectively,
\begin{subequations}
  \begin{align}
    \gamma_{ji}(\omega) = &2\pi \delta_{j,i} \int_0^\infty d\omega_k \, J(\omega_k) \left[ \delta(\omega-\omega_k) \left( 1 + N_{BE}(\omega_k) \right) + \delta(\omega+\omega_k) N_{BE}(\omega_k) \right]  \\
    S_{ji}(\omega) = &\delta_{j,i} \text{P} \int_0^\infty d\omega_k \, J(\omega_k) \left[ \frac{ 1 + N_{BE}(\omega_k) }{\omega-\omega_k} + \frac{ N_{BE}(\omega_k)}{\omega+\omega_k} \right]
  \end{align}
\end{subequations}
On integrating the $\delta(\omega\pm\omega_k)$ functions we have the real part of the rates
\begin{align}
    \label{gammaT}
    \gamma_{ji}(\omega) = 2\pi \delta_{j,i} \left[ J(\omega)  \left( 1 + N_{BE}(\omega) \right) + J(-\omega) N_{BE}(-\omega) \right] = \gamma^{(p)}(\omega) \delta_{j,i} \, ,
\end{align}
where we have defined the diagonal rate $\gamma^{(P)}(\omega)$ implicitly. The imaginary parts $S_{mn}(\omega)$ of the rates induce a constant renormalization of the site energies and thus they can be neglected.

\subsection{Lindblad Master Equations on the system eigenbasis}

The Lindblad Master Equations may be written back to the Schrödinger picture as
\begin{equation}
    \frac{d\hat{\rho}_S}{dt} = -\frac{\mbox{i}}{\hbar}\left[ \hat{H}_S, \hat{\rho}_S \right] + {\cal L}_T[\hat{\rho}_S]
\end{equation}
where the dissipator ${\cal D}_L[\hat{\rho}_S]$ in the Lindblad case~\eqref{redfield2} is

\begin{equation}
      {\cal L}_T[\hat{\rho}_S] = \sum_{\omega} \gamma^{(p)}(\omega) \sum_{i} \left[ \hat{A}_i(\omega) \hat{\rho}_S \hat{A}_i^\dag(\omega) -\frac{1}{2} \left\{ \hat{A}_i^\dag(\omega) \hat{A}_i(\omega), \hat{\rho}_S \right\} \right] \, .
\end{equation}

We now proceed to write the Master Equation on the eigenbasis of $\hat{H}_S$, i.e. we use the expression~\eqref{Asigma} to compute the terms $\bra{E_m} {\cal L}_T \ket{E_n}$.

For simplicity, let us assume that the coefficients $C_m(i)=\braket{i|E_m}$ are all real numbers. Since $\hat{H}_S$ is hermitian, it is always possible to find an eigenbasis that satisfies this requirement.

The dynamics of the populations is given by
\begin{equation}
    \label{DLindpop}
   \bra{E_m} {\cal L}_T \ket{E_m}= \sum_{n} \left( T_{m n} \rho_{n n} - T_{n m} \rho_{m m} \right)
\end{equation}
while for the coherences ($m \neq n$) we have
\begin{equation}
\label{DLindcoh}
  \bra{E_m} {\cal L}_T \ket{E_n} = - \Gamma_{m n} \rho_{m n}
\end{equation}
and
\begin{equation}
  \bra{0} {\cal L}_T \ket{E_m} = - \Gamma_{0m} \rho_{0m} \, .
\end{equation}
Defining the coefficients
\begin{equation}
  \label{lambR}
  \Lambda_{m n} = \sum_i |C_m(i)|^2 |C_n(i)|^2\, ,
\end{equation}

we derive the relaxation rates for the populations, expressed as
\begin{equation}
    \label{TLind}
  T_{m n} = \gamma^{(p)}[(E_n-E_m)/\hbar] \Lambda_{m n}.
\end{equation}

The Lindblad Master Equation for the Hamiltonian eigenstate populations reads explicitly: 
\begin{align}
   \frac{d\rho_{mm}}{dt}= &\sum_{n} \left\{ 2\pi \left[ J(\omega_n-\omega_m)  \left( 1 + N(\omega_n-\omega_m) \right) + J(\omega_m-\omega_n) N(\omega_m-\omega_n) \right]\Lambda_{mn} \rho_{nn} \right. \nonumber \\
   &\left.- 2\pi \left[ J(\omega_m-\omega_n)  \left( 1 + N(\omega_m-\omega_n) \right) + J(\omega_n-\omega_m) N(\omega_n-\omega_m) \right]\Lambda_{mn} \rho_{nn} \right\}~,
\end{align}
where we chose $J(\omega)=J(\omega_n-\omega_m)=k_{vib}\omega$ as the phonon spectral density already used  by some of the authors of this manuscript in~\citenum{mattiotti2021bio}.

\section{Multichromophoric transfer rates MC-FRET}
\label{SI:fret}

In this section a deeper study of the MC-FRET is given. MC-FRET has already been used in literature to model the incoherent energy transfer between different aggregates~\cite{jianshumcfretnjp,jianshumcfretjcp,strumpfer2012quantum} and it is computed by using the  F\"orster rate $K_{nm}$ defined in Eq.~\eqref{forster} in the main text.   
$K_{nm}$ depends mainly on two quantities: the overlap between the emission spectrum of the donor and the absorption spectrum of the acceptor, and the coupling strength between the eigenstates of the two aggregates. The former depends on the parameter $\Gamma_\phi$.

An aggregate absorption spectrum is given by~\cite{renger2002relation,babcock2024ultraviolet,mattiotti2022efficient}: 
\begin{equation}
\label{abs}
A(E) \propto \sum_n \mu^2|D_n|^2 A_n(E),
\end{equation}
where $|D_n|^2$ and $A_n(E)$ are, respectively, the dipole strength and the normalized lineshape for each $n$ eigenstate.

The emission spectrum, on the other hand, is
\begin{equation}
\label{fluo}
F(E) \propto \sum_{m} \mu^2|D_m|^2 F_m(E),
\end{equation}
where the emission lineshapes are multiplied by the thermal populations $p_m$, see Eq.~\eqref{eq:weights}  in the main text, namely
\begin{equation}
F_m(E) = p_m A_m(E).
\end{equation}

For high temperature and short bath correlation time~\cite{mattiotti2022efficient,jianshumcfretnjp}, we can neglect the phonon-induced Stokes and anti-Stokes shifts and approximate all the absorption lines as Lorentzians:
\begin{equation}
A_n(E) = \frac{2\Gamma_\phi}{\Gamma_\phi^2 + (E - E_n)^2}
\label{line}
\end{equation}
peaked at the eigenstate energy $E_n$ and with a dephasing-induced linewidth $\Gamma_\phi$. Note that the normalization condition gives $\int_{-\infty}^{+\infty}A_n(E)dE=2\pi$.

Fig.~\ref{overlap} shows the overlap between the emission and absorption spectra of MT, PD and RD  cylinders and baseplate respectively, assuming a Lorentzian linewidth of $500 \ \mbox{cm}^{-1}$ for both cylinders and baseplate spectra. The choice of the linewidth for the emission and absorption spectra is consistent with experimental data. See Ref.~\citenum{valzelli2024large} for a  more detailed comparison between experimental and numerical spectra in GSB antenna complexes, where a similar linewidth has been used. 

\begin{figure}[!ht]
    \centering
    \includegraphics[width=0.9\linewidth]{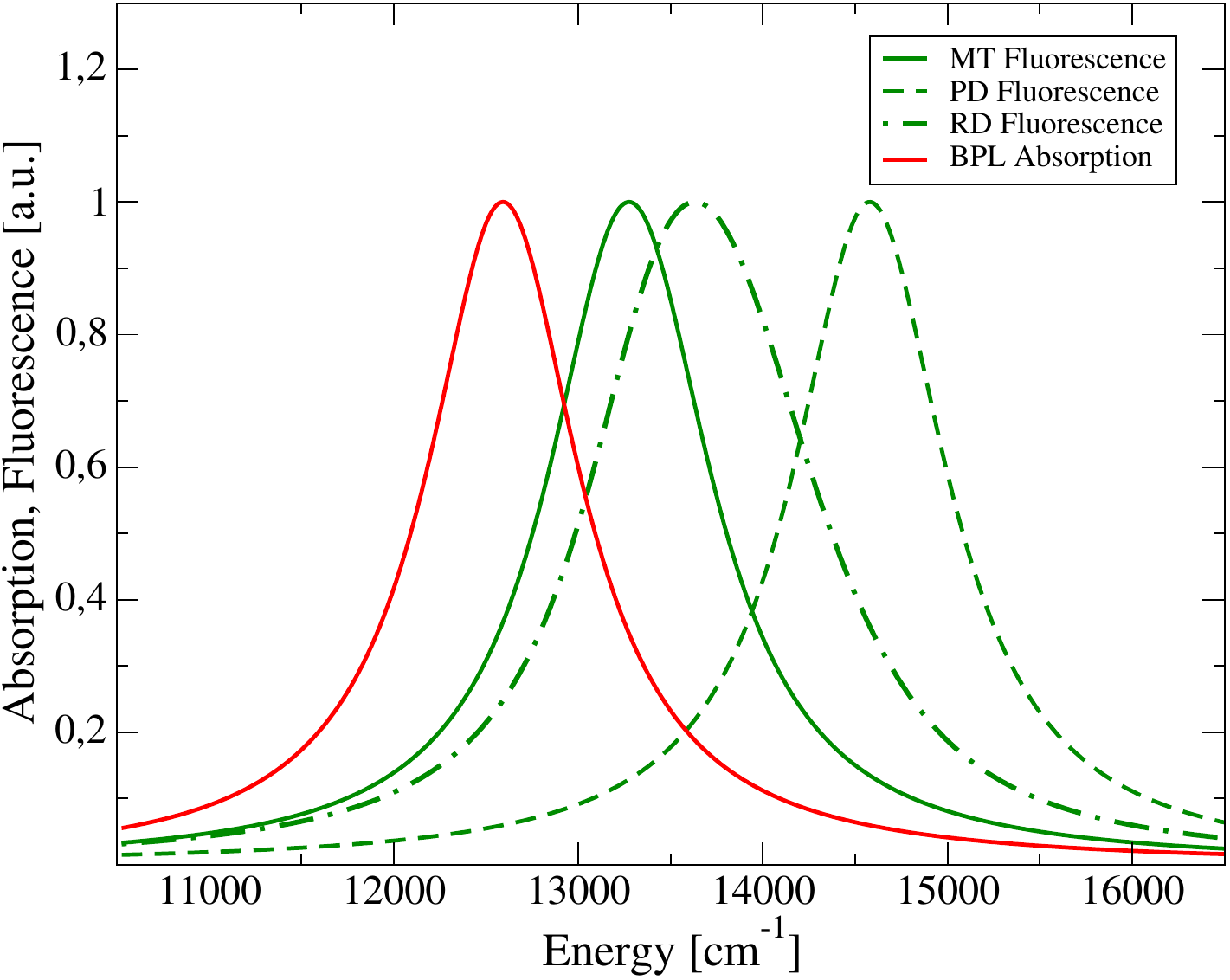}
    \caption{ {\it Emission and absorption spectra in MT, PD, RD cylinders and baseplate.} Normalized emission spectra for the donor aggregates in green color, see Eq.~\eqref{fluo}. Continuous line for the MT model, dashed line for the PD model and dashed-dotted line for the RD model. The red continuous line represents the absorption spectrum of the baseplate, see Eq.~\eqref{abs}. Here a Lorentzian lineshape with linewidth of $500 \ \mbox{cm}^{-1}$ has been considered for all the models. }
    \label{overlap}
\end{figure}

Furthermore the couplings $\Omega_{n,m}$ between cylinder and baseplate eigenstates are studied. If $\hat{H}_{DH}$ is the Hamiltonian of the entire system (cylinder $+$ baseplate) written on the site basis, see Sec.~\ref{m:ham} in the main text, the coupling strength between the $m^{th}$ eigenstate of the cylinder $\ket{E_m}$ and the $n^{th}$ eigenstate of the baseplate $\ket{E_n}$ reads:
\begin{equation}
\label{omega}
    \Omega_{n,m}= \bra{E_m}\hat{H}_{DH}\ket{E_n} =\sum_{i \in C,j \in B} C_m^*(i) C_n(j) H_{DH}(i,j),
\end{equation}
where $C_m(i)$ and $C_n(j)$ are the projections of the eigenstates on the sites basis, while $H_{DH}(i,j)$ is the coupling computed with the Frenkel Hamiltonian between sites $i$ and $j$ belonging respectively to the cylinder and baseplate.

Finally, the following calculations justify the choice of the parameter $\Gamma_\phi = 1000 \ \mbox{cm}^{-1}$, where $\Gamma_\phi/\hbar$ is the dephasing rate, in Eq.~\eqref{forster} in the main text as the sum of the absorption and emission spectra linewidths.

The MC-FRET rate is usually expressed as
\begin{equation}
K_{D,A} = \sum_{m \in D} \sum_{n \in A}  \frac{|\Omega_{n,m}|^2}{2\pi \hbar} \int_{-\infty}^{+\infty} F_m(E)A_n(E)dE= \sum_{m \in D} \sum_{n \in A} p_m \, K_{n,m},
\label{I.15}
\end{equation}
where $\Omega_{n,m}=\langle E_m | H_{DH} | E_n \rangle$ is the Hamiltonian matrix element between the $m$ donor eigenstate and the $m$ acceptor eigenstate computed in Eq.~\eqref{omega}.

Under the assumption made before for the absorption and emission lineshapes, the overlap integral in equation~\eqref{I.15} is analytically computed. If $\Gamma_d$ and $\Gamma_a$ are the linewidths for the donor and acceptor aggregates respectively, the overlap integral reads
\begin{equation}
    \label{eq:overlap-integral}
\int_{-\infty}^{+\infty} F_m(E)A_n(E)dE = p_m  \int_{-\infty}^{+\infty} \frac{4 \Gamma_d \Gamma_a}{[\Gamma_d^2 + (E-E_m)^2][\Gamma_a^2 + (E-E_n)^2]}dE.   
\end{equation}
The integral in Eq.~\eqref{eq:overlap-integral} can be solved using Jordan's lemma and the final expression is as follows:
\begin{equation}
    \label{eq:overlap-final}
    \int_{-\infty}^{+\infty} F_m(E)A_n(E)dE = p_m\frac{4\pi(\Gamma_d+\Gamma_a)}{(\Gamma_d+\Gamma_a)^2+(E_m-E_n)^2}.
\end{equation}
Eq.~\eqref{eq:overlap-final} shows that the overlap integral of the two Lorentzian functions with $\Gamma_d$ and $\Gamma_a$ linewidths is still a Lorentzian function peaked at $E_n-E_m$ and with a total dephasing-induced linewidth  which is the sum of $\Gamma_d$ and $\Gamma_a$. 

Therefore, if  $\Gamma_\phi = \Gamma_a + \Gamma_d $, we can express the MC-FRET rate in equation~\eqref{I.15} as

\begin{equation}
    \label{mc-fret}
K_{D,A} = \sum_{m \in D}\sum_{n \in A} p_m K_{n,m},    
\end{equation}
where the transfer rates between a $m$ donor eigenstate and a $n$ acceptor eigenstate are 
\begin{equation}
K_{n,m} = \frac{|\langle E_m | H_{DH} | E_n\rangle|^2}{\hbar} \cdot
\frac{2\Gamma_\phi}{\Gamma_\phi^2 + (E_m -E_n)^2}.
\label{I.16}
\end{equation}

Note that these rates are symmetric, $K_{n,m} = K_{m,n}$, causing  Eq.~\eqref{I.16} to break detailed balance. Since in the GSB aggregates that we consider here usually the donor and acceptor eigenvalues are not resonant, we correct these rates by using 
Eq.~\eqref{I.16} only for
energetically downward transitions, otherwise  taking $K_{n,m}=K_{m,n}e^{-(E_n-E_m)/k_BT}$. This method has already been used in Refs.~\citenum{baghbanzadeh2016geometry,kassal2} in order to describe excitation energy transfer in Purple bacteria light-harvesting aggregates.
In the main text (see Eq.~\eqref{forster}) the transfer rates $K_{n,m}$  have been given by adding this correction. As a consequence, if both donor and acceptor aggregates are made of one molecule, the MC-FRET gives forward and backward rates which are detailed  balance.   
\clearpage

\section{Parameters of the model}
\label{SI:parameters}

In this section a brief discussion of the main parameters adopted in our simulations has been provided in order to give a validation of the model under consideration. 
\begin{itemize}
    \item We kept the intrinsic radiative decay rate of each molecule  constant and we determined the inter-molecular couplings according to the radiative non-hermitian Hamiltonian shown Eq.~\eqref{eq:ham} in the main text, strictly dependent on the geometry of the system. The consistency of our modeling  has been verified by reproducing experimental spectroscopic data of absorption and emission spectra for GSB light-harvesting complexes, see Fig.~\ref{fig:abs-emis} and our previous work~\citenum{valzelli2024large}.   
\item The dephasing strength $\Gamma_{\Phi}$ used to determine the FRET rates between eigenstates belonging to different aggregates was tuned to match the linewidth of the experimental emission and absorption spectra. However, a $30\%$ variation of $\Gamma_{\Phi}$ has been considered in order to assess the parameter's sensitivity.   This is shown in Fig.~\ref{fig:change-GF}, where the trapped current per RC (panel A) and the internal efficiency (panel B) as a function of $\Gamma_{\Phi}$ are represented for a cylinder (MT model) with 6000 Bchl {\it c} coupled to a baseplate with 2184 Bchl {\it a} molecules.  The foundings shown in Fig.~\ref{fig:change-GF} result in only a small discrepancy compared to the value used to fit the spectra ($\Gamma_{\Phi}=1000 \ \mbox{cm}^{-1}$, which represents the sum of the absorption and emission linewidths).
\item Finally,  the non-radiative decay rate has been kept constant, adopting a standard value consistent with existing literature for these systems~\cite{baghbanzadeh2016geometry}. While the parameter $k_{vib}$ in the spectral density $J(\omega)$ has been chosen in order to ensure thermal relaxation within each aggregate in a few picoseconds.  
\end{itemize}

\begin{figure}
    \centering
    \includegraphics[width=\linewidth]{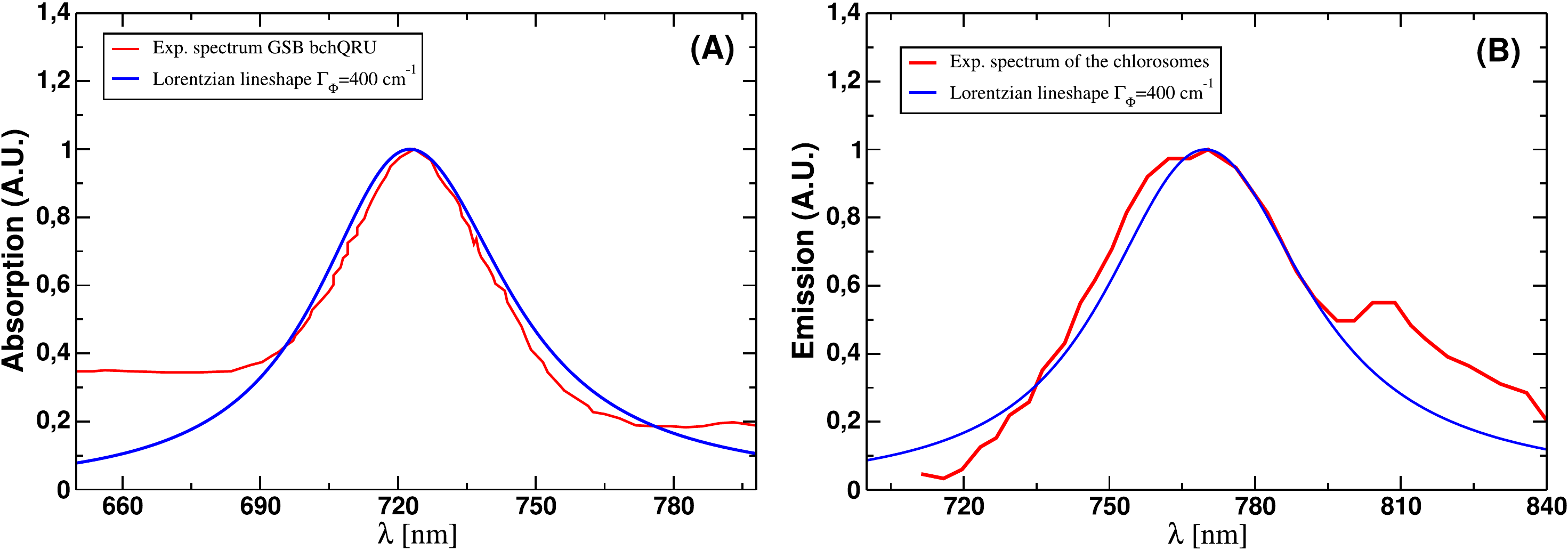}   
    \caption{{\it Absorption and emission spectra in GSB.} The figure shows the comparison between the experimental and numerical absorption and emission spectra for the GSB model, see panels (A) and (B), respectively. Panel (A):  experimental and numerical normalized absorption spectra for GSB bchQRU triple mutant are compared. The experimental absorption spectrum of whole cells of GSB bchQRU mutant (red solid line) taken from Ref.~\citenum{chew2007}  is compared with the numerical spectrum for a single cylinder with 180 rings (blue solid line), assuming a Lorentzian lineshape with homogeneous broadening $\Gamma_{\Phi} = 400 \ \mbox{cm}^{-1}$.  Panel (B): experimental and numerical normalized emission spectra for GSB are compared. The experimental emission spectrum of chlorosomes of GSB (red solid line) taken from Ref.~\citenum{malina2021superradiance}  is compared with the numerical spectrum for a single cylinder with 180 rings (blue solid line), assuming a Lorentzian lineshape with homogeneous broadening $\Gamma_{\Phi} = 400 \ \mbox{cm}^{-1}$ and room temperature.}
    \label{fig:abs-emis}
\end{figure}

\begin{figure}
    \centering
    \includegraphics[width=\linewidth]{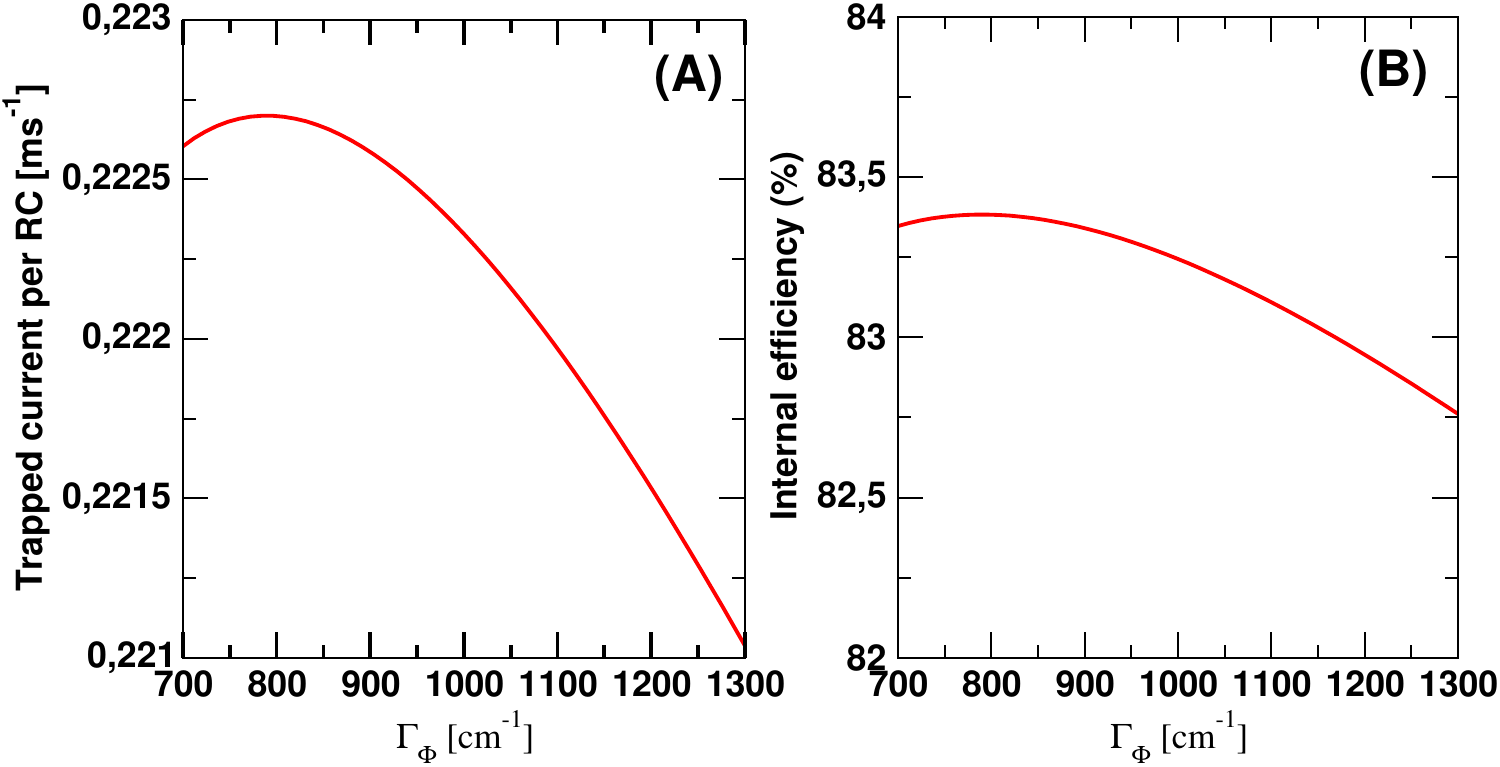}
    \caption{{\it Trapped current and efficiency in MT model vs dephasing linewidth $\Gamma_{\Phi}$.} The figure shows the trapped current (panel A) and the internal efficiency (panel B) as a function of the dephasing linewidth $\Gamma_{\Phi}$ for a cylinder (MT model) with 6000 Bchl {\it c} coupled to a baseplate with 2184 Bchl {\it a} molecules.   } 
    \label{fig:change-GF}
\end{figure}

\clearpage

\section{Size-dependent energy transfer efficiency and trapped current in GSB light-harvesting architectures}
\label{SI:size-dependence}

In this section a deeper study of the energy transfer efficiency and trapped current dependence on the system size is addressed. In our calculations we investigated two different geometries in order to describe natural models: a full-scale system comprising more than $10^5$ Bchl {\it c} molecules arranged on three adjacent concentric cylinders, that mimics the natural light-harvesting architectures, and a smaller model consisting of a single cylindrical aggregate of 6000 BChl {\it c} molecules (MT model).  Our results provided in the main text (Figs.~\ref{f:cyl} and~\ref{f:chlorosome}, panel B) demonstrate that both models achieve internal efficiencies between $70\%$ and $85\%$, consistent with literature values. However, the single cylinder model has a much smaller trapped current which fails to match the RC closure rate. Indeed, the single-cylinder system yields a trapped current an order of magnitude lower, resulting in sub-optimal operating conditions. Only the full chlorosome model generates a trapped current comparable to the RC closure rate, thereby optimizing the energy transfer process (Fig.~\ref{f:chlorosome}, panel A). These findings are summarized in Fig.~\ref{fig:size}, where the trapped current and the internal efficiency have been shown for the entire chlorosome and for a single cylinder (MT) with two different dimensions coupled to a baseplate.

\begin{figure}
    \centering
    \includegraphics[width=\linewidth]{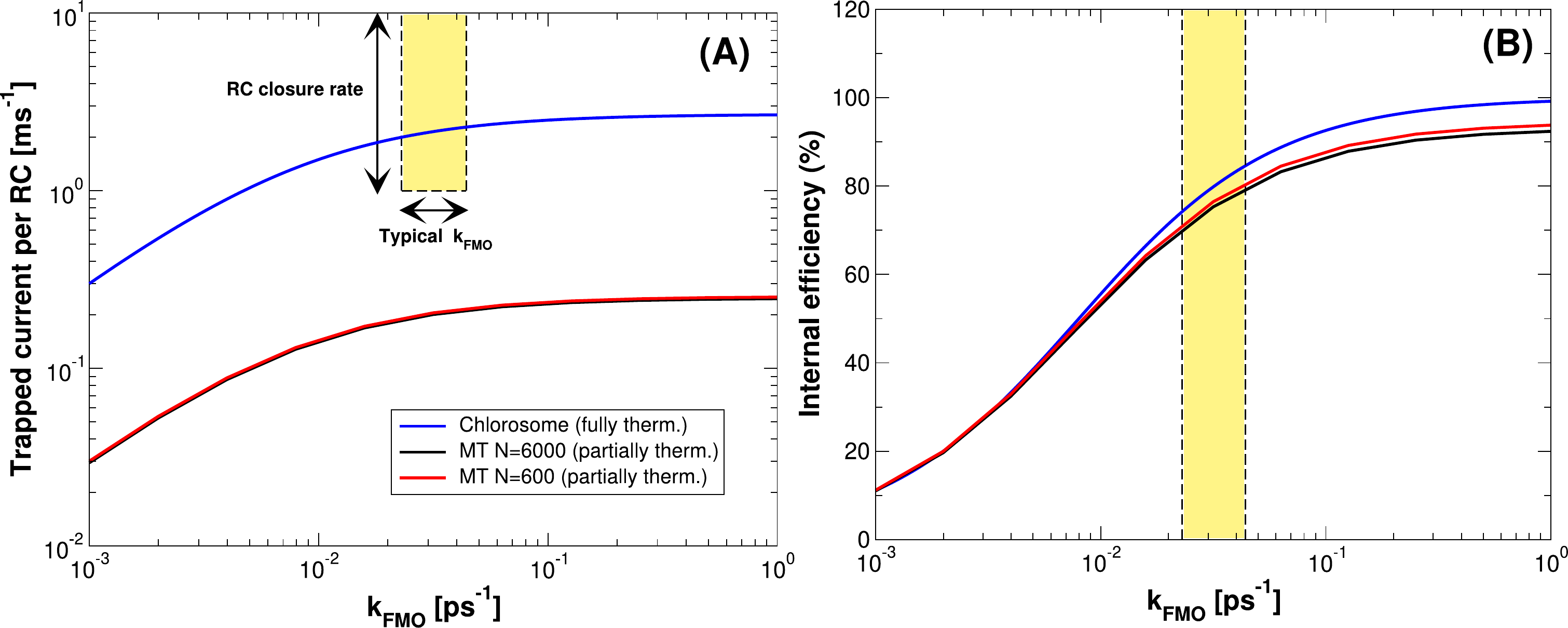}
    \caption{{\it Study of the size-dependence of the energy transfer and trapped current in GSB light-harvesting complexes.} The figure shows the trapped current per RC (panel A) and the internal efficiency (panel B) as a function of the $k_{FMO}$ trapping rate. The yellow windows represent the regime where GSB typically work. In both panels the comparison between the entire chlorosome (blue line) and a single cylindrical MT model coupled to a baseplate has been represented. For the single MT cylinder two different sizes have been considered: the one already studied in the manuscript containing N=6000 Bchl {\it c} (black line)  and a smaller one where N=600 Bchl {\it c} molecules are arranged on a cylinder made of 10 rings (red line). For the entire chlorosome the full rate equations approach has been used to compute both trapped current and internal efficiency, while for the single MT cylinders the partially thermalized model has been employed.  
}
    \label{fig:size}
\end{figure}

\clearpage

\section{ Analysis of TDMs distribution in a cone} 
\label{SI:cone}

In the main text we clearly demonstrate, through explicit comparisons between the MT, PD and RD models, that light-harvesting performance is strongly influenced by the orientation of transition dipole moments, with the natural geometry yielding the highest trapped current and internal efficiency. 
In this section we provide a further study in order to investigate and identify a tolerance range for the $\beta$ angle. A MT cylinder with 6000 Bchl \textit{c} coupled to a baseplate with 2184 Bchl \textit{a} molecules has been considered. In this model the $\beta$ angle is no longer kept constant, but randomly and uniformly distributed in a cone around the main direction of the original TDMs. Fig.~\ref{fig:cone} shows how the solid angle cone aperture ($\delta$) (from zero to $4\pi$) affects both the internal efficiency and the trapped current. For each $\delta$ angle the average values of trapped current and internal efficiency have been determined over 10 realizations. The obtained results show that both internal efficiency and trapped current start to drop as the cone aperture increases, reaching the RD model for  $\delta > 2\pi$ (see the area between the two black dashed lines which represents the range values obtained with the RD model). Specifically the trapped current is  decreased  by  $4\%$ and the efficiency by the $1.6\%$ for $\delta=\pi (2-\sqrt{2})$. These findings suggest that while the system is optimized for the $\beta$ close to $55^\circ$, it maintains a degree of robustness against moderate angular fluctuations. 

\begin{figure}[ht!]
    \centering
    \includegraphics[width=\linewidth]{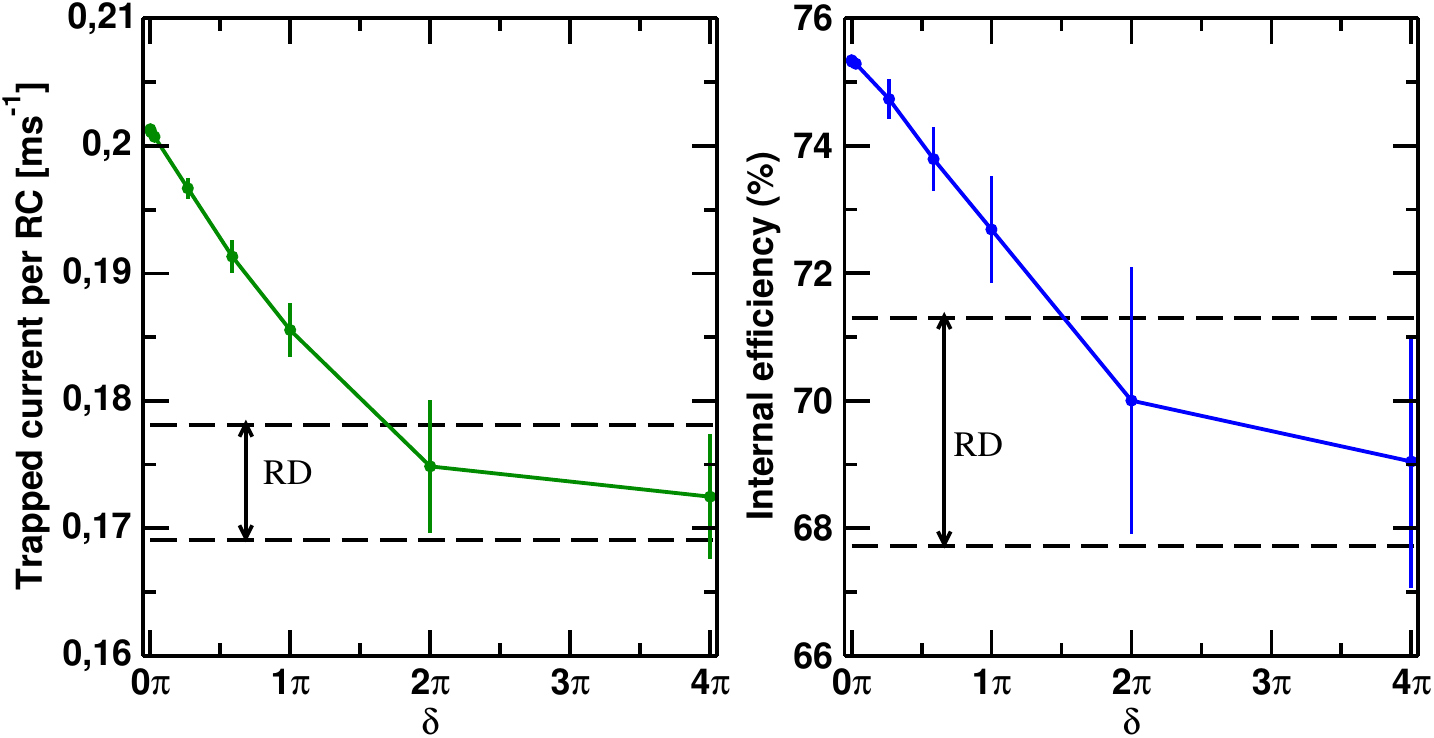}
    \caption{{\it Analysis of trapped current and internal efficiency assuming a TDMs distribution in a cone.} The figure shows the trapped current (panel A) and the internal efficiency (panel B) as a function of the cone aperture $\delta$ for a system comprising a single cylinder (MT model) with 6000 Bchl \textit{c} coupled to a baseplate with 2184 Bchl \textit{a} molecules.    The cone aperture is defined by the solid angle $0\le\delta \le 4\pi$. Specifically, the case $\delta=0$ corresponds to the MT model, where all the TDMs form  an angle $\beta=55^\circ$ with respect to the cylinder axis. Conversely, $\delta=4\pi$  represents a uniform random distribution within a sphere and it corresponds to the RD model. For each $\delta$ angle the average values of trapped current and internal efficiency have been determined over 10 realizations. The two dashed black lines in both panel A and B represent the range values obtained with the RD model.}  
    \label{fig:cone}
\end{figure}

\clearpage

\section{Comparison between MT and WT models}
\label{SI:comp}

GSB light-harvesting systems studied in the main text belong to the family of the mutant type model, obtained by genetic modifications of the system found in nature, which is the wild type model (WT). Experimental studies reveal that the WT chlorosomes are much more heterogeneous than chlorosomes from the MT model, furthermore  the half-band width of the $Q_y$ absorption maximum of the BChl {\it c} aggregates is largest for WT chlorosomes isolated from cells grown at low light intensity and it is also much broader than for chlorosomes of the MT model~\cite{ganapathy2009,gunther2016}.

In this section we propose a comparison between the mutant type (MT) and the wild type (WT)  models for the GSB light-harvesting single-wall nanotubes.  In this manuscript the geometry of the WT and MT models has been determined starting from the 2-dimensional Bravais lattice and wrapping it up according to two different rolling vectors, which are mutually perpendicular, see Ref.~\citenum{gunther2016} for a description of the Bravais lattice. The result is that in the MT model BChls are organized into equal, horizontal and coaxial rings, while in the WT model BChls are organized into vertical chains, originating a helical structure. A brief explanation of the geometry of the WT model is provided in Fig.~\ref{fig:WTmodel}, while a wider explanation of the geometry of all the single-walled models can be found in Ref.~\citenum{macroscopic}.   Here we compare the trapped current and internal efficiency for both MT and WT models, computed by using the partially thermalized rate equations approach described in Sec.~\ref{model2} in the main text. Fig.~\ref{fig:comp} shows the trapped current (panel A) and the internal efficiency (panel B) as a function of the FMO trapping rate. Our foundings demonstrate that MT and WT models have a similar behavior: both of them show a trapped current between $1-3 \times 10^{-1} \ \mbox{ms}^{-1}$ and an internal efficiency between $70-80\%$, see the yellow box that represents the typical FMO trapping rate range. These results confirm that  the realistic models (both WT and MT) exploit excitation energy transfer efficiently and they are able to funnel almost all the absorbed excitation  to the RCs with higher performance than the other  mathematical models.

\begin{figure}[ht!]
    \centering
    \includegraphics[scale=0.2]{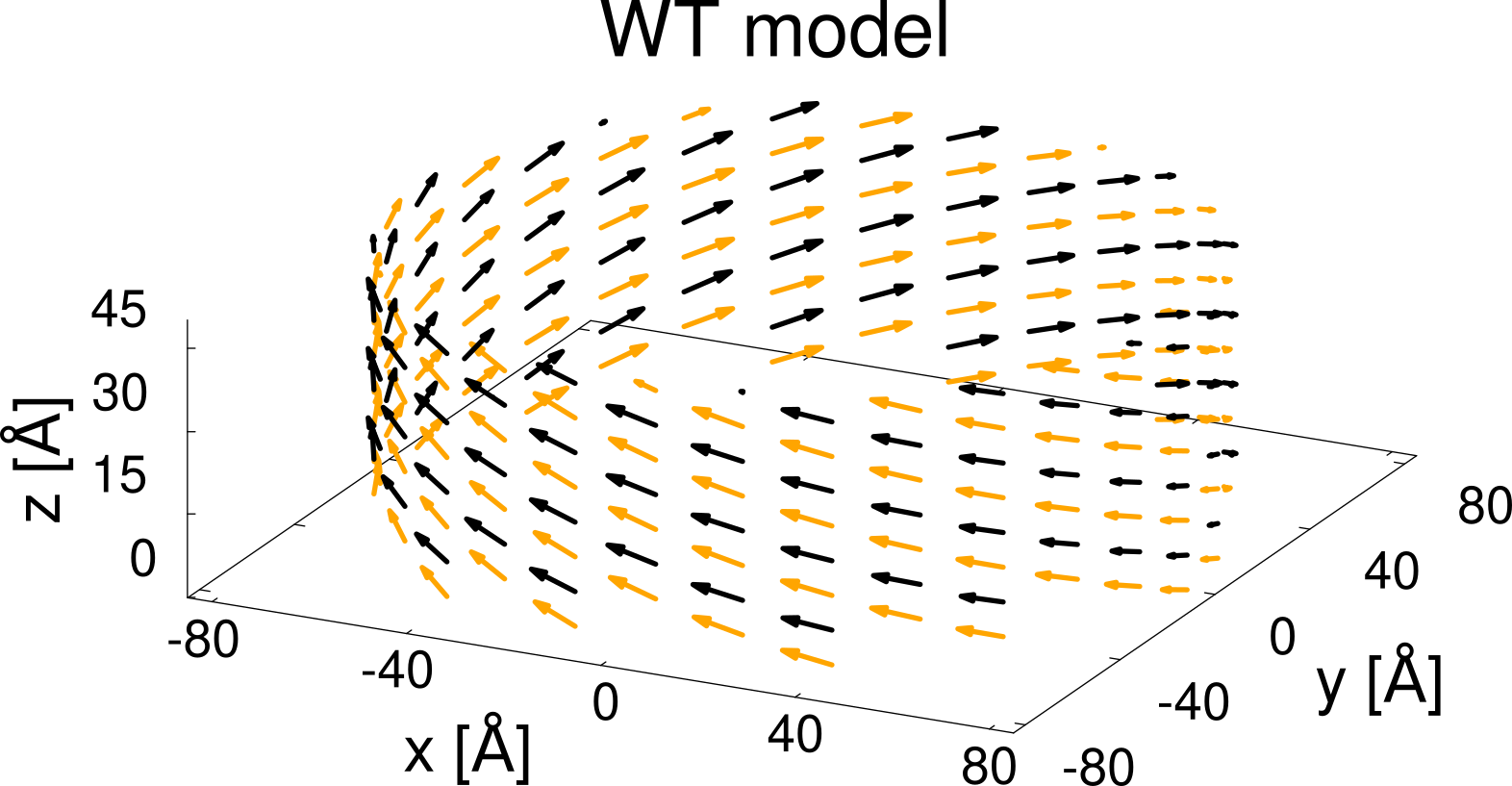}
    \caption{{\it Geometry of the wild type (WT) model.} The figure represents a section of the wild type antenna complex. Positions and orientations of the TDMs associated to each BChl {\it c} molecule are represented by orange and black arrows. For the sake of clarity we show only 30 dipoles per ring instead of 60 as we considered in this paper. Moreover the distances along the z-axis are enhanced by a factor of 5 with respect to the distances on the $x–y$ axes. The WT model can be thought as organized into vertical chains to originate a helical structure. Also in the WT model there is the alternation $\pm 4^\circ$  between consecutive dipoles on the same chain, here represented by the alternation between black and orange arrows. For more details about the structure of the WT model see Ref.~\citenum{macroscopic}. }
    \label{fig:WTmodel}
\end{figure}

\begin{figure}[ht!]
    \centering
    \includegraphics[width=\columnwidth]{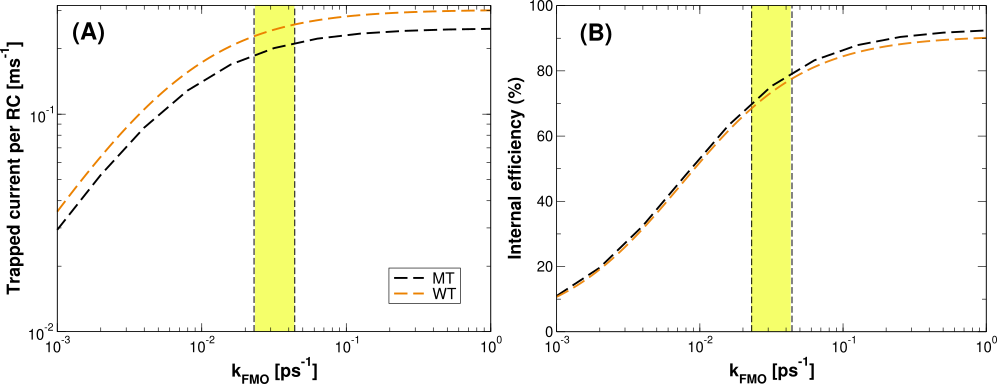}
    \caption{{\it Comparison between WT and MT models.} Trapped current (panel A) and internal efficiency (panel B) as a function of the $k_{FMO}$ trapping rate for MT and WT models coupled to a dimeric baseplate. The results have been obtained with the partially thermalized model using Eqs.~\eqref{curr3} and~\eqref{eff3} for the trapped current and the internal efficiency, respectively. For the MT model we consider a cylinder made of $6000$ BChl {\it c} molecules and a dimeric baseplate containing $2184$ BChl {\it a} molecules. For the WT model a cylinder with $6000$ BChl {\it c} molecules and a dimeric baseplate containing $1608$ BChl {\it a} molecules have been considered. More details about the size of the aggregates and the number of BChl molecules are provided in Tab.~\ref{table_size}. The yellow window between the two dashed lines represent the region where the $k_{FMO}$ trapping rate typically works in GSB species.}
    \label{fig:comp}
\end{figure}

\end{suppinfo}

\end{document}